\documentclass[seceq]{jpsj3}
\usepackage{times}
\usepackage{amsmath}
\usepackage{amssymb}
\usepackage{bm}% bold math

\def\runauthor{Fusayoshi J. \textsc{Ohkawa}}
\begin{document}
\title{Resonating-Valence-Bond Liquid in Low Dimensions}
\author{Fusayoshi J. \textsc{Ohkawa}\hskip-1pt\thanks{E-mail address: fohkawa@mail.sci.hokudai.ac.jp}\hskip-1pt\thanks{Professor emeritus}}
\inst{Division of Physics, Hokkaido University, Sapporo 060-0810, Japan}
\recdate{November 26, 2013; accpeted September 3, 2014; published online November 11, 2014}
\abst{
The Hubbard model in $D$ dimensions, with the on-site repulsion $U$ and the transfer integral between nearest neighbors $-t/\sqrt{D}$, is studied on the basis of the Kondo-lattice theory.
If $U/|t|\hskip-2pt\gg\hskip-2pt 1$, $|n\hskip-1pt -\hskip-1pt 1| \hskip-2pt\lesssim\hskip-2pt |t|/(DU)$, where $n$ is the number of electrons per unit cell, and $D$ is so small that $|J|/D \hskip-2pt\gg\hskip-2pt k_{\rm B}T_c$, where $J\hskip-2pt=\hskip-2pt -4t^2/U$ and $T_c$ is $0\hskip2pt{\rm K}$ for $D\hskip-2pt=\hskip-2pt 1$ and is the highest critical temperature among possible ones for $D\hskip-2pt\ge\hskip-2pt 2$, a low-$T$ phase where $T_c\hskip-2pt<\hskip-2pt T\hskip-2pt \ll\hskip-2pt |J|/(k_{\rm B}D)$ is a frustrated electron liquid.
Since the liquid is stabilized by the Kondo effect in conjunction with the resonating-valence-bond (RVB) mechanism, it is simply the RVB electron liquid; in one dimension, it is also the Tomonaga-Luttinger liquid.
The Kondo energy of the RVB liquid is $k_{\rm B}T_{\rm K}\hskip-2pt=\hskip-2pt O(|J|/D)$; its effective Fermi energy is $O(k_{\rm B}T_{\rm K})$.
A midband appears on the chemical potential between the upper and lower Hubbard bands; the Hubbard gap is a pseudogap. 
As regards the density of states per unit cell of the midband,
its bandwidth is $O(k_{\rm B}T_{\rm K})$ or $O(|J|/D)$, its peak height is $O(1/U)$, and its spectral weight is $O\bigl[t^2/(DU^2)\bigr]$.
Since the midband almost disappears in the Heisenberg limit,
the RVB electron liquid in the Heisenberg limit is simply the RVB spin liquid.
The RVB electron and spin liquids adiabatically continue to each other. 
Since local moments form in a high-$T$ phase where $T\hskip-1pt\gtrsim\hskip-1pt T_{\rm K}$, the high-$T$ phase is simply the Mott insulator.
}
%
%

%%%%%%%%%%%%%%%%%%%%%%%%%%%%%%%%%%%%%%%%
\maketitle
\section{Introduction}
\label{SecIntroduction}
A strong electron correlation is one of the most important issues in condensed-matter physics. 
The Hubbard model is one of the simplest effective Hamiltonians for studying the correlation.
The Hubbard model in the Heisenberg limit is equivalent to the Heisenberg model.
In the Heisenberg model on the triangular lattice, the resonating-valence-bond (RVB) mechanism is crucial for the stabilization of a frustrated spin liquid, in which no symmetry is broken; the frustrated spin liquid is simply the RVB spin liquid.\cite{fazekas}
The RVB mechanism is crucial for not only the triangular lattice but also other types of lattice in sufficiently low dimensions, and in not only the Heisenberg model but also the Hubbard model.\cite{Anderson-SC}
If no symmetry is broken in a strongly correlated electron liquid, the electron liquid is frustrated, as the RVB spin liquid.
If a frustrated electron liquid is stabilized by the RVB mechanism, it is simply an RVB electron liquid.
It is interesting to study how crucial the RVB mechanism is for the stabilization of a frustrated electron liquid in low dimensions, particularly the Tomonaga-Luttinger (TL) liquid in one dimension.\cite{Luttinger-liq,solyom}

The Hubbard model in one dimension is of particular interest, because no symmetry is broken in it,\cite{mermin} and the Bethe-ansatz solution is available for it.\cite{lieb-wu}
We consider the thermodynamic limit of $N\hskip-2pt\rightarrow \hskip-2pt+\infty$ and $L\hskip-2pt\rightarrow \hskip-2pt+\infty$, where $N$ and $L$ are the numbers of electrons and unit cells, respectively.
According to the Bethe-ansatz solution,\cite{lieb-wu} if the on-site repulsion $U$ is nonzero once, a gap opens in the spectrum of a single-particle excitation in the half-filled ground state with $N\hskip-2pt=\hskip-2pt L$; however, no gap opens in any non-half-filled ground state with $N\hskip-2pt\ne\hskip-2pt L$.
When $T\hskip-2pt=\hskip-2pt 0\hskip2pt$K, if and only if $N\hskip-2pt =\hskip-2pt L$, a metal-insulator (MI) transition occurs at $U\hskip-2pt =\hskip-2pt 0$ as a function of $U$.
Since no symmetry is broken, the MI transition is never due to a broken symmetry. 
Lieb and Wu argue that the MI transition at $U\hskip-2pt =\hskip-2pt 0$ is the Mott transition.\cite{lieb-wu}
On the other hand, it is expected 
that the Mott transition will be possible only at a sufficiently large $U$ such that $U$ is as large as the bandwidth, or $U\hskip-2pt=\hskip-2pt O(|t|)$, where $|t|$ is the strength of the transfer integral between nearest neighbors, because it is due to the strong electron correlation. \cite{mott,hubbard1,hubbard3,brinkman}

If $U/|t|\hskip-1pt\gg\hskip-1pt 1$, the gap $\epsilon_{\rm G}(U)$ given by the Bethe-ansatz solution is as large as the Hubbard gap;\cite{hubbard1,hubbard3} $\epsilon_G(U)\hskip-1pt =\hskip-1pt U\hskip-1pt -\hskip-1pt O(|t|)$.
However, $\epsilon_{\rm G}(U)$ is so singular at $U\hskip-1pt =\hskip-1pt 0$ as a function of $U$ that it cannot be expanded in terms of $U$, as the ground-state energy\cite{takahashi}.
If $U/|t|\hskip-1pt \ll\hskip-1pt 1$, the gap is extremely small.
It is doubtful whether the extremely small gap for $U/|t|\hskip-1pt\ll \hskip-1pt 1$ is the Hubbard gap. Therefore, it is doubtful whether the MI transition at $U=0$ in one dimension is the Mott transition.

Unless $U/|t|\hskip-2pt=\hskip-2pt +\infty$, the residual entropy per unit cell is zero or infinitesimal, depending on $N$ and $L$,\cite{ComKramers} in the thermodynamic limit; the third law of thermodynamics holds for any finite $U/|t|$, even for the insulating ground state with $N\hskip-2pt =\hskip-2pt L$.
On the other hand, if $U/|t|\hskip-2pt=\hskip-2pt +\infty$, the half-filled ground state is the prototype of the Mott insulator in not only one dimension but also higher dimensions; its residual entropy is $k_{\rm B} \ln 2$ per unit cell.
According to Brinkman-Rice theory,\cite{brinkman} the Mott transition occurs at $U_c\hskip-2pt=\hskip-2pt O(|t|)$ and the $T$-linear specific-heat coefficient diverges as $U\hskip-2pt\rightarrow\hskip-2pt U_c\hskip-2pt-\hskip-2pt0$. If no symmetry is broken even for $U\hskip-2pt\ge\hskip-2pt U_c$, the divergence means that the residual entropy per unit cell is nonzero for $U\hskip-2pt\ge\hskip-2pt U_c$, or for the insulating ground state.
In a previous paper,\cite{FJO-MottIns} it is argued that if the ground state is the Mott insulator with no symmetry broken, the third law of thermodynamics is broken in it. 
In the present paper, the insulating ground state in one dimension is called Lieb and Wu's insulator in order to distinguish it from the Mott insulator.

The number of electrons $N$ is a natural variable in the canonical ensemble.
If $N\hskip-2pt = \hskip-2pt L$, the ground state is Lieb and Wu's insulator;
if $N\hskip-2pt\ne \hskip-2pt L$, it is the TL liquid.\cite{Luttinger-liq,solyom}
On the other hand, the chemical potential $\mu$ is a natural variable in the grand canonical ensemble. The average number of electrons per unit cell as a function of $\mu$ is denoted by $n(\mu)$; it corresponds to $N/L$. 
The Bethe-ansatz solution is given for the Hubbard model on a bipartite lattice, which is symmetric under the particle-hole transformation for a particular chemical potential $\mu_0$;
$n(\mu_0)\hskip-2pt =\hskip-2pt 1$ for any $T$.
If $|\mu\hskip-2pt -\hskip-2pt \mu_0|\hskip-2pt <\hskip-2pt (1/2)\epsilon_{\rm G}(U)$, it is expected that the ground state will be Lieb and Wu's insulator and $n(\mu)\hskip-3pt =\hskip-3pt 1$ for it.
If $|\mu\hskip-2pt -\hskip-2pt \mu_0|\hskip-2pt\ge \hskip-2pt(1/2)\epsilon_{\rm G}(U)$, the ground state is the TL liquid; and $n(\mu)\hskip-2pt\ne \hskip-2pt 1$ for any $T\hskip-2pt\ge\hskip-2pt0\hskip2pt$K.
If $T\hskip-2pt >\hskip-2pt 0\hskip2pt$K once, even if $n(\mu)\hskip-2pt = \hskip-2pt 1$, metallic configurations with $N\hskip-2pt\ne\hskip-2pt L$ more or less contribute to physical properties statistically averaged in the grand canonical ensemble; thus, in-gap states have to appear even if the ground state is an insulator. 
An insulator in which a complete gap opens is possible only on the line of $T=0\hskip2pt$K and $|\mu\hskip-1pt -\hskip-1pt \mu_0|\hskip-2pt <\hskip-2pt (1/2)\epsilon_{\rm G}(U)$ in the $T$-$\mu$ phase plane. 
If $T$ is sufficiently low and $|\mu\hskip-1pt -\hskip-1pt \mu_0|\hskip-2pt \ge\hskip-2pt (1/2)\epsilon_{\rm G}(U)$, the TL liquid is stabilized.
The phase in the region of $T\hskip-2pt >\hskip-2pt 0\hskip2pt$K and $|\mu\hskip-1pt -\hskip-1pt \mu_0|\hskip-2pt <\hskip-2pt (1/2)\epsilon_{\rm G}(U)$ is an {\it intermediate} phase.
It is interesting to study whether the {\it intermediate} phase
is insulating or metallic, and how crucial the RVB mechanism is for the stabilization of the {\it intermediate} phase as well as the TL liquid.

All the single-single site terms are rigorously considered in the supreme single-site approximation (S$^3$A).\cite{Metzner,Muller-H1,Muller-H2,Janis}
The Kondo-lattice theory (KLT) is a perturbative theory based on S$^3$A to include multisite terms \cite{Mapping-1,Mapping-2,Mapping-3}
and is also a $1/D$ expansion theory, where $D$ is the dimensionality.
The RVB mechanism is a multisite effect and is a higher-order effect in $1/D$.
The present paper is an extension of previous study;\cite{FJO-MottIns,toyama} the RVB mechanism is explicitly considered on the basis of KLT. 
The main purpose of the present paper is to show that the RVB liquid can be stabilized in sufficiently low dimensions owing to the Kondo effect in conjunction with the RVB mechanism; the RVB liquid in one dimension is also the TL liquid, or the RVB-TL liquid. 
The present paper is organized as follows:
Preliminary is given in Sect.\hskip2pt\ref{SecPreliminary}.
KLT is reviewed in Sect.\hskip2pt\ref{SecKLtheory}.
The RVB liquid is studied on the basis of KLT in Sect.\hskip2pt\ref{SecRVB}.
Discussion is given in Sect.\hskip2pt\ref{SecDiscussion}.
Conclusions are given in Sect.\hskip2pt\ref{SecConclusion}.
A sum rule is proved in Appendix\ref{AppSumRule}.
An application of the sum rule is given in Appendix\ref{AppAppSumRule}.
An equality is proved in Appendix\ref{AppEq}. 
It is examined in Appendix\ref{AppDMFT} whether, if symmetry breaking is ignored, S$^3$A is rigorous in the limit $1/D\rightarrow0$.

\section{Preliminary}
\label{SecPreliminary}
\subsection{Hubbard model}
\label{SecHubbardModel}
We consider the Hubbard model in $D$ dimensions, i.e., on a chain, square, cubic, or hypercubic lattice:
\begin{align}\label{EqHubbardModel}
\mathcal{H} = 
\epsilon_d \hskip-1pt \sum_{i\sigma}n_{i\sigma}
- \frac{t}{\sqrt{D}} \hskip-2pt \sum_{\left<ij\right>\sigma}
\hskip-3pt d_{i\sigma}^\dag d_{j\sigma}^{\phantom{\dag}}
\hskip-1pt + U \hskip-1pt \sum_{i}n_{i\uparrow}n_{i\downarrow},
\end{align}
where $d_{i\sigma}^\dag$ and $d_{i\sigma}^{\phantom{\dag}}$ are the creation and annihilation operators of an electron with spin $\sigma$ on the $i$th unit cell, $n_{i\sigma}\hskip-2pt=\hskip-2pt d_{i\sigma}^\dag d_{i\sigma}^{\phantom{\dag}}$, $\epsilon_d$ is the band center, $-t/\sqrt{D}$ is the transfer integral between nearest neighbors and $t\hskip-2pt >\hskip-2pt0$ is assumed, 
$\left<ij\right>$ stands for a pair of nearest neighbors, and $U$ is the on-site repulsion.
The number of unit cells is $L$.
The thermodynamic limit $L\hskip-2pt\rightarrow\hskip-2pt+\infty$ is assumed.
The periodic boundary condition is assumed.
When $U\hskip-2pt=\hskip-2pt0$, the dispersion relation of an electron is given by
\begin{align}\label{EqEk}
E({\bm k}) =\epsilon_d -2t\varphi_D({\bm k}),
\end{align}
where ${\bm k}=(k_1,k_2, \cdots, k_D)$ is the wave number, 
and
\begin{align}\label{EqVarPhiD}
\varphi_D({\bm k})=\frac1{\sqrt{D}}\sum_{\nu=1}^{D} \cos(k_\nu a),
\end{align}
where $a$ is the lattice constant.
Because of the factor $1/\sqrt{D}$ in the transfer integral, 
the effective bandwidth of $E({\bm k})$ is $O(|t|)$ for any $D$; the absolute bandwidth is $4\sqrt{D}|t|$.

The number of electrons per unit cell is given by
$n(\mu)\hskip-2pt=\hskip-2pt\bigl<\mathcal{N}\bigr>\hskip-1pt/L$, 
where $\mathcal{N}\hskip-2pt =\hskip-2pt\sum_{i\sigma}\hskip-1pt d_{i\sigma}^\dag d_{i\sigma}^{\phantom{\dag}}$ and $\left<\cdots\right>$ stands for the thermal average in the grand canonical ensemble.
If the chemical potential $\mu$ lies at
\begin{align}\label{EqMu0}
\mu_0 = \epsilon_d+(1/2)U,
\end{align}
the Hubbard model is symmetric and half-filled; $n(\mu_0)\hskip-2pt=\hskip-2pt 1$ for any $T$.

%%%%%%%%%%%%%%%%%%%%%%%%%%%%%%%%%%%%%%%%%%%%%%%%%%%%%%%%%%%%%%
\subsection{Bethe-ansatz solution for one dimension}
\label{SecOneDimension}
\subsubsection{Effect of $O(1/L)$ due to electron correlation}
\label{SecO(1/L)}
We consider one dimension.
Since $\left[\mathcal{H},\mathcal{N}\right]\hskip-2pt =\hskip-2pt 0$,
a many-body eigenstate is specified by the number of electrons $N$:
$\mathcal{N} \left|N\alpha \right>\hskip-2pt =\hskip-2pt N \left|N\alpha \right>$
and
$\mathcal{H} \left|N\alpha \right>\hskip-2pt =\hskip-2pt E_{N\alpha} \left|N\alpha \right>$,
where $\alpha$ is a set of quantum numbers.
In the thermodynamic limit, an infinite number of bosonic excitations, which do not change $N$, are possible, so that
many-body eigenvalues $E_{N\alpha}$ are continuously distributed in the range of
\begin{align}\label{EqEa-Region}
E_{N\mathrm{g}}\le E_{N\alpha}<+\infty,
\end{align}
where $E_{N\mathrm{g}}$ is the energy of the ground state or the ground-state multiplet, or $\left|N\mathrm{g} \right>$.
There is no upper limit for $E_{N\alpha}$.

Many-body eigenstates in the presence of a thermal and/or electron reservoir are more or less different from those in the absence of it. We assume that this difference can be ignored.

We assume that $T=0\hskip2pt$K in the canonical ensemble.
The retarded Green function in the site representation is given by
\begin{align}
R_{ij\sigma}^{(N)}(\varepsilon +{i}0) &=
\frac1{v_N} \sum_{\mathrm{g}} R_{ij\sigma}^{(N\mathrm{g})}(\varepsilon+{i}0), 
\end{align}
where the summation is over the ground-state multiplet and $v_{N}$ is its degeneracy, and
\begin{align}\label{EqZij} 
\hskip-10pt
R_{ij\sigma}^{(\hskip-0.5pt N\alpha)}\hskip-1pt (\hskip-0.5pt z\hskip-0.5pt ) \hskip-2pt &= \hskip-2pt
\sum_{\beta}\hskip-1pt \Biggl\{\hskip-2pt
\frac{\hskip-1.5pt
\bigl<\hskip-1pt N\alpha\big|a_{i\sigma}^{\phantom{\dag}}\big|( \hskip-0.5pt N
\hskip-2pt + \hskip-2pt 1 \hskip-0.5pt)\beta\bigr>\hskip-0.5pt
\bigl<\hskip-1pt(\hskip-0.5pt N \hskip-2pt + \hskip-2pt 1 \hskip-0.5pt )\beta \big|a_{j\sigma}^\dag\big|N\alpha\bigr>}
{z- \left[E_{(N+1)\beta} -E_{N\alpha}\right]}
%%%
\nonumber \\ & \hskip0pt 
+ \hskip-2pt \frac{
\bigl<\hskip-1pt N\alpha\big|a_{j\sigma}^{\dag}\big|(N \hskip-2pt - \hskip-2pt 1)\beta\bigr>\hskip-0.5pt
\bigl<\hskip-1pt(N \hskip-2pt - \hskip-2pt 1)\beta \big|a_{i\sigma}^{\phantom{\dag}}\big|N\alpha\bigr>}
{z - \left[E_{N\alpha} \hskip-1pt - \hskip-1pt E_{(N-1)\beta}\right]}
\hskip-2.5pt \Biggr\}.
\end{align}
The site-diagonal Green function is given by
\begin{align}
R_{\sigma}^{(N)}(\varepsilon+{i}0) &=
\frac1{v_N}\sum_{\mathrm{g}}\int_{-\infty}^{+\infty}\hskip-10pt d\varepsilon^\prime
\frac{D_{N\mathrm{g}}(\varepsilon^\prime)}{\varepsilon+{i}0 -\varepsilon^\prime},
\end{align}
\begin{align}\label{EqDNA1}
\hskip-5pt
D_{N\alpha}(\hskip-0.5pt \varepsilon\hskip-0.5pt) & \hskip-2pt = \hskip-2pt
\sum_{\beta}\hskip-1pt 
\left|\hskip-1pt\bigl< \hskip-1.5pt
(\hskip-0.5pt N \hskip-2pt + \hskip-2pt 1\hskip-0.5pt )\beta\right|
\hskip-2pt a_{i\sigma}^\dag\hskip-2pt
\left|N \alpha\bigr>\hskip-1pt\right|^2 \hskip-2pt
\delta\hskip-2pt\left[ \varepsilon \hskip-2pt - \hskip-2pt
E_{(\hskip-0.5pt N \hskip-1pt + \hskip-1pt 1\hskip-0.5pt)\beta} \hskip-2pt + \hskip-2pt E_{N\alpha}\right] \hskip-2pt 
\nonumber \\ & \hskip-15pt
+ \hskip-2pt\sum_{\beta}\hskip-1pt
\left|\hskip-1pt\bigl< \hskip-1.5pt 
(\hskip-0.5pt N \hskip-2pt - \hskip-2pt 1\hskip-0.5pt)\beta\right|\hskip-2pt a_{i\sigma}^{\phantom{\dag}} \hskip-2pt 
\left|N \alpha\bigr>\hskip-1pt \right|^2 \hskip-2pt
\delta\hskip-2pt \left[\varepsilon \hskip-2pt - \hskip-2pt E_{N\alpha} \hskip-2pt + \hskip-2pt 
E_{(\hskip-0.5pt N\hskip-1pt -\hskip-1pt 1\hskip-0.5pt)\beta}\right] \hskip-2pt .
\end{align}
Neither $R_{\sigma}^{(N)}(\varepsilon+{i}0)$ nor $D_{N\alpha}(\varepsilon)$ depends on the unit cell. 
The density of states per unit cell is given by
\begin{align}\label{EqRhoCanonical}
\rho_N(\varepsilon) & \hskip-1pt =\hskip-1pt
- \frac1{\pi} {\rm Im} R_{\sigma}^{(N)}(\varepsilon \hskip-1pt +\hskip-1pt {i}0)
\hskip-1pt=\hskip-1pt \frac1{v_N}\hskip-1pt\sum_{\mathrm{g}}\hskip-1pt D_{N\mathrm{g}}(\varepsilon).
\end{align}

Two types of {\it Fermi level} are defined:
$\mu_{+}(N)\hskip-2pt =\hskip-2pt E_{(N+1)\mathrm{g}}\hskip-2pt-\hskip-2pt E_{N\mathrm{g}}$ 
for the addition of an electron
and $\mu_{-}(N)\hskip-2pt=\hskip-2pt E_{N\mathrm{g}}\hskip-2pt-\hskip-2pt E_{(N-1)\mathrm{g}}$
for the removal of an electron.
According to the Bethe-ansatz solution,\cite{lieb-wu} 
if $N\ne L$, $\bigl[\mu_{+}(N)-\mu_{-}(N)\bigr]\rightarrow 0$ as $L\rightarrow+\infty$. 
No gap opens in $\rho_{N\ne L}(\varepsilon)$. The ground state is the TL liquid for any non-half filling.\cite{Luttinger-liq,solyom}
Either $\rho_{N\ne L}(\varepsilon)$ or $D_{N\ne L\hskip1pt\alpha}(\varepsilon)$ is more or less nonzero for any $\varepsilon$.\cite{comWidth}

On the other hand, if and only if $N=L$, 
\begin{align}\label{EqLiebWuGap}
\epsilon_{\rm G}(U) &= \mu_{+}(L)- \mu_{-}(L)
\end{align}
is nonzero even in the limit $L\hskip-2pt\rightarrow\hskip-2pt+\infty$; 
%\begin{align}
$\mu_{\pm}(L)\hskip-2pt=\hskip-2pt\mu_0\hskip-2pt \pm\hskip-2pt (1/2)\epsilon_{\rm G}(U)$.
%\end{align}
Then, $\epsilon_{\rm G}(U)$ is simply a gap in $\rho_L(\varepsilon)$.
The half-filled ground state is Lieb and Wu's insulator. 
It is easy to see that
$\rho_{L}(\varepsilon)\hskip-2pt =\hskip-2pt 0$ and $D_{L\mathrm{g}}(\varepsilon)\hskip-2pt =\hskip-2pt 0$ for $|\varepsilon\hskip-2pt-\hskip-2pt\mu_{0}|\hskip-2pt<\hskip-2pt(1/2)\epsilon_{\rm G}(U)$.
We define $\Delta E_{L\alpha}\hskip-2pt=\hskip-2pt E_{L\alpha}\hskip-2pt -\hskip-2pt E_{L\mathrm{g}}$.
If $\Delta E_{L\alpha}\hskip-2pt <\hskip-2pt (1/2)\epsilon_{\rm G}(U)$, then $D_{L\alpha}(\varepsilon)\hskip-2pt=\hskip-2pt0$ for $|\varepsilon-\mu_0|\hskip-2pt<\hskip-2pt(1/2)\epsilon_{\rm G}(U)\hskip-1pt-\hskip-1pt \Delta E_{L\alpha}$, but $D_{L\alpha}(\varepsilon)\hskip-2pt>\hskip-2pt0$ for $|\varepsilon\hskip-1pt-\hskip-1pt\mu_0|\ge(1/2)\epsilon_{\rm G}(U)\hskip-2pt-\hskip-2pt \Delta E_{L\alpha}$.
If $\Delta E_{L\alpha}\hskip-2pt\ge\hskip-2pt (1/2)\epsilon_{\rm G}(U)$, then $D_{L\alpha}(\varepsilon)\hskip-2pt>\hskip-2pt0$ for any $\varepsilon$.
Thus, $D_{L\alpha}(\varepsilon)\hskip-2pt>\hskip-2pt 0$ at least for certain $\alpha$'s, regardless of $\varepsilon$.

As a function of $N$, $\rho_{N}(\varepsilon)$ is {\it discontinuous} between $N=L$ and $N=L\pm 1$, or between Lieb and Wu's insulator and the TL liquid.
This {\it discontinuous} behavior is different from the {\it continuous} behavior between Wilson's band insulator and metal.
The opening of $\epsilon_{\rm G}(U)$ and the {\it discontinuous} behavior of $\rho_{N}(\varepsilon)$ are different appearances of an effect due to the difference in electron correlation between $N=L$ and $N=L\pm 1$.
The effect is an effect of $O(1/N)$ or $O(1/L)$ due to electron correlation.

%%%%%%%%%%%%%%%%%%%%%%%%%%%%%%%%%%%%%%%%%%%%%%%%%%%%%%%%%%%
\subsubsection{Rigidity of Lieb and Wu's insulator}
\label{SecRigidity}
If many-body eigenstates of the Bethe-ansatz solution
are used, the thermal Green function is given by
\begin{align}\label{EqRjj}
& 
R_{ij\sigma}({i}\varepsilon_l) = 
\sum_{N\alpha}p_{N\alpha}(T) 
R_{ij\sigma}^{(N\alpha)}({i}\varepsilon_l+\mu) ,
\end{align}
where $\varepsilon_l=(2l+1)\pi k_{\rm B}T$, with $l$ being an integer, $R_{ij\sigma}^{(N\alpha)}(z)$ is defined by Eq.\hskip2pt(\ref{EqZij}), $\mu$ is the chemical potential, and
%$p_{N\alpha}(T)$ is the probability function defined by
%
\begin{align}\label{EqpN}
p_{N\alpha}(T) =
\frac{\exp[-(E_{N\alpha}-\mu N)/(k_{\rm B}T)]}
{\displaystyle \sum_{N^\prime\alpha^\prime}\exp[-(E_{N^\prime\alpha^\prime}-\mu N^\prime)/(k_{\rm B}T)]}. 
\end{align}
The site-diagonal thermal Green function is given by
\begin{align}\label{EqSiteDiagonalR}
R_{\sigma}({i}\varepsilon_l) &=
\sum_{N\alpha}p_{N\alpha}(T) \hskip-3pt
\int_{-\infty}^{+\infty} \hskip-10pt d\varepsilon^\prime
\frac{D_{N\alpha}(\varepsilon^\prime) }{{i}\varepsilon_l+\mu - \varepsilon^\prime}.
\end{align}
%
%which does not depend on the unit-cell index.
The density of states per unit cell is given by
\begin{align}\label{EqRho}
\hskip-5pt
\rho_\mu(\varepsilon) & \hskip-2pt = \hskip-2pt
- \frac1{\pi}{\rm Im} R_{\sigma}(\varepsilon \hskip-2pt + \hskip-2pt {i}0)
%\nonumber \\ &
\hskip-2pt=\hskip-2pt
\sum_{N \alpha}p_{N\alpha}(T)D_{N\alpha}(\varepsilon \hskip-2pt +\hskip-2pt \mu).
\end{align}
The average number of electrons per unit cell is given by
\begin{align}\label{EqNumber2}
n(\mu) &=
\frac1{L}\sum_{N\alpha}p_{N\alpha}(T)N
%\nonumber \\ &
= \hskip-2pt\int_{-\infty}^{+\infty} \hskip-10pt d\varepsilon 
\frac{\rho_\mu(\varepsilon)}
{e^{\varepsilon/(k_{\rm B}T)} \hskip-1pt + \hskip-1pt 1}.
%f_+(\varepsilon) \rho_\mu(\varepsilon),
\end{align}

We assume that $|\mu-\mu_{0}|< (1/2)\epsilon_{\rm G}(U)$. If $T=0\hskip2pt$K, 
\begin{align}\label{EqRhoC-G}
\rho_\mu(\varepsilon)=\rho_L(\mu+\varepsilon).
\end{align} 
Then, $\rho_\mu(\varepsilon)\hskip-2pt =\hskip-2pt 0$ for $|\varepsilon \hskip-1pt+\hskip-1pt\mu \hskip-1pt-\hskip-1pt\mu_0|\hskip-2pt < \hskip-2pt(1/2)\epsilon_{\rm G}(U)$. 
The ground state is Lieb and Wu's insulator. It follows that
%Then,
%
%
\begin{align}\label{EqR-Lieb-Wu}
R_\sigma(\varepsilon+{i}0) &=
-\frac1{\pi} \int_{-\infty}^{+\infty}\hskip-8pt 
d\varepsilon^\prime
\frac{\rho_L(\mu+\varepsilon^\prime)}{\varepsilon-\varepsilon^\prime+{i}0}.
\end{align}
Since $\rho_L(\varepsilon)$ does not depend on $\mu$, neither $\rho_\mu(\varepsilon-\mu)$ nor $R_\sigma(\varepsilon-\mu+{i}0)$ depends on $\mu$. Lieb and Wu's insulator is {\it rigid} against the movement of $\mu$,\cite{comRigidity} as Wilson's band insulator.
This is simply because many-body eigenstates in the absence of a reservoir are used and the reservoir effect is only implicitly treated through the statistical average or the probability function $p_{N\alpha}(T)$.
The relevance of this treatment is critically examined in Sect.\hskip2pt\ref{SecNatureLW}.

The static homogeneous charge susceptibility or the compressibility is given by
\begin{align}\label{EqHS-ChargeS}
\chi_c(0,0) = dn(\mu)/d\mu. 
\end{align}
Because of the {\it rigidity}, $\chi_c(0,0)=0$ for $|\mu \hskip-1pt-\hskip-1pt\mu_0|\hskip-2pt < \hskip-2pt(1/2)\epsilon_{\rm G}(U)$, or for Lieb and Wu's insulator.

%%%%%%%%%%%%%%%%%%%%%%%%%%%%%%%%%%%%%%%%%%%%%%%%%%%%%%%%%%%
\subsubsection{Discontinuous MI transition at $T=0\hskip2pt${\rm K}}
\label{SecMI-transiiton}
We assume that $T\hskip-2pt =0\hskip-2pt\hskip2pt$K in the grand canonical ensemble. An MI transition occurs at $\mu\hskip-2pt=\hskip-2pt\mu_0\hskip-1pt \pm\hskip-1pt (1/2)\epsilon_{\rm G}(U)$ as a function of $\mu$.
If $|\mu\hskip-1pt -\hskip-1pt \mu_{0}|\hskip-2pt<\hskip-2pt (1/2)\epsilon_{\rm G}(U)$,
the ground state is Lieb and Wu's insulator and no Fermi {\it surface} exists. Since no density wave appears, no folding of the Brillouin zone occurs. The absence of the Fermi {\it surface} means that its {\it volume} or size is absolutely zero in either the particle or hole picture.\cite{comFermiSF}
If $|\mu\hskip-1pt -\hskip-1pt \mu_{0}|\hskip-2pt>\hskip-2pt (1/2)\epsilon_{\rm G}(U)$,
the ground state is the TL liquid and the Fermi {\it surface} exists. According to the Fermi-surface sum rule,\cite{Luttinger1,Luttinger2} the sizes of the Fermi {\it surface} are $2|k_{\rm F}|\hskip-2pt=\hskip-2pt(\pi/a)n(\mu)$ and $2|k_{\rm F}|\hskip-2pt =\hskip-2pt (\pi/a)[2\hskip-1pt -\hskip-1pt n(\mu)]$ in the electron and hole pictures, respectively, where $k_{\rm F}$ is the Fermi wave number in each picture.
For example, if $\mu\hskip-2pt=\hskip-2pt\mu_0 \hskip-1pt+\hskip-1pt (1/2)\epsilon_{\rm G}(U)\hskip-1pt -\hskip-1pt 0^+$, the size of the Fermi {\it surface} is zero in either picture and $\rho_\mu(\varepsilon)\hskip-2pt=\hskip-2pt0$ for $-\epsilon_{\rm G}(U)\hskip-2pt<\hskip-2pt\varepsilon\hskip-2pt <\hskip-2pt 0$.
If $\mu\hskip-2pt=\hskip-2pt\mu_0\hskip-1pt+\hskip-1pt (1/2)\epsilon_{\rm G}(U)+0^+$, the size of the Fermi {\it surface} is $2|k_{\rm F}|\hskip-2pt=\hskip-2pt\pi/a$ in either picture and $\rho_\mu(\varepsilon)\hskip-2pt>\hskip-2pt0$ even for $-\epsilon_{\rm G}(U)\hskip-2pt<\hskip-2pt\varepsilon\hskip-2pt<\hskip-2pt 0$.
These discontinuous behaviors mean that the MI transition is discontinuous.

The compressibility $\chi_c(0,0)$ is also discontinuous at $\mu\hskip-2pt =\hskip-2pt \mu_0\hskip-2pt \pm\hskip-2pt (1/2)\epsilon_{\rm G}(U)$ at least in the limit $U/|t|\hskip-2pt \rightarrow\hskip-2pt +\infty$, as studied below.
As preliminary, we consider many-body eigenstates in the limit $U/|t|\hskip-2pt \rightarrow\hskip-2pt +\infty$.
If $N\hskip-2pt \le\hskip-2pt L$, no double occupancy at a unit cell is allowed.
Every many-body eigenstate is specified by the set of $N$ pairs of wave number and spin, $\{k\sigma\}\hskip-2pt =$ $\{k_1\sigma_1,$ $k_2\sigma_2$, $\cdots$, $k_N\sigma_N\}$.
Its eigenfunction is the direct product of the charge part, which is
a single Slater determinant of noninteracting $N$ spinless fermions, and the spin part, which is a product of spin functions of noninteracting $N$ spins:
%
%\begin{subequations}
\begin{align}
&\hskip-5pt\left|N\hskip-2pt\left\{k\sigma\right\}\hskip-1pt\right> \hskip-2pt =\hskip-3pt
%\frac1{\sqrt{N!}}
\sum_{\left\{xs\right\}}\hskip-2pt
\Phi_{\left\{k\sigma\right\}}\hskip-3pt \left(\hskip-1.5pt \left\{x s\right\}\hskip-1.5pt \right)
a_{x_1\hskip-1pt s_1}^\dag a_{x_2\hskip-1pt s_3}^\dag\cdots a_{x_N\hskip-1pt s_N}^\dag \hskip-1pt\left|0\right>,
\\
& \Phi_{\left\{k\sigma\right\}}\hskip-2pt \left(\hskip-1pt
\left\{x s\right\}\hskip-1pt\right) \hskip-2pt = \hskip-2pt
\frac1{\sqrt{N!}}\hskip-1pt \sum_{P}(-1)^P 
\prod_{i=1}^{N} \hskip-1pt
\frac1{\sqrt{L}} e^{{i}k_{P_i}x_i}\hskip-2pt
\prod_{j=1}^{N} \hskip-1pt\chi_{\sigma_{j}}
\hskip-1pt (s_{j} ),
\end{align}
%\end{subequations}
%
where $\left|0\right>$ is the vacuum where no electron exists, $\left\{x\sigma\right\}\hskip-2pt=\hskip-2pt\bigl\{(x_1,s_1),(x_2,s_2),\cdots,(x_N,s_N)\bigr\}$ with $0\hskip-2pt\le\hskip-2pt x_1\hskip-2pt<\hskip-2pt x_2\hskip-2pt<\hskip-2pt \cdots\hskip-2pt <\hskip-2pt x_N\hskip-2pt\le\hskip-2pt La$ is the set of the position and spin coordinates of $N$ electrons, $P\hskip-2pt=\hskip-2pt(P_1,P_2,\cdots,P_N)$ is an $N$-permutation, $(-1)^P\hskip-2pt=\hskip-2pt1$ for an even $P$ while $(-1)^P\hskip-2pt=\hskip-2pt -1$ for an odd $P$, and $\chi_\sigma(s)$ is the spin function for $S=1/2$. Its eigenenergy is given by
$E_{\left\{k\sigma\right\}}\hskip-2pt=\hskip-2pt\sum_{i=1}^{N}E(k_i)$.
If $N\hskip-2pt=\hskip-2pt L$, the spinless fermion band is completely filled and the electron state is the prototype of the Mott insulator. If $N\hskip-2pt \ne \hskip-2ptL$, the spinless fermion band is partially filled and the electron state is an exotic metal such that a complete charge-spin separation occurs in it and its ground-state degeneracy is $2^N$.
If the hole picture is taken, 
a similar argument is possible for $N\ge L$.

We consider the limit $U/|t|\hskip-2pt \rightarrow\hskip-2pt +\infty$ in the grand canonical ensemble.
If $n(\mu)\hskip-2pt =\hskip-2pt 1$, which corresponds to $N= L$, $\chi_c(0,0)=0$.
If $n(\mu)\hskip-2pt \simeq\hskip-2pt 1$ but $n(\mu)\hskip-2pt \ne\hskip-2pt 1$, which corresponds to $N\hskip-2pt \simeq\hskip-2pt L$ but $N\hskip-2pt \ne\hskip-2pt L$, $\chi_c(0,0)\hskip-2pt \propto\hskip-2pt 1/|1-n(\mu)|$. Thus, $\chi_c(0,0)$ diverges as $n(\mu)\hskip-2pt \rightarrow\hskip-2pt 1$, i.e., $\chi_c(0,0)$ diverges as $\mu\hskip-2pt \rightarrow\hskip-2pt \mu_0\hskip-2pt \pm\hskip-2pt (1/2)\epsilon_{\rm G}(U)\pm 0$; this divergence is due to the band-edge van Hove singularity in the one-dimensional dispersion relation of the noninteracting spinless fermion.
The compressibility $\chi_c(0,0)$ is discontinuous at $\mu\hskip-2pt =\hskip-2pt \mu_0\hskip-2pt \pm\hskip-2pt (1/2)\epsilon_{\rm G}(U)$ as a function of $\mu$ at least in the limit $U/|t|\hskip-2pt \rightarrow\hskip-2pt +\infty$.

If $U/|t|$ is finite and $n(\mu)\hskip-2pt\simeq \hskip-2pt1$ but $n(\mu)\hskip-2pt\ne\hskip-2pt 1$, the Fermi {\it surface} as large as $2|k_{\rm F}|\hskip-2pt\simeq\hskip-2pt (\pi/a)$ exists in either the particle or hole picture.
The existence of such a {\it large} Fermi {\it surface} implies that $\chi_c(0,0)$ is more or less nonzero.
Moreover, the divergence of $\chi_c(0,0)$ in the limit $U/|t|\hskip-2pt \rightarrow \hskip-2pt+\infty$ implies that the increase in $\chi_c(0,0)$ occurs as $\mu\hskip-2pt \rightarrow\hskip-2pt\mu_0\hskip-2pt \pm\hskip-2pt (1/2)\epsilon_{\rm G}(U)\hskip-2pt \pm\hskip-2pt 0$ at least for a sufficiently large $U/|t|$.
It is interesting to determine whether $\chi_c(0,0)$ continuously becomes zero as $\mu\hskip-2pt \rightarrow\hskip-2pt \mu_0\hskip-2pt \pm(1/2)\hskip-2pt \epsilon_{\rm G}(U)\pm 0$ or discontinuously becomes zero at $\mu\hskip-2pt =\hskip-2pt\mu_0\hskip-2pt \pm\hskip-2pt (1/2)\epsilon_{\rm G}(U)$, and whether $\chi_c(0,0)$ increases as $\mu\hskip-2pt \rightarrow\hskip-2pt \mu_0\hskip-2pt \pm\hskip-2pt (1/2)\epsilon_{\rm G}(U)\hskip-2pt \pm\hskip-2pt 0$.

%%%%%%%%%%%%%%%%%%%%%%%%%%%%%%%%%%%%%%%%%%%%%%%%%%%%%%%%%%%
\subsubsection{In-gap states at $T>0\hskip2pt${\rm K}}
\label{SecInGapStates}
We assume that $T\hskip-2pt >\hskip-2pt 0\hskip2pt$K.
Then, $p_{N\alpha}(T)$ defined by Eq.\hskip2pt(\ref{EqpN}) is more or less nonzero.
If $N\hskip-2pt\ne\hskip-2pt L$, $D_{N\alpha}(\varepsilon)$ defined by Eq.\hskip2pt(\ref{EqDNA1}) is nonzero; even if $N\hskip-2pt=\hskip-2pt L$, $D_{L\alpha}(\varepsilon)\hskip-2pt>\hskip-2pt0$ for certain $\alpha$'s, regardless of $\varepsilon$. 
The density of states $\rho_{\mu}(\varepsilon)$ given by Eq.\hskip2pt(\ref{EqRho}) is more or less nonzero,\cite{comWidth} even for $|\varepsilon\hskip-1pt+\hskip-1pt\mu\hskip-1pt-\hskip-1pt\mu_0|\hskip-2pt<\hskip-2pt(1/2)\epsilon_{\rm G}(U)$. 
It is straightforward to show that the density of states is also more or less nonzero in the canonical ensemble.

The gap opens only in the {\it exactly} half-filled case at $T=0\hskip2pt$K.
If the filling is non-half or $T\hskip-2pt>\hskip-2pt0\hskip2pt$K once, the effect of $O(1/L)$ disappears or becomes insufficient for the gap to open, in either the canonical or grand canonical ensembles.

\section{Kondo-Lattice Theory}
\label{SecKLtheory}
\subsection{Single-site properties of the Hubbard model}
\label{SecSinglel-site}
\subsubsection{Mapping to the Anderson model}
\label{SecMapCondition}
In Sect.\hskip2pt\ref{SecKLtheory}, we review KLT,\cite{Mapping-1,Mapping-2,Mapping-3,toyama} 
and reformulate it in a form appropriate for the present study.
We assume that $T\hskip-2pt >\hskip-2pt T_c$, where $T_c$ is $0\hskip2pt$K for $D\hskip-2pt=\hskip-2pt 1$ and is the highest critical temperature among possible ones for $D\hskip-2pt\ge\hskip-2pt 2$.
If $T\hskip-2pt>\hskip-2pt T_c$, $\rho_{\mu}(\varepsilon)$ is more or less nonzero at least for $|\varepsilon|\hskip-2pt<\hskip-2pt|t|$,\cite{comWidth} and even in one dimension.
%, as studied in Sect.\hskip2pt\ref{SecInGapStates}.
{\it The fact that $\rho_{\mu}(\varepsilon)$ is necessarily nonzero for $T\hskip-2pt>\hskip-2pt T_c$ is crucial in the present study}.

{\it If \hskip1pt$T \hskip-2pt >\hskip-2pt T_c$\hskip1pt, no symmetry is broken and no gap opens; thus, there is no doubt on the validity of the perturbative treatment in terms of \hskip1pt$U$ based on the Feynman-diagram method.}\cite{ComFeynmanMethod}
We consider a connected and irreducible Feynman diagram for a physical property.
The diagram is composed of electron lines, each of which stands for $R_{ij\sigma}({i}\varepsilon_l)$, and interaction lines, each of which stands for $U$.
The site-diagonal $R_{ii\sigma}({i}\varepsilon_l)$ is simply denoted by $R_{\sigma}({i}\varepsilon_l)$.
If only site-diagonal $R_\sigma({i}\varepsilon_l)$'s appear in the diagram, it is a single-site diagram; 
if at least a site-off-diagonal $R_{i\ne j\hskip1pt\sigma}({i}\varepsilon_l)$ appears in the diagram, it is a multisite diagram.
The diagram can be classified into a single-site or multisite diagram.
The physical property is decomposed into the single-site term, which is the sum of all the single-site diagrams, and the multisite term, which is the sum of all the multisite diagrams. 

The self-energy in the site representation is defined by
\begin{align}\label{EqBetheSal}
R_{ij\sigma}({i}\varepsilon_l\hskip-1pt) &
\hskip-1pt =\hskip-1pt
R_{ij\sigma}^{(0)}\hskip-1pt({i}\varepsilon_l\hskip-0.5pt)
\hskip-1pt + \hskip-2pt \sum_{i^\prime\hskip-0.5pt j^\prime}\hskip-0.5pt
R_{ii^\prime\hskip-0.5pt\sigma}^{(0)}\hskip-0.5pt({i}\varepsilon_l\hskip-0.5pt)
\Sigma_{i^\prime\hskip-0.5pt j^\prime\hskip-0.5pt\sigma}({i}\varepsilon_l\hskip-0.5pt)R_{j^\prime\hskip-0.5pt j\sigma}({i}\varepsilon_l\hskip-0.5pt), 
\end{align}
where $\Sigma_{ij\sigma}({i}\varepsilon_l)$ is the self-energy and 
\begin{align}
R_{ij\sigma}^{(0)}({i}\varepsilon_l) &=
\frac1{L}\sum_{\bm k}e^{{i}{\bm k}\cdot({\bm R}_i-{\bm R}_{j})}
\frac1{{i}\varepsilon_l +\mu -E({\bm k})}
\end{align}
is the Green function for $U=0$; ${\bm R}_i$ is the position of the $i$th unit cell.
The self-energy $\Sigma_{ij\sigma}({i}\varepsilon_l)$ is decomposed into the single-site $\delta_{ij}\Sigma_\sigma({i}\varepsilon_l)$ and the multisite $\Delta\Sigma_{ij\sigma}({i}\varepsilon_l)$:
\begin{align}\label{EqSelf-Rrep}
\Sigma_{ij\sigma}({i}\varepsilon_l) =
\delta_{ij}\Sigma_\sigma({i}\varepsilon_l) + \Delta\Sigma_{ij\sigma}({i}\varepsilon_l).
\end{align}
The single-site $\Sigma_\sigma({i}\varepsilon_l)$ does not depend on the unit cell.

The site-diagonal $R_{\sigma}({i}\varepsilon_l)$ and the single-site $\Sigma_\sigma({i}\varepsilon_l)$ are local properties. It is possible to map them to their corresponding local properties of an appropriate impurity model. The appropriate impurity model is the Anderson model.\cite{Mapping-1,Mapping-2,Mapping-3}

We consider the Anderson model defined by
\begin{align}\label{EqAndersonModel}
\tilde{\mathcal{H}}&=
\tilde{\epsilon}_d \sum_{\sigma}n_{d\sigma}
+ \sum_{{\bm k}\sigma}E_c({\bm k})c_{{\bm k}\sigma}^\dag c_{{\bm k}\sigma}^{\phantom{\dag}}+ \tilde{U} n_{d\uparrow}n_{d\downarrow}
\nonumber \\ & \quad 
+ \frac1{\tilde{L}}\sum_{{\bm k}\sigma}\left(
V_{\bm k} c_{{\bm k}\sigma}^\dag d_\sigma^{\phantom{\dag}}
+V_{\bm k}^* d_{\sigma}^\dag c_{{\bm k}\sigma}^{\phantom{\dag}}
\right),
\end{align}
where $n_{d\sigma}\hskip-2pt=\hskip-2pt d_{\sigma}^\dag d_{\sigma}^{\phantom{\dag}}$, $\tilde{\epsilon}_d$ is the level of $d$ electrons, $E_c({\bm k})$ is the dispersion relation of conduction electrons, $\tilde{U}$ is the on-site repulsion, $\tilde{L}$ is the number of unit cells, and $V_{\bm k}$ is the hybridization matrix between conduction and $d$ electrons. 
In the present paper, the temperature of the reservoir for the Anderson model is denoted by $\tilde{T}$ and treated as a parameter independent of $T$ for the Hubbard model.
The Green function for $d$ electrons is given by
%
%\begin{subequations}
\begin{align}\label{EqAndersonH}
&\tilde{G}_\sigma({i}\tilde{\varepsilon}_l) =
\frac1{\displaystyle 
{i}\tilde{\varepsilon}_l +\tilde{\mu} - \tilde{\epsilon}_d
-\tilde{\Sigma}_\sigma({i}\tilde{\varepsilon}_l) 
- \tilde{\Gamma}({i}\tilde{\varepsilon}_l)},
\\ \label{EqAndersonGam}
& \hskip35pt
\tilde{\Gamma}({i}\tilde{\varepsilon}_l) =
\frac1{\pi} \int_{-\infty}^{+\infty} \hskip-10pt d\varepsilon^\prime
\frac{\Delta(\varepsilon^\prime)}{{i}\tilde{\varepsilon}_l - \varepsilon^\prime} , 
\\ \label{EqAndersonD}
& \Delta(\varepsilon) 
\hskip-2pt = \hskip-2pt
-{\rm Im}\tilde{\Gamma}(\varepsilon \hskip-2pt + \hskip-2pt {i}0)
\hskip-2pt=\hskip-2pt
\frac{\pi}{\tilde{L}}\hskip-1pt
 \sum_{\bm k}\left|V_{\bm k}\right|^2
\hskip-1pt
\delta\bigl[\varepsilon \hskip-2pt + \hskip-2pt 
\tilde{\mu} \hskip-2pt - \hskip-2pt E_c({\bm k})\bigr],
\end{align}
%\end{subequations}
where $\tilde{\varepsilon}_l=(2l+1)\pi k_{\rm B}\tilde{T}$, with $l$ being an integer, and 
$\tilde{\Sigma}_\sigma({i}\tilde{\varepsilon}_l)$ is the self-energy for $d$ electrons. 
The Anderson model is essentially uniquely characterized by the four {\it parameters} $\tilde{T}$, $\tilde{U}$, $\tilde{\epsilon}_d-\tilde{\mu}$, and $\Delta(\varepsilon)$; none of the arbitrariness of $\tilde{\epsilon}_d$, $\tilde{\mu}$, $V_{\bm k}$, or $E_c({\bm k})$ is crucial. 
The four {\it parameters} have to be determined to satisfy an appropriate mapping condition. 

Since $\tilde{\Sigma}_\sigma({i}\tilde{\varepsilon}_l)$ is a local term, only $\tilde{U}$ and $\tilde{G}_\sigma({i}\tilde{\varepsilon}_l)$ appear in any Feynman diagram for $\tilde{\Sigma}_\sigma({i}\tilde{\varepsilon}_l)$ of the Anderson model;
only $U$ and $R_{\sigma}({i}\varepsilon_l)$ appear in any Feynman diagram for the single-site $\Sigma_\sigma({i}\varepsilon_l)$ of the Hubbard model.
Provided that
\begin{subequations}\label{EqMap10}
\begin{align} \label{EqMap11} 
%\varepsilon_l =\tilde{\varepsilon}_l \quad & \mbox{or}\quad 
& 
T =\tilde{T}, \hskip10pt
%\\ & \hskip15pt \label{EqMap12} 
U =\tilde{U}, 
\\ \label{EqMap13} &
R_{\sigma}({i}\varepsilon_l) = \tilde{G}_\sigma({i}\tilde{\varepsilon}_l) 
\end{align}
\end{subequations}
are satisfied, it immediately follows that
\begin{align}\label{EqMapSigma}
\Sigma_\sigma({i}\varepsilon_l) = \tilde{\Sigma}_\sigma({i}\tilde{\varepsilon}_l).
\end{align}
Equation\hskip2pt(\ref{EqMap10}) is the appropriate mapping condition. At least
\begin{align}\label{EqMapEpsilon}
\epsilon_d- \mu = \tilde{\epsilon}_d-\tilde{\mu}
\end{align}
has to be satisfied in order that Eq.\hskip2pt(\ref{EqMap13}) can be satisfied.
%According to Eqs.\hskip2pt(\ref{EqMap11}) and (\ref{EqMap13}), 
Then, it immediately follows that\cite{Mapping-3}
\begin{align}\label{EqMapDelta0}
\Delta(\varepsilon) &=
{\rm Im} \left[\tilde{\Sigma}_\sigma(\varepsilon+{i}0) + 1/R_{\sigma}(\varepsilon+{i}0) \right].
\end{align}
If all of Eqs.\hskip2pt(\ref{EqMap11}) and (\ref{EqMapEpsilon}), and 
\begin{align}\label{EqMapDelta}
\Delta(\varepsilon) &=
{\rm Im} \left[\Sigma_\sigma(\varepsilon+{i}0) + 1/R_{\sigma}(\varepsilon+{i}0)\right]
\end{align}
are satisfied, Eqs.\hskip2pt(\ref{EqMapSigma}) and (\ref{EqMapDelta0}) are satisfied. 
We can substitute Eq.\hskip2pt(\ref{EqMapDelta}) for Eq.\hskip2pt(\ref{EqMapDelta0}).
The set of Eqs.\hskip2pt(\ref{EqMap11}), (\ref{EqMapEpsilon}), and (\ref{EqMapDelta}) is the mapping condition; Eq.\hskip2pt(\ref{EqMapDelta}) is a practical mapping condition because the others are simple. 

It should be noted that $\Delta(\varepsilon)$ given by Eq.\hskip2pt(\ref{EqMapDelta}) depends on the temperature $T$ of the reservoir for the Hubbard model.
The mapped Anderson model itself includes $T$ as a {\it parameter}.

{\it There is no doubt on the possibility that, if the Hubbard model is solved once, the four parameters $\tilde{T}$, $\tilde{U}$, $\tilde{\epsilon}_d\hskip-1pt-\hskip-1pt\tilde{\mu}$, and $\Delta(\varepsilon)$ of the Anderson model can be uniquely determined from the mapping condition.}
All pairs of corresponding properties are exactly equal to each other between the Hubbard and Anderson models; e.g.,
$\Sigma_\sigma(\varepsilon\hskip-1pt+\hskip-1pt{i}0)\hskip-1pt =\hskip-1pt \tilde{\Sigma}_\sigma(\varepsilon\hskip-1pt+\hskip-1pt{i}0)$,
$R_{\sigma}(\varepsilon\hskip-1pt+\hskip-1pt{i}0)\hskip-1pt =\hskip-1pt \tilde{G}_\sigma(\varepsilon\hskip-1pt+\hskip-1pt{i}0)$,
$n(\mu)\hskip-1pt=\hskip-1pt\tilde{n}(\tilde{\mu})$,
$\rho_\mu(\varepsilon)\hskip-1pt =\hskip-1pt \tilde{\rho}(\varepsilon)$, 
and so on, where $\tilde{n}(\tilde{\mu})\hskip-1pt =\hskip-1pt \bigl<n_{d\uparrow}\hskip-1pt +\hskip-1pt n_{d\downarrow}\bigr>$ and 
\begin{align}\label{EqRhoAnderson}
\tilde{\rho}(\varepsilon) = - (1/\pi){\rm Im} \tilde{G}_\sigma(\varepsilon+{i}0).
\end{align}
If $\mu=\mu_0$, the Anderson model is also symmetric.

The Green function for the Hubbard model in the wave-number representation is given by
\begin{align}\label{EqGreenK}
G_{\sigma}({i}\varepsilon_l,{\bm k}) &=
\frac1{L}\sum_{ij}e^{-{i}{\bm k}\cdot({\bm R}_i-{\bm R}_{j})} R_{ij\sigma}({i}\varepsilon_l)
\nonumber \\ 
\phantom{G_{\sigma}({i}\varepsilon_l,{\bm k})} &=
\frac1{{i}\varepsilon_l +\mu- E({\bm k})- \Sigma_{\sigma}({i}\varepsilon_l,{\bm k})},
\\ 
\Sigma_{\sigma}({i}\varepsilon_l,{\bm k}) & =
\Sigma_\sigma({i}\varepsilon_l)+
\Delta\Sigma_{\sigma}({i}\varepsilon_l,{\bm k}),
\\ 
\Delta\Sigma_{\sigma}({i}\varepsilon_l, {\bm k}) &=
\frac1{L} \sum_{ij}
e^{-{i}{\bm k}\cdot({\bm R}_i -{\bm R}_{j})}
\Delta\Sigma_{ij\sigma}({i}\varepsilon_l).
\end{align}
Here, $\Sigma_{\sigma}({i}\varepsilon_l,{\bm k})$, $\Sigma_\sigma({i}\varepsilon_l)$, and $\Delta\Sigma_{\sigma}({i}\varepsilon_l, {\bm k})$ are the total, single-site, and multisite self-energies, respectively.

The theory reviewed and reformulated above is KLT.\cite{Mapping-1,Mapping-2,Mapping-3,toyama}
Multisite terms have to be self-consistently considered with the Anderson model to be mapped.
If no multisite term is considered in KLT, it is reduced to S$^3$A.\cite{Metzner,Muller-H1,Muller-H2,Janis} 
Either the dynamical mean-field theory \cite{georges,RevModDMFT} (DMFT)
or the dynamical coherent-potential approximation \cite{dcpa} (DCPA) is also S$^3$A.

\subsubsection{Nonzero and finite $\Delta(\varepsilon)$ of the Anderson model}
\label{SecProof}
The purpose of Sect.\hskip2pt\ref{SecProof} is to show that if $T>T_c$, 
\begin{align}\label{EqNonzeroFiniteDelta}
0<\Delta(\varepsilon)<+\infty
\end{align}
has to be satisfied in any self-consistent solution of KLT, in which $\rho_\mu(\varepsilon)\hskip-2pt >\hskip-2pt 0$ is necessarily satisfied.
We refer to a previous paper.\cite{toyama} We define two real functions:
\begin{subequations}\label{EqSSYZ}
\begin{align}
\label{EqYn} & \hskip10pt
Y_n(\varepsilon) = \frac1{L} \sum_{\bm k}
\frac{S_1^n(\varepsilon,{\bm k} )}
{S_1^2(\varepsilon,{\bm k}) +S_2^2(\varepsilon,{\bm k})},
\\ \label{EqZn} & \hskip10pt
Z_n(\varepsilon) = \frac1{L} \sum_{\bm k}
\frac{S_2^n(\varepsilon,{\bm k} )}
{S_1^2(\varepsilon,{\bm k}) +S_2^2(\varepsilon,{\bm k})},
\end{align}
\end{subequations}
where $S_1(\varepsilon,{\bm k})\hskip-2pt =\hskip-2pt {\rm Re}\hskip1pt [1/G_\sigma(\varepsilon\hskip-2pt +\hskip-2pt {i}0,{\bm k})]$ and $S_2(\varepsilon,{\bm k})\hskip-2pt =\hskip-2pt {\rm Im}\hskip1pt [1/G_\sigma(\varepsilon\hskip-2pt +\hskip-2pt {i}0,{\bm k})]$.
It follows that
%\begin{subequations}
\begin{align}
\label{EqYZ0} & \hskip28pt
Y_0(\varepsilon) = Z_0(\varepsilon),
\\ \label{EqRYZ1} & 
R_{\sigma}(\varepsilon + {i}0) 
%= \frac1{L} \sum_{\bm k} G_\sigma(\varepsilon +{i}0,{\bm k}) 
= Y_1(\varepsilon) - {i}Z_1(\varepsilon),
\\ \label{EqYZ1} & \hskip20pt
Z_1(\varepsilon) = \pi \rho_\mu(\varepsilon)>0,
\\ \label{EqYZ2} & \hskip25pt
Y_2(\varepsilon) + Z_2(\varepsilon)=1.
\end{align}
%\end{subequations}
%
Since either of
\begin{subequations}\label{EqInequality}
\begin{align}\label{EqInequalityY}
\frac1{L} \hskip-2pt \sum_{\bm k}\hskip-2pt
\frac{\left[x\hskip-1pt +\hskip-1pt S_1(\varepsilon,{\bm k})\right]^2}
{S_1^2(\varepsilon,\hskip-1pt{\bm k}) \hskip-2pt +\hskip-2pt S_2^2(\varepsilon,\hskip-1pt{\bm k})} 
&\hskip-1pt=\hskip-1pt 
Y_0(\varepsilon) x^2 \hskip-2pt + \hskip-2pt 2 Y_1
(\varepsilon) x \hskip-2pt +\hskip-2pt Y_2(\varepsilon),
\\ \label{EqInequalityZ}
\frac1{L} \hskip-2pt \sum_{\bm k}\hskip-2pt
\frac{\left[x\hskip-1pt +\hskip-1pt S_2(\varepsilon,{\bm k})\right]^2}
{S_1^2(\varepsilon,\hskip-1pt{\bm k}) \hskip-2pt +\hskip-2pt S_2^2(\varepsilon,\hskip-1pt{\bm k})} 
&\hskip-1pt=\hskip-1pt 
Y_0(\varepsilon) x^2 \hskip-2pt + \hskip-2pt 2 Z_1
(\varepsilon) x \hskip-2pt +\hskip-2pt Z_2(\varepsilon),
\end{align}
\end{subequations}
cannot be negative for any real $x$,\cite{comInequlaities} it follows that 
\begin{subequations}\label{EqYZ1P}
\begin{align}\label{EqY1P}
&-Y_1^2(\varepsilon) + Y_0(\varepsilon)Y_2(\varepsilon) \ge 0,
\\ \label{EqZ1P}
&-Z_1^2(\varepsilon) + Z_0(\varepsilon)Z_2(\varepsilon) \ge 0.
\end{align}
\end{subequations}
The perturbative analysis in terms of $U$ is useful for $U/|t|\hskip-2pt \ll\hskip-2pt 1$; that in terms of $J\hskip-2pt =\hskip-2pt -4t^2/U$ based on KLT is useful for $U/|t|\hskip-2pt \gg\hskip-2pt 1$, as demonstrated in Sect.\hskip2pt\ref{SecRVB}.
In either analysis, it is easy to find terms that give a nonzero contribution to ${\rm Im}\Sigma_\sigma(\varepsilon\hskip-1pt+\hskip-1pt {i}0)$ or ${\rm Im}\Delta\Sigma_\sigma(\varepsilon\hskip-1pt+\hskip-1pt {i}0,{\bm k})$, provided that $T\hskip-2pt >\hskip-2pt 0 \hskip2pt$K and $\rho_\mu(\varepsilon)\hskip-2pt>\hskip-2pt0$.
If $T\hskip-1pt >\hskip-1pt T_c$, then $\rho_\mu(\varepsilon)\hskip-1pt >\hskip-1pt 0$, so that 
\begin{align}\label{EqS2>0}
0<-{\rm Im}\Sigma_\sigma(\varepsilon+{i}0) 
< -{\rm Im}\Sigma_\sigma(\varepsilon+{i}0,{\bm k}).
\end{align}
Then,
$Y_1(\varepsilon)$ is finite and $Z_1(\varepsilon)$ is nonzero and finite, so that
\begin{align}\label{EqY1Z1}
0< Y_1^2(\varepsilon) + Z_1^2(\varepsilon) <+\infty.
\end{align}

First, we show that $\Delta(\varepsilon)$ cannot be positively divergent.
If Eq.\hskip2pt(\ref{EqRYZ1}) is used, 
Eq.\hskip2pt(\ref{EqMapDelta}) can be described as 
\begin{align}\label{EqDelta1}
\Delta(\varepsilon) & \hskip-1pt=\hskip-1pt 
{\rm Im}\Sigma_\sigma(\varepsilon\hskip-1pt+\hskip-1pt{i}0) 
\hskip-1pt+\hskip-1pt 
Z_1(\varepsilon)/\bigl[Y_1^2(\varepsilon) 
\hskip-1pt+\hskip-1pt Z_1^2(\varepsilon)\bigr].
\end{align}
In general, ${\rm Im}\Sigma_\sigma(\varepsilon\hskip-1pt+{i}0)\hskip-2pt\le \hskip-2pt 0$.
According to Eq.\hskip2pt(\ref{EqYZ1}) or (\ref{EqY1Z1}), 
$Z_1(\varepsilon)/\bigl[Y_1^2(\varepsilon)\hskip-1pt +\hskip-1pt Z_1^2(\varepsilon)\bigr]\hskip-1pt<\hskip-1pt +\infty$. Then, 
$\Delta(\varepsilon)\hskip-2pt <\hskip-2pt +\infty$.

Next, we show that $\Delta(\varepsilon)$ has to be nonzero and positive.
If Eq.\hskip2pt(\ref{EqYZ2}) is used,
Eq.\hskip2pt(\ref{EqDelta1}) can be described as 
%
%\begin{subequations}
\begin{align}\label{EqDelta2}
& \hskip10pt
\Delta(\varepsilon) 
= X(\varepsilon)/\bigl[Y_1^2(\varepsilon) + Z_1^2(\varepsilon)\bigr],
%%%%
\\ \label{EqX}
& X(\varepsilon) = 
{\rm Im}\Sigma_\sigma( \varepsilon + {i}0 )
\left[
Y_1^2(\varepsilon) + Z_1^2(\varepsilon)\right]
\nonumber \\ & \hskip35pt
+ Z_1(\varepsilon) 
\bigl[Y_2( \varepsilon) + Z_2(\varepsilon)\bigr].
\end{align}
%\end{subequations}
%
%(\ref{EqS2}), (\ref{EqYn}), 
According to Eqs.\hskip2pt(\ref{EqZn}) and (\ref{EqS2>0}), 
\begin{align}\label{EqZ>0}
Z_1(\varepsilon) >
-{\rm Im}\Sigma_\sigma (\varepsilon+{i}0) Z_0(\varepsilon) .
\end{align}
According to Eqs.\hskip2pt(\ref{EqYZ0}), (\ref{EqX}), and (\ref{EqZ>0}), 
\begin{align}\label{EqXiLarger0}
X(\varepsilon) &>
-{\rm Im}\Sigma_\sigma(\varepsilon+{i}0)
\bigl\{\left[- Y_1^2(\varepsilon) + Y_0(\varepsilon)Y_2(\varepsilon)\right]
\nonumber \\ & \hskip20pt
+\left[ -Z_1^2(\varepsilon) + Z_0(\varepsilon)Z_2(\varepsilon)
\right]\bigr\}.
\end{align}
According to Eqs.\hskip2pt(\ref{EqYZ1P}), (\ref{EqS2>0}), and (\ref{EqXiLarger0}), it follows that
$X(\varepsilon)\hskip-2pt >\hskip-2pt 0$.
According to $X(\varepsilon)\hskip-2pt >\hskip-2pt 0$ and Eqs.\hskip2pt(\ref{EqY1Z1}) and (\ref{EqDelta2}), it follows that
$\Delta(\varepsilon)\hskip-2pt >\hskip-2pt 0$.
Thus, Eq.\hskip2pt(\ref{EqNonzeroFiniteDelta}) has to be satisfied.
Since $\Delta(\varepsilon)\hskip-2pt <\hskip-2pt +\infty$,
$\tilde{\Gamma}(\varepsilon\hskip-1pt +\hskip-1pt{i}0)$ can have no pole on the real axis in any self-consistent solution of KLT for $T\hskip-2pt >\hskip-2pt T_c$.

%%%%%%%%%%%%%%%%%%%%%%%%%%%%%%%%%%%%%%%%%%%%%%%%%%
\subsubsection{Polarization and vertex functions in the spin channel}
\label{SecPolFunc}
Here, we refer to previous papers.\cite{FJO-Qint,FJO-supJ0,FJO-supJ}
The bosonic energy is denoted by $\omega_l=2l\pi k_{\rm B}T$, with $l$ being an integer.
In the wave-number representation, the irreducible polarization function $\pi_s({i}\omega_l,{\bm q})$ in the spin channel is decomposed into the single-site $\pi_s({i}\omega_l)$ and the multisite $\Delta\pi_s({i}\omega_l,{\bm q})$:
\begin{align}
\pi_s({i}\omega_l,{\bm q}) =
\pi_s({i}\omega_l) + \Delta\pi_s({i}\omega_l,{\bm q}).
\end{align}
The single-site $\pi_s({i}\omega_l)$ is equal to the local $\tilde{\pi}_s({i}\omega_l)$ of the Anderson model:
$\pi_s({i}\omega_l)\hskip-2pt=\hskip-2pt\tilde{\pi}_s({i}\omega_l)$.
The spin susceptibilities of the Anderson and Hubbard models are given by%
%\cite{ComFactor}
%
\begin{subequations}\label{EqSusEaxct}
\begin{align}
\tilde{\chi}_s({i}\omega_l)& =
2\tilde{\pi}_s({i}\omega_l)/[1-U\tilde{\pi}_s({i}\omega_l)],
\\ 
\chi_s({i}\omega_l,{\bm q}) &=
2\pi_s({i}\omega_l,{\bm q})/[1-U\pi_s({i}\omega_l,{\bm q})],
\end{align}
\end{subequations}
respectively. Here, the conventional factor $(1/4)\mathrm{g}^2\mu_{\rm B}^2$ is not included, where $\mathrm{g}$ is the $\mathrm{g}$ factor and $\mu_{\rm B}$ is the Bohr magneton. The susceptibility $\tilde{\chi}_s({i}\omega_l)$ of the Anderson model includes no contribution from the polarization of conduction electrons.

The Kondo temperature or energy is defined by
\begin{align}\label{EqDefTK1}
\bigl[\tilde{\chi}_s(0;T)\bigr]_{\tilde{T}= 0\hskip2pt{\rm K}}&=1/\bigl[k_{\rm B}T_{\rm K}(T)\bigr],
\end{align}
where $\tilde{T}\hskip-2pt=\hskip-2pt 0\hskip2pt{\rm K}$ means that the temperature of the reservoir for the Anderson model is the absolute zero Kelvin; $T$ of the reservoir for the Hubbard model or the {\it parameter} $T$ is explicitly shown because $\tilde{\chi}_s(0;T)$ and $T_{\rm K}(T)$ depend on $T$. 
Although this definition of $T_{\rm K}$ is different from Wilson's \cite{wilsonKG} by a numerical factor, $k_{\rm B}T_{\rm K}$ is still a measure of the magnitude of stabilization energy. 
Since $\tilde{\rho}(\varepsilon)\hskip-2pt=\hskip-2pt\rho_\mu(\varepsilon)\hskip-2pt >\hskip-2pt0$ and $0\hskip-2pt <\hskip-2pt \Delta(\varepsilon)\hskip-2pt <\hskip-2pt +\infty$ for $T\hskip-2pt >\hskip-2pt T_c$, 
$k_{\rm B}T_{\rm K}\hskip-2pt >\hskip-2pt 0$ and never $k_{\rm B}T_{\rm K}\hskip-2pt =\hskip-2pt 0$ in any self-consistent solution of KLT for $T\hskip-2pt >\hskip-2pt T_c$.

The formulation so far is valid for any finite $U/|t|$.
The formulation in the following part is only valid for $U/|t|\gg 1$.
If $U/|t|\gg1$, then $k_{\rm B}T_{\rm K}/U \ll 1$;
\begin{subequations}\label{EqHigh1/U}
\begin{align}
&\tilde{\chi}_s({i}\omega_l) = O[1/(k_{\rm B}T_{\rm K})],
\\
&\chi_s({i}\omega_l,{\bm q}) = O[1/(k_{\rm B}T_{\rm K})],
\end{align}
\end{subequations}
for $T\lesssim T_{\rm K}$ and $|\omega_l| \lesssim k_{\rm B}T_{\rm K}$.
According to Eqs.\hskip2pt(\ref{EqSusEaxct}) and (\ref{EqHigh1/U}), 
$U\tilde{\pi}_s({i}\omega_l)=1+O(k_{\rm B}T_{\rm K}/U)$, 
$U\pi_s({i}\omega_l,{\bm q})=1+O(k_{\rm B}T_{\rm K}/U)$,
and $U^2\Delta \pi_s({i}\omega_l,{\bm q})=O\bigl[(k_{\rm B}T_{\rm K}/U)^0\bigr]$. Then, 
\begin{align}
& \label{EqUPiChi} \hskip20pt
U[1 - U\tilde{\pi}_s({i}\omega_l)] 
= 2/\tilde{\chi}_s({i}\omega_l),
\\ \label{EqChiKondo} &
\chi_s({i}\omega_l,{\bm q}) =
\frac{\tilde{\chi}_s({i}\omega_l)}
{1 - (1/4)I_s({i}\omega_l,{\bm q})\tilde{\chi}_s({i}\omega_l)}, 
\end{align}
for $T\lesssim T_{\rm K}$ and $|\omega_l| \lesssim k_{\rm B}T_{\rm K}$, where 
\begin{align}\label{EqExchI}
I_s({i}\omega_l,{\bm q}) &= 
2U^2\Delta\pi({i}\omega_l,{\bm q}).
\end{align}
Here, the terms of $O(k_{\rm B}T_{\rm K}/U)$ are ignored.

Equation\hskip2pt(\ref{EqChiKondo}) is consistent with the physical picture of Kondo lattices that local spin fluctuations interact with each other with an intersite exchange interaction; $I_s({i}\omega_l,{\bm q})$ is the intersite exchange interaction.
The N\'{e}el temperature $T_{\rm N}$ is determined from Eq.\hskip2pt(\ref{EqChiKondo}):
\begin{subequations}\label{EqTN-Q}
\begin{align}
T_{\rm N} = \max\bigl[T_{\rm N}({\bm q})\bigr],
\end{align}
where $T_{\rm N}({\bm q})$ as a function of ${\bm q}$ is defined by
\begin{align}
\bigl[1-(1/4)I_s(0,{\bm q})\tilde{\chi}_s(0)\bigr]_{T=T_{\rm N}({\bm q})}=0.
\end{align}
\end{subequations}
According to the definition of $T_c$, if Eq.\hskip2pt(\ref{EqTN-Q}) gives $T_{\rm N}$, $T_c\ge T_{\rm N}$; if not, $T_c\ge 0\hskip2pt$K.

The reducible and irreducible three-point vertex functions in the spin channel are decomposed into single-site and multisite terms. The single-site terms can be mapped to the local vertex functions of the mapped Anderson model.
If they are denoted by
$\tilde{\Lambda}_s({i}\varepsilon_l,{i}\varepsilon_l+{i}\omega_{l^\prime};{i}\omega_{l^\prime})$ and $\tilde{\lambda}_s({i}\varepsilon_l,{i}\varepsilon_l+{i}\omega_{l^\prime};{i}\omega_{l^\prime})$, 
\begin{align}\label{EqVertexR-IR1}
\tilde{\Lambda}_s({i}\varepsilon_l,{i}\varepsilon_l+{i}\omega_{l^\prime};{i}\omega_{l^\prime})&=
\frac{\tilde{\lambda}_s({i}\varepsilon_l,{i}\varepsilon_l+{i}\omega_{l^\prime};{i}\omega_{l^\prime})}
{1-U\tilde{\pi}_s({i}\omega_{l^\prime})}.
\end{align}
If Eq.\hskip2pt(\ref{EqUPiChi}) is used, 
\begin{align}\label{EqVertexR-IR2}
U\tilde{\lambda}_s({i}\varepsilon_l,{i}\varepsilon_l \hskip-1pt +\hskip-1pt {i}\omega_{l^\prime};{i}\omega_{l^\prime}\hskip-1pt) & \hskip-1pt = \hskip-1pt
\frac{2}{\tilde{\chi}_s({i}\omega_{l^\prime})}
\tilde{\Lambda}_s({i}\varepsilon_l,{i}\varepsilon_l \hskip-1pt +\hskip-1pt {i}\omega_{l^\prime};{i}\omega_{l^\prime}\hskip-1pt) .
\end{align}

%%%%%%%%%%%%%%%%%%%%%%%%%%%%%%%%%%%%%%%%%%%
\subsection{Perturbation scheme to include multisite terms}
\subsubsection{Unperturbed state}
\label{SecNormalFL}
In the Anderson model of Eq.\hskip2pt(\ref{EqAndersonModel}), the Fermi surface of the conduction band is defined by $\tilde{\mu}= E_{c}({\bm k}_{\rm F})$, where ${\bm k}_{\rm F}$ is the Fermi wave number. According to Eq.\hskip2pt(\ref{EqAndersonD}), $\Delta(0)>0$ is a sufficient condition for the existence of the Fermi surface. 
According to Eq.\hskip2pt(\ref{EqNonzeroFiniteDelta}), the Fermi surface exists.
Then, the ground state of the Anderson model is the normal Fermi liquid because of the Kondo effect. The normal Fermi liquid is characterized by nonzero $T_{\rm K}(T)$, which depends on $T$ of the reservoir for the Hubbard model, or the {\it parameter} $T$. 
For convenience, $\tilde{T}$ of the reservoir for the Anderson model is treated as being independent of the {\it parameter} $T$, although eventually $\tilde{T}=T$ has to be assumed. In the following part, we assume that $0\hskip2pt{\rm K}\le \tilde{T}\lesssim T_{\rm K}(T)$ and $T_c<T\ll T_{\rm K}(T)$, and the {\it parameter} $T$ is explicitly shown.

We introduce an infinitesimal Zeeman energy into the Anderson model:
%
%\begin{align}\label{EqZeeman}
$\tilde{\mathcal{H}}_Z = - 
\tilde{h}(n_{d\uparrow}-n_{d\downarrow})$,
%\end{align}
%
where $\tilde{h}=0^+$.
The self-energy can be expanded in such a way that \cite{yamada1,yamada2,shiba}
\begin{align}\label{EqExpansionAM}
\tilde{\Sigma}_\sigma(\varepsilon \hskip-2pt + \hskip-2pt {i}0;T) &\hskip-2pt =\hskip-2pt
\tilde{\Sigma}_0(T)
\hskip-2pt + \hskip-2pt 
\bigl[1 \hskip-2pt - \hskip-2pt\tilde{\phi}_1(T) \bigr] \varepsilon
\hskip-2pt + \hskip-2pt \sigma \bigl[1 \hskip-2pt - \hskip-2pt \tilde{\phi}_s(T) \bigr] \tilde{h}
\nonumber \\ & \hskip8pt 
- \hskip-1pt {i} \Bigl[\hskip-1pt \tilde{\phi}_{21}(T)\varepsilon^2
\hskip-3pt + \hskip-2pt\tilde{\phi}_{22}(T) (k_{\rm B}\tilde{T})^2\hskip-1pt \Bigr]
\hskip-1pt /\hskip-1pt
\bigl[\pi\Delta(\hskip-0.5pt 0;\hskip-1pt T)\hskip-1pt\bigr]
\nonumber \\ & \hskip8pt
- \hskip-1pt \tilde{\phi}_{2}(T) \varepsilon^2/\bigl[\pi\Delta(0;T)\bigr]
+\cdots.
\end{align}
In general, 
$\tilde{\phi}_1(T)\hskip-2pt\ge\hskip-2pt 1$, $\tilde{\phi}_s(T)\hskip-2pt\ge\hskip-2pt 1$, $\tilde{\phi}_{21}(T)\hskip-2pt\ge\hskip-2pt 0$, $\tilde{\phi}_{22}(T)\hskip-2pt\ge \hskip-2pt 0$, and $\tilde{\phi}_{2}(T)\hskip-2pt\gtreqless\hskip-2pt 0$; if $\mu\hskip-2pt=\hskip-2pt\mu_0$, $\tilde{\phi}_{2}(T)\hskip-2pt=\hskip-2pt 0$.
If $\tilde{U}/\bigl[\pi\Delta(0;T)\bigr]\hskip-2pt \gtrsim\hskip-2pt 1$ and $\tilde{n}(\tilde{\mu})\hskip-2pt\simeq\hskip-2pt 1$, $\tilde{\phi}_1(T)\hskip-2pt \gg\hskip-2pt 1$, $\tilde{\phi}_{21}(T)\hskip-2pt \gg\hskip-2pt 1$, and $\tilde{\phi}_{22}(T)\hskip-2pt \gg\hskip-2pt 1$.

For convenience, we define three ratios:
\begin{subequations}
\begin{align}\label{EqWilsonRatio}
&\tilde{W}_s(T) = \tilde{\phi}_s(T)/\tilde{\phi}_1(T), 
\\ &
\tilde{W}_{21}(T) =\tilde{\phi}_{21}(T)/\tilde{\phi}_1^2(T), 
\\ &
\tilde{W}_{22}(T) =\tilde{\phi}_{22}(T)/\tilde{\phi}_1^2(T). 
\end{align}
\end{subequations}
Any of them is $O(1)$, even if $\tilde{U}/\bigl[\pi\Delta(0;T)\bigr]\hskip-2pt \gtrsim\hskip-2pt 1$.
The ratio $\tilde{W}_s(T)$ is nothing but the Wilson ratio.\cite{wilsonKG}
If $\Delta(\varepsilon;T)$ does not depend on $\varepsilon$, $\tilde{W}_s(T)\hskip-2pt=\hskip-2pt2$ in the $s$-$d$ model or the $s$-$d$ limit of the Anderson model.\cite{wilsonKG,yamada1,yamada2} Thus, it is expected that, if $\tilde{U}/[\pi\Delta(0;T)]\hskip-2pt\gtrsim\hskip-2pt 1$ and $\tilde{n}(\tilde{\mu})\hskip-2pt\simeq\hskip-2pt 1$, then $\tilde{W}_s(T)\hskip-2pt\simeq\hskip-2pt 2$ for the mapped Anderson model, whose $\Delta(\varepsilon;T)$ depends on $\varepsilon$.

The Fermi-liquid relation is available for the normal Fermi liquid. \cite{Luttinger1,Luttinger2,yamada1,yamada2,shiba}
If $\tilde{T}= 0\hskip2pt$K and no polarization of conduction electrons occurs, %it follows that
the static susceptibility of the Anderson model is given by
\begin{align}\label{EqSusAM}
\tilde{\chi}_s(0;T) 
= 2 \tilde{\phi}_s(T)\tilde{\rho}(0;T),
\end{align}
where $\tilde{\rho}(0;T)$ is the density of states of the Anderson model.
If $\Delta(\varepsilon;T)$ depends on $\varepsilon$,
the polarization of conduction electrons occurs, in general; 
the static susceptibility of the Anderson model is approximately given by Eq.\hskip2pt(\ref{EqSusAM}). Since $T_{\rm K}(T)$ is defined by Eq.\hskip2pt(\ref{EqDefTK1}), it follows that 
\begin{align}\label{EqRhoNum1}
1/\tilde{\rho}(0; T) \simeq
2\tilde{\phi}_s(T)k_{\rm B}T_{\rm K}(T) 
\simeq 4 \tilde{\phi}_1(T) k_{\rm B}T_{\rm K}(T).
\end{align}
Since an electron liquid in the Hubbard model can be characterized by $\rho_\mu(0;T)\hskip-2pt =\hskip-2pt\tilde{\rho}(0; T)$, $\tilde{\phi}_1(T)$, and $k_{\rm B}T_{\rm K}(T)$, 
Eq.~(\ref{EqRhoNum1}) is useful in the present paper.

The bosonic energy for $\tilde{T}$ is denoted by $\tilde{\omega}_{l}=2\pi l k_{\rm B}\tilde{T}$, with $l$ being an integer.
According to the Ward relation,\cite{ward} 
\begin{align}\label{EqWard1}
\tilde{\Lambda}_s({i}\tilde{\varepsilon}_l,{i}\tilde{\varepsilon}_l\hskip-1pt +\hskip-1pt {i}\tilde{\omega}_{l^\prime};{i}\tilde{\omega}_{l^\prime};T)& \hskip-1pt =\hskip-1pt
1\hskip-1pt-\hskip-1pt \lim_{h\rightarrow 0}\frac{d\hskip2pt}{d \tilde{h}}
\hskip-1pt\sum_{\sigma}\frac{\sigma}{2} 
%\nonumber \\ & \hskip-20pt \times
\tilde{\Sigma}_\sigma({i}\tilde{\varepsilon}_l;T),
\end{align}
for $\tilde{\omega}_{l^\prime}=0$.
According to Eqs.\hskip2pt(\ref{EqExpansionAM}) and (\ref{EqWard1}),
\begin{align}\label{EqWard2}
\tilde{\Lambda}_s({i}\tilde{\varepsilon}_l,{i}\tilde{\varepsilon}_l+{i}\tilde{\omega}_{l^\prime};{i}\tilde{\omega}_{l^\prime};T)
&= \tilde{\phi}_s(T) ,
\end{align}
for $\tilde{\omega}_{l^\prime}=0$ and $|\varepsilon_l|/(k_{\rm B}T_{\rm K})\rightarrow 0$. 
According to Eqs.\hskip2pt(\ref{EqVertexR-IR2}) and (\ref{EqWard2}), 
\begin{align}\label{EqUlambda}
U\tilde{\lambda}_s({i}\tilde{\varepsilon}_{l},{i}\tilde{\varepsilon}_{l}+{i}\tilde{\omega}_{l^\prime};{i}\tilde{\omega}_{l^\prime};T)=2\tilde{\phi}_s(T)/\tilde{\chi}_s({i}\tilde{\omega}_{l^\prime};T),
\end{align}
for $\tilde{\omega}_{l^\prime}=0$ and $|\varepsilon_l|/(k_{\rm B}T_{\rm K})\rightarrow 0$.

Every single-site property depends on the {\it parameter} $T$.
If $T_c<T\ll T_{\rm K}(T)$, the {\it parameter} $T$ dependence is so small that it can be ignored, except in the case of $D=2$ and $n(\mu)\simeq 1$, as discussed in Sect.\hskip2pt\ref{SecDiscussionRVB}. 
In the following part, $\tilde{T}=T$ is assumed, and the {\it parameter} $T$ is not shown.

If Eq.\hskip2pt(\ref{EqExpansionAM}) is used, the Green function is given by
\begin{subequations}\label{EqGreenKL}
\begin{align}\label{EqGreenKL1}
& \hskip50pt
G_\sigma ({i}\varepsilon_l,{\bm k}) = (1/\tilde{\phi}_1)\mathrm{g}_\sigma ({i}\varepsilon_l,{\bm k}),
\\ \label{EqGreenKL2}
& \hskip-3pt
\mathrm{g}_\sigma (\hskip-0.5pt {i}\varepsilon_l,{\bm k}\hskip-0.5pt) 
\hskip-1.5pt = \hskip-1.5pt
\frac1{{i}\varepsilon_l \hskip-1pt + \hskip-1pt \mu^*
\hskip-2pt - \hskip-1.5pt \left[E({\bm k}) 
\hskip-1.5pt + \hskip-1.5pt \Delta\Sigma_\sigma({i}\varepsilon_l,{\bm k})\right]\hskip-2pt/\hskip-1pt\tilde{\phi}_1
\hskip-2pt - \hskip-1.5pt 
\tilde{\gamma}_{\rm K}(\hskip-0.5pt{i}\varepsilon_l\hskip-0.5pt) } ,
\\ & \label{EqGreenKL3} \hskip50pt
\mu^* = \bigl(\mu-\tilde{\Sigma}_0\bigr)/\tilde{\phi}_1,
\\ & \label{EqGreenKL4} \hskip3pt
\tilde{\gamma}_{\rm K}(\hskip-0.5pt {i}\varepsilon_l\hskip-0.5pt) \hskip-1pt = \hskip-2pt - \hskip-1pt 
{i}\frac{\varepsilon_l}{|\varepsilon_l|}
\hskip-1pt\bigl[
\tilde{W}_{21}(\hskip-0.5pt{i}\varepsilon_l\hskip-0.5pt)^2 
\hskip-2pt +\hskip-2pt \tilde{W}_{22} (\hskip-0.5pt k_{\rm B}T)^2\bigr]
\frac{\tilde{\phi}_1}{\pi\Delta(0)}.
\end{align}
\end{subequations}
In Eq.\hskip2pt(\ref{EqGreenKL2}), $\tilde{h}$ and $- \tilde{\phi}_2\varepsilon^2/\bigl[\pi\Delta(0)\bigr]$ are ignored. 
The Green function given here is accurate for $|\varepsilon_l|\ll k_{\rm B}T_{\rm K}$ and $T\ll T_{\rm K}$; it can be approximately used for $|\varepsilon_l|\lesssim k_{\rm B}T_{\rm K}$ and $T\lesssim T_{\rm K}$ with sufficient accuracy.

% for a qualitative analysis in the present paper.

If $\tilde{\phi}_1 \gg 1$, the density of states $\tilde{\rho}(\varepsilon)$ of the Anderson model has a three-peak structure with the Kondo peak between two subpeaks;
the bandwidth and spectral weight of the Kondo peak are $O(k_{\rm B}T_{\rm K})$ and $1/\tilde{\phi}_1$, respectively.\cite{yamada1,yamada2}
Since $\rho_\mu(\varepsilon)=\tilde{\rho}(\varepsilon)$, the density of states $\rho_\mu(\varepsilon)$ of the Hubbard model also has a three-peak structure with a midband between the upper and lower Hubbard bands, or within the Hubbard gap that is a pseudogap;
the bandwidth and spectral weight of the midband are also $O(k_{\rm B}T_{\rm K})$ and $1/\tilde{\phi}_1$, respectively.
In this case, the Green function given by Eq.\hskip2pt(\ref{EqGreenKL}) can only describe the midband but the upper or lower Hubbard band.

Equation\hskip2pt(\ref{EqGreenKL2}) can also be described as
\begin{align}
&\mathrm{g}_\sigma ({i}\varepsilon_l,{\bm k}) =
\mathrm{g}_\sigma^{(0)} ({i}\varepsilon_l,{\bm k})+
(1/\tilde{\phi}_1) \Delta\Sigma_\sigma({i}\varepsilon_l,{\bm k}) \mathrm{g}_\sigma ({i}\varepsilon_l,{\bm k}) ,
\\ 
&\mathrm{g}_\sigma^{(0)} ({i}\varepsilon_l,{\bm k}) =
\frac1{{i}\varepsilon_l+\mu^*- (1/\tilde{\phi}_1)E({\bm k})
- \tilde{\gamma}_{\rm K}({i}\varepsilon_l) } .
\end{align}
In KLT, $\mathrm{g}_\sigma^{(0)} ({i}\varepsilon_l,{\bm k})$ is the {\it unperturbed} Green function, which is determined using the mapped Anderson model; then, the multisite $\Delta\Sigma_\sigma({i}\varepsilon_l,{\bm k})$ of the Hubbard model has to be self-consistently calculated with the mapped Anderson model to satisfy the mapping condition given in Sect.\hskip2pt\ref{SecMapCondition}.

%%%%%%%%%%%%%%%%%%%%%%%%%%%%%%%%%%%%%%%%%%%%%%%%%%%%%%%%%%%%%%%%%
%\subsection{Intersite exchange inetractiom}
\subsubsection{Superexchange interaction}
\label{SecJs}
We assume that $U/|t|\gg 1$.
The intersite $I_s({i}\omega_l,{\bm q})$ given by Eq.\hskip2pt(\ref{EqExchI}) can be decomposed into three terms: 
\begin{align}\label{EqThreeJ}
I_s({i}\omega_l,{\bm q})= J_s({i}\omega_l,{\bm q})+ J_Q({i}\omega_l,{\bm q}) - \Lambda({i}\omega_l,{\bm q}). 
\end{align}
Here, $J_s({i}\omega_l,{\bm q})$ is the superexchange interaction, which arises from the virtual exchange of a pair excitation of an electron in the upper Hubbard band and a hole in the lower Hubbard band,\cite{FJO-supJ0,FJO-supJ} and $J_Q({i}\omega_l,{\bm q})$ is an exchange interaction due to the virtual exchange of an electron-hole pair excitation within the midband.\cite{satoh,miyai}
The last term or $-\Lambda({i}\omega_l,{\bm q})$ is the sum of all the remaining terms, or the so-called mode-mode coupling term; because it suppresses magnetic instability, it is defined in such a way that the minus sign appears for it.

We refer to previous papers\cite{FJO-supJ0,FJO-supJ} to derive the superexchange interaction.
The band splits into the upper and lower Hubbard bands.
Since Hubbard's theory is under the single-site approximation (SSA),\cite{hubbard1,hubbard3} it can be approximately used to describe the high-energy properties of the Anderson model; the local Green function is given by 
\begin{subequations}\label{EqLocalG-H}
\begin{align}\label{EqLocalG-H1}
\tilde{G}_\sigma({i}\varepsilon_l) &=
\frac1{{i}\varepsilon_l+\mu-\epsilon_d + \sigma \tilde{h}
- \tilde{\Sigma}_\sigma({i}\varepsilon_l) }
\\ \label{EqLocalG-H2} &
= \frac{1-\tilde{n}_{-\sigma}\bigl(\mu,\tilde{h}\bigr)}{{i}\varepsilon_l+\mu-\epsilon_d}
+ \frac{\tilde{n}_{-\sigma}\bigl(\mu,\tilde{h}\bigr)}{{i}\varepsilon_l+\mu-\epsilon_d-U},
\end{align}
\end{subequations}
for $|\varepsilon_l| \gg k_{\rm B}T_{\rm K}$, where $\tilde{n}_\sigma\bigl(\mu,\tilde{h}\bigr)$ is the number of localized electrons with spin $\sigma$ in the presence of the infinitesimal Zeeman energy $\tilde{h}$; in Eq.\hskip2pt(\ref{EqLocalG-H2}), $\tilde{h}$'s in the denominators are ignored because they are not crucial.
It follows that
\begin{align}\label{EqChiSSA}
\tilde{\chi}_s(0) = \lim_{\tilde{h}\rightarrow 0}
\frac{d\hskip5pt}{d\tilde{h}}
%(d/d\tilde{h})
\bigl[\tilde{n}_{\uparrow}\bigl(\mu,\tilde{h}\bigr)-\tilde{n}_{\downarrow}\bigl(\mu,\tilde{h}\bigr)\bigr].
\end{align}
%
%is used in Eq.\hskip2pt(\ref{EqLocalG-H}), 
According to Eqs.\hskip2pt(\ref{EqVertexR-IR2}), (\ref{EqWard1}), (\ref{EqLocalG-H}), and (\ref{EqChiSSA}),
\begin{align}\label{EqVertexHigh}
U\tilde{\lambda}_s({i}\varepsilon_l,{i}\varepsilon_l+{i}\omega_{l^\prime};{i}\omega_{l^\prime}) &= 
- \frac1{\tilde{G}_\sigma^2({i}\varepsilon_l)}
\nonumber \\ & \hskip-60pt \times 
\left(
\frac1{{i}\varepsilon_l+\mu-\epsilon_d}
+\frac1{{i}\varepsilon_l+\mu-\epsilon_d-U}\right),
\end{align}
for $\omega_{l^\prime}=0$ and $|\varepsilon_l| \gg k_{\rm B}T_{\rm K}$.

Since the superexchange interaction is a second-order effect in $-t/\sqrt{D}$, 
according to Eq.\hskip2pt(\ref{EqExchI}), 
%
%\begin{subequations}
\begin{align}\label{EsSiteJs}
& \hskip20pt
J_s({i}\omega_l,{\bm q}) =
\frac1{L}\sum_{\left<ij\right>} e^{{i}{\bm q}\cdot({\bm R}_i-{\bm R}_{j})}J_{\left<ij\right>}({i}\omega_l),
\\ &
J_{\left<ij\right>}({i}\omega_l) \hskip-2pt=\hskip-2pt
 2 k_{\rm B}T \hskip-1pt \sum_{l^\prime}
 U^2\tilde{\lambda}_s^2({i}\varepsilon_{l^\prime}\hskip-1pt,{i}\varepsilon_{l^\prime}\hskip-2pt +\hskip-2pt {i}\omega_l;{i}\omega_{l}\hskip-0.5pt)
\hskip-2pt \left(\hskip-2pt -t/\hskip-1pt\sqrt{D}\right)^2 
\nonumber \\ & \hskip50pt \times
R_{ii\sigma}^2({i}\varepsilon_{l^\prime})R_{jj\sigma}^2({i}\varepsilon_{l^\prime}+{i}\omega_l).
\end{align}
%\end{subequations}
% where $\left<ij\right>$ stands for a pair of nearest neighbors.
If Eq.\hskip2pt(\ref{EqLocalG-H}) is used for $R_{ii\sigma}({i}\varepsilon_l)$
and Eq.\hskip2pt(\ref{EqVertexHigh}) is used for 
$U\tilde{\lambda}_s({i}\varepsilon_{l^\prime},{i}\varepsilon_{l^\prime}\hskip-1pt +\hskip-1pt {i}\omega_l;{i}\omega_{l})$, its static part between nearest neighbors is given by $J_{\left<ij\right>}(0)\hskip-1pt =\hskip-1pt J/D$, where
\begin{align}\label{EqJ_ij}
J=-4t^2/U.
\end{align}
This is simply given in previous papers\cite{FJO-supJ0,FJO-supJ} and is in agreement with that derived from the conventional theory.\cite{Js-mech-pert}

%$J_{\left<ij\right>}(\omega+{i}0)\rightarrow 0$ as $\omega\rightarrow+\infty$, and

Since $J_{\left<ij\right>}(\omega+{i}0)$ is analytical in the upper-half complex plane, $J_{\left<ij\right>}({i}\omega_l)$ can be generally described as
\begin{align}
J_{\left<ij\right>}({i}\omega_l) \hskip-1pt = \hskip-1pt
\frac{J}{D} \hskip-3pt
\int_{0}^{+\infty} \hskip-13pt dx X_J(x)
\hskip-2pt \left(\frac1{{i}\omega_l
\hskip-1pt +\hskip-1pt x} \hskip-1pt -\hskip-1pt
\frac1{{i}\omega_l
\hskip-1pt - \hskip-1pt x}\right),
\end{align}
where $X_J(x)$ satisfies
\begin{align}
\int_{0}^{+\infty} \hskip-10pt dx \frac{X_J(x)}{x} =\frac1{2}.
\end{align}
Since $X_J(x)$ has a peak at $x\simeq U$, it is assumed that
$X_J(x) = (1/2)U\delta(x-U)$.
Then, 
\begin{align}\label{EqJs}
J_s({i}\omega_l,{\bm q}) &=
\frac{2J}{\sqrt{D}} \varphi_D({\bm q})\hskip1pt
\frac{U}{2}\hskip-1pt\left(\hskip-1pt
\frac1{{i}\omega_l\hskip-1pt +\hskip-1pt U}
\hskip-1pt - \hskip-1pt \frac1{{i}\omega_l \hskip-1pt - \hskip-1pt U}\hskip-1pt\right),
\end{align}
where $\varphi_D({\bm q})$ is defined by Eq.\hskip2pt(\ref{EqVarPhiD}).
In the static limit $|\omega_l|/U\rightarrow 0$, Eq.\hskip2pt(\ref{EqJs}) is reduced to
\begin{align}\label{EqJs2}
J_s({i}\omega_l,{\bm q}) =
\frac{2J}{\sqrt{D}}\varphi_D({\bm q})
= \frac{2J}{D}\sum_{\nu=1}^{D}\cos(q_\nu a).
\end{align}
The superexchange interaction
$J_s({i}\omega_l,{\bm q})$ is of higher order in $1/D$ for almost all ${\bm q}$'s and is of the zeroth order in $1/D$ only for particular ${\bm q}$'s, e.g., ${\bm q}$'s on the line between $(\pi/a)(0, \pm1, \cdots,\pm1)$ and $(\pi/a)(\pm1, \pm1, \cdots, \pm1)$.

%%%%%%%%%%%%%%%%%%%%%%%%%%%%%%%%%%%%%%%%%%%%%%%%%%%%%%%%%%%
\subsubsection{Mutual interaction due to spin fluctuations}
\label{SecMutualInt}
The mutual interaction due to spin fluctuations is given by
\begin{align}\label{EqI*1}
\Gamma_{\rm sf}({i}\omega_{l},{\bm q}; {i}\varepsilon_{l_1},{i}\varepsilon_{l_2}) &=
U^2\tilde{\lambda}_s({i}\varepsilon_{l_1},{i}\varepsilon_{l_1}+{i}\omega_{l};{i}\omega_{l}) 
\nonumber \\ & \hskip10pt \times
\tilde{\lambda}_s({i}\varepsilon_{l_2},{i}\varepsilon_{l_2}-{i}\omega_{l};-{i}\omega_{l})
\nonumber \\ & \hskip10pt \times
\left[\chi_s({i}\omega_{l},{\bm q})-\tilde{\chi}_s({i}\omega_{l})\right],
\end{align}
where ${i}\varepsilon_{l_1}$ and ${i}\varepsilon_{l_2}$ are the energies of incoming electrons, ${i}\varepsilon_{l_1}+{i}\omega_{l}$ and ${i}\varepsilon_{l_2}-{i}\omega_{l}$ are those of outgoing electrons, and ${i}\omega_{l}$ is the transfer energy. 
Since the single-site part is considered in the {\it unperturbed} state, it is subtracted in Eq.\hskip2pt(\ref{EqI*1}) in order to avoid double counting. 
It follows that\cite{FJO-supJ,satoh,miyai}
\begin{align}
& \label{EqChiI*1}
\chi_s({i}\omega_l,{\bm q})-\tilde{\chi}_s({i}\omega_l) =
(1/4)\tilde{\chi}^2({i}\omega_l) I_s^*({i}\omega_l,{\bm q}),
\\ & \label{EqChiI*2} 
I_s^*({i}\omega_l,{\bm q})=
\frac{I_s({i}\omega_l,{\bm q})}{1 - (1/4)I_s({i}\omega_l,{\bm q})\tilde{\chi}_s({i}\omega_l)}.
\end{align}
If Eq.\hskip2pt(\ref{EqUlambda}) is used, it follows that
%(\ref{EqChiI*1}) and (\ref{EqChiI*2}), 
\begin{align}
\Gamma_{\rm sf}({i}\omega_{l},{\bm q}; {i}\varepsilon_{l_1},{i}\varepsilon_{l_2}) 
=\tilde{\phi}_s^2 I_s^*({i}\omega_l,{\bm q}),
\end{align}
for $|\omega_l|\lesssim k_{\rm B}T_{\rm K}$, $|\varepsilon_{l_1}|\lesssim k_{\rm B}T_{\rm K}$, and $|\varepsilon_{l_2}|\lesssim k_{\rm B}T_{\rm K}$.
The single-site $\tilde{\phi}_s$ appears as a type of three-point vertex function.

If the mapped Anderson model is solved and the single-site $\tilde{\Sigma}_\sigma({i}\varepsilon_l)$ and $\tilde{\chi}_s({i}\omega_l)$ are given, the multisite $\Delta\Sigma_\sigma({i}\varepsilon_l,{\bm k})$ and $\Delta\pi({i}\omega_l,{\bm q})$ can be perturbatively calculated in terms of the intersite $I_s({i}\omega_l,{\bm q})$ on the basis of KLT.
Since the single-site terms are considered in the Anderson model, only multisite terms have to be considered in order to avoid double counting.
The intersite $I_s({i}\omega_l,{\bm q})$ has to be treated as a {\it bare} intersite exchange interaction, and 
the single-site $\tilde{\phi}_s$ has to be treated as a {\it bare} vertex function; $I_s^*({i}\omega_l,{\bm q})$ is the renormalized intersite exchange interaction, which is enhanced or screened by intersite spin fluctuations, depending on ${\bm q}$.

The intersite $I_s({i}\omega_l,{\bm q})$ is of higher order in $1/D$ for almost all ${\bm q}$'s except for particular ${\bm q}$'s; e.g., 
$I_s(0,{\bm Q})$, where ${\bm Q}$ is the ordering wave number determined using Eq.\hskip2pt(\ref{EqTN-Q}), is of the zeroth order in $1/D$ and corresponds to the conventional Weiss mean field.
Thus, KLT is a perturbative theory in terms of $I_s({i}\omega_l,{\bm q})$ and also a $1/D$ expansion theory.

%%%%%%%%%%%%%%%%%%%%%%%%%%%%%%%%%%%%%%%%%%%%%%%%%%%%%%%%%%%%%%%%%
\section{RVB Liquid in Low Dimensions}
\label{SecRVB}
\subsection{RVB self-energy} 
\label{SecFock}
We assume that 
\begin{align}\label{EqCondition}
U/|t|\hskip-2pt \gg\hskip-2pt 1, \hskip3pt |n(\mu)\hskip-2pt-\hskip-2pt 1|\hskip-2pt\ll\hskip-2pt 1, \hskip3pt 
T_c\hskip-2pt <\hskip-2pt T\hskip-2pt \ll\hskip-2pt |J|/(k_{\rm B}D).
\end{align}
We consider only the superexchange interaction $J_s({i}\omega_l,{\bm q})$ in the intersite $I_s({i}\omega_l,{\bm q})$.
% given by Eq.\hskip2pt(\ref{EqExchI}) or (\ref{EqThreeJ}).
There are two types of self-energy of the first order in $J_s({i}\omega_l,{\bm q})$: the Hartree-type and Fock-type self-energies.
Since the Hartree-type self-energy is included in the conventional Hartree term, which is one of the single-site terms and is considered in the {\it unperturbed} state, it should not be considered in order to avoid double counting.
%
%\begin{align}\label{EqHartree}
%\Sigma_\sigma^{(J)}({i}\varepsilon_l) =
%k_{\rm B}T \sum_{l^\prime}
%e^{{i}\varepsilon_{l^\prime} 0^+} \tilde{\phi}_s \frac{1}{4}J_s(0,0)
%R_{-\sigma}({i}\varepsilon_{l^\prime}).
%\end{align}
The Fock-type self-energy is given by\cite{RVB-FJO}
\begin{align}\label{EqFock}
\Delta\Sigma_\sigma^{\rm (RVB)}({i}\varepsilon_l,{\bm k}) &=
\frac{k_{\rm B}T}{L}
\sum_{l^\prime{\bm p}\sigma^\prime}
\tilde{\phi}_s^2 \frac{1}{4}J_s({i}\varepsilon_l-{i}\varepsilon_{l^\prime},{\bm k}-{\bm p})
\nonumber \\ & \quad \times
\bigl({\bm \sigma}^{\sigma\sigma^\prime}
\hskip-4pt\cdot\hskip-1pt{\bm \sigma}^{\sigma^\prime\hskip-1pt\sigma}\bigr)
G_{\sigma^\prime}({i}\varepsilon_{l^\prime},{\bm p}),
\end{align}
where ${\bm \sigma}=(\sigma_x,\sigma_y,\sigma_z)$ is the Pauli matrix.
What is considered for the Fock-type self-energy is simply the RVB mechanism.\cite{plain-vanilla} We call it the RVB self-energy.

%
%\begin{align}
%$\Sigma_\sigma^{(U)}({i}\varepsilon_l)=k_{\rm B}T \sum_{l^\prime} e^{{i}\varepsilon_{l^\prime} 0^+}U R_{-\sigma}({i}\varepsilon_{l^\prime})$,
%\end{align}
%
%as a part of it.

If Eqs.\hskip2pt(\ref{EqGreenKL}) and (\ref{EqJs}) 
%
%\begin{align}\label{EqJsKP}
%\varphi_D({\bm k} - {\bm p}) &=
%\frac1{\sqrt{D}}\hskip-1pt \sum_{\nu=1}^{D}\bigl[
%\cos(k_\nu a)\cos(p_\nu a)
%\nonumber \\ & \hskip50pt 
%+ \sin(k_\nu a)\sin(p_\nu a)\bigr],
%\end{align}
%
are used, it follows that
\begin{align}\label{EqFock1}
\Delta\Sigma_\sigma^{\rm (RVB)}({i}\varepsilon_l,{\bm k}) &=
\tilde{\phi}_1 \frac{3}{4}\tilde{W}_s^2 \frac{J}{D}\Xi_D({i}\varepsilon_l)\varphi_D({\bm k}),
\end{align}
where $\tilde{W}_s$ is the Wilson ratio, $J=-4t^2/U$, and 
\begin{align}
\Xi_D({i}\varepsilon_l) &=
\frac1{L}\sum_{{\bm p}}\varphi_D({\bm p})
%\nonumber \\ & \hskip-20pt \times
 \Biggl\{\frac1{2}U\Bigl[
f_{-}(U)\mathrm{g}_\sigma({i}\varepsilon_l+U,{\bm p}) 
\nonumber \\ & \hskip80pt
- f_{-}(-U)\mathrm{g}_\sigma({i}\varepsilon_l-U,{\bm p}) \Bigr]
\nonumber \\ & \hskip10pt
- \frac1{\pi}\int_{-\infty}^{+\infty} \hskip-18pt d\epsilon f_{+}(\epsilon)
\hskip0.5pt \frac1{2} U\hskip-2pt
\left(\hskip-1pt\frac1{{i}\varepsilon_l \hskip-1pt - \hskip-1pt \epsilon \hskip-1pt + \hskip-1pt U }
\hskip-1pt -\hskip-1pt \frac1{{i}\varepsilon_l \hskip-1pt -\hskip-1pt \epsilon \hskip-1pt -\hskip-1pt U}\hskip-2pt\right)\hskip-1pt
\nonumber \\ & \hskip80pt \times 
{\rm Im} \mathrm{g}_\sigma(\epsilon \hskip-1pt + \hskip-1pt {i}0,{\bm p}) \Biggr\},
\end{align}
where $\mathrm{g}_\sigma(\epsilon+{i}0,{\bm p})$ is given by Eq.\hskip2pt(\ref{EqGreenKL2}) and
\begin{align}
f_{\pm}(\epsilon)=1/\bigl[e^{\epsilon/(k_{\rm B}T)} \pm 1\bigr].
\end{align}
It is easy to confirm that
\begin{align}
\lim_{|\varepsilon_l|/U\rightarrow +\infty}\Delta\Sigma_\sigma^{\rm (RVB)}({i}\varepsilon_l,{\bm k})=0,
\end{align}
which is crucial to prove the sum rule in Appendix\ref{AppSumRule}.
In the static limit $|\varepsilon_l|/U \rightarrow 0$, $\Xi_D({i}\varepsilon_l)$ is simply given by
\begin{align}\label{EqXiD}
\Xi_D & \hskip-1.5pt = \hskip-1.5pt
\frac1{L}\hskip-1pt \sum_{{\bm p}}\hskip-1pt \varphi_D({\bm p})
\hskip-3pt \int_{-\infty}^{+\infty}\hskip-15pt d\epsilon 
f_+(\epsilon) \hskip-2pt \left(\hskip-2pt -\frac1{\pi}\hskip-1pt\right) \hskip-2pt
{\rm Im}\hskip1pt \mathrm{g}_\sigma(\epsilon\hskip-1.5pt +\hskip-1.5pt {i}0,{\bm p}).
\end{align}
If $|\varepsilon_l|/U \ll 1$, Eq.\hskip2pt(\ref{EqXiD}) can be used for $\Xi_D({i}\varepsilon_l)$ with sufficient accuracy. Then,
\begin{align}\label{EqRVB-Self}
\Delta\Sigma_\sigma^{\rm (RVB)}({i}\varepsilon_l,{\bm k}) & =
\tilde{\phi}_1 \frac{3}{4}\tilde{W}_s^2 \frac{J}{D} \Xi_D \varphi_D({\bm k}).
\end{align}
If Eq.\hskip2pt(\ref{EqJs2}) is used instead of Eq.\hskip2pt(\ref{EqJs}),
Eq.\hskip2pt(\ref{EqRVB-Self}) is simply derived instead of Eq.\hskip2pt(\ref{EqFock1}).

\subsection{Parameters characterizing the RVB liquid}
%\subsection{Fermi-liquid parameters}
\label{SecFLparameter}

If Eq.\hskip2pt(\ref{EqRVB-Self}) is used, Eq.\hskip2pt(\ref{EqGreenKL2}) is simply given by
\begin{align}
& \label{EqGr-KL} 
\mathrm{g}_{\sigma} ({i}\varepsilon_l,{\bm k}) =
\frac1{{i}\varepsilon_l+\mu^*-\xi({\bm k})
-\tilde{\gamma}_{\rm K}({i}\varepsilon_l)},
\end{align}
where $\mu^*$ and $\tilde{\gamma}_{\rm K}({i}\varepsilon_l)$ are given by Eqs.\hskip2pt(\ref{EqGreenKL3}) and (\ref{EqGreenKL4}), respectively,
%
%\begin{subequations}
\begin{align}
\label{EqGr-KL2} & \hskip0pt
\xi({\bm k}) = -2t^* \varphi_D({\bm k}),
\\&\label{EqGr-KL3}
t^* = 
%(1/\tilde{\phi}_1)t - (1/2) c_J (J/D) 
%\nonumber \\ & \phantom{t^*} =
t\bigl[ (1/\tilde{\phi}_1) + 2c_J |t|/(DU)\bigr],
\\&\label{EqGr-KL4}
c_J = (3/4) \tilde{W}_s^2 \Xi_D.
\end{align}
%\end{subequations}
%%
The density of states at the chemical potential is given by 
\begin{align}\label{EqRhoNum2}
\rho_\mu(0) & \hskip-1pt = \hskip-1pt
- \frac1{\pi L} \hskip-1pt \sum_{\bm k}\hskip-1pt 
\frac1{\tilde{\phi}_1} 
{\rm Im} \mathrm{g}_{\sigma}(+{i}0,{\bm k})
\hskip-1pt =\hskip-1pt 
O\left(\frac1{\tilde{\phi}_1|t^*|}\right).
\end{align}
From Eqs.\hskip2pt(\ref{EqRhoNum1}) and (\ref{EqRhoNum2}),
it follows that
\begin{align}\label{EqTKNum2}
k_{\rm B}T_{\rm K} =O(|t^*|).
\end{align}
Electrons in the midband can be described by $\mu^*$, $t^*$ or $k_{\rm B}T_{\rm K}$, $\tilde{\phi}_1$, and $c_J$.
In principle, they have to be self-consistently evaluated with each other as a function of $T$ and $\mu$ to satisfy the mapping condition given in Sect.\hskip2pt\ref{SecMapCondition}.
However, they can be approximately evaluated, as studied below.

According to the Fermi-surface sum rule, \cite{Luttinger1,Luttinger2} 
\begin{align}\label{EqFS-SumRule}
n(\mu) = \frac{2}{L}\sum_{\bm k}
\int_{-\infty}^{+\infty} \hskip-10pt d\epsilon f_+(\epsilon)\delta\bigl[\epsilon+\mu^*-\xi({\bm k})\bigr],
\end{align}
for $T=0\hskip2pt$K.
If $T=0\hskip2pt$K and $n(\mu)=1$, then $\mu^*=0$, which is required by the particle-hole symmetry.
Then, $|\Xi_{1}|=1/\pi=0.31831\cdots$, $|\Xi_{2}|=2\sqrt{2}/\pi^2=0.28658\cdots$, $\cdots$, and $|\Xi_{\infty}|=1/(2\sqrt{\pi})=0.283095\cdots$; 
i.e., $|\Xi_{D}|\simeq 1/3$ for any $D$. 
If $T_c<T\ll T_{\rm K}$, Eq.\hskip2pt(\ref{EqFS-SumRule}) can be approximately used, but with sufficient accuracy. 
If $n(\mu)\simeq 1$, it follows that
%\begin{subequations}
\begin{align}
|\mu^*|\ll |t^*|, \quad
%\\ &
|\Xi_{D}|\simeq 1/3, \quad
%\\ &
c_J\simeq 1.
\end{align}
%\end{subequations}
%
If $|\Xi_{D}|= 1/3$ and $\tilde{W}_s=2$ are assumed, then $c_J =1$.

The asymptotic behavior of $1/\tilde{\phi}_1$ as $U/|t|\hskip-2pt \rightarrow\hskip-2pt+\infty$ is studied in Appendix\ref{AppAppSumRule}: If $n(\mu)=1$, $1/\tilde{\phi}_1\hskip-2pt =\hskip-2pt O\bigl[t^2/(DU^2)\bigr]$, 
as shown in Eq.\hskip2pt(\ref{EqAsympFinal}). 
According to Gutzwiller's theory,\cite{Gutzwiller1,Gutzwiller2,Gutzwiller3} if $U/|t|\hskip-2pt \rightarrow \hskip-2pt +\infty$ and $|n(\mu)-1|\hskip-2pt \ll\hskip-2pt 1$, $1/\tilde{\phi}_1\hskip-2pt =\hskip-2pt O(|n(\mu)\hskip-1pt -\hskip-1pt 1|)$.
If $U/|t|\hskip-2pt\gg\hskip-2pt 1$ and $|n(\mu)\hskip-1pt -\hskip-1pt 1|\hskip-2pt\ll\hskip-2pt 1$, it is reasonable to assume that 
\begin{align}\label{EqSpeculation}
1/\tilde{\phi}_1 = \max\bigl\{O\bigl[t^2/(DU^2)\bigr] , O\bigl[|n(\mu)-1|\bigr]\bigr\}.
\end{align}

%
%%%%%%%%%%%%%%%%%%%%%%%%%%%%%%%%%%%%%%%%%%%%%%%%%%%%%%%%%%%%%
\subsection{Possible types of electron liquid}
\label{SecPossible}
If Eq.\hskip2pt(\ref{EqCondition}) is satisfied, the spectral weight of the midband, which is $1/\tilde{\phi}_1$, is much smaller than unity, and the bandwidth of the midband, which is $O(|t|^*)$, is much smaller than the bare bandwidth, which is $O(|t|)$. 
We consider four typical cases of Eq.\hskip2pt(\ref{EqCondition}), where $t^2/(DU^2)\hskip-2pt \ll\hskip-2pt |t|/(DU)$. 
First, we consider the case of
\begin{align}\label{EqUregion1}
t^2/(DU^2) \ll |t|/(DU) \lesssim |n(\mu)-1| ,
\end{align}
where $n(\mu)\hskip-2pt\ne\hskip-2pt 1$ is assumed.
According to Eq.\hskip2pt(\ref{EqUregion1}) together with Eqs.\hskip2pt(\ref{EqGr-KL3}), (\ref{EqRhoNum2}), (\ref{EqTKNum2}), and (\ref{EqSpeculation}),
\begin{subequations}\label{EqUregion11}
\begin{align}\label{EqUregion111}
& 1/\tilde{\phi}_1 = O(|n(\mu)-1|), 
\\& \label{EqUregion112}
\rho_\mu(0) = O(1/|t|),
\\ & \label{EqUregion113}
k_{\rm B}T_{\rm K} =O\bigl[|t| \cdot |n(\mu)-1|\bigr].
\end{align}
\end{subequations}
The single-site effect considered by Gutzwiller's theory is more crucial than the RVB mechanism; $\rho_\mu(0)$ weakly increases as $U$ increase.
If $|t|/(DU)\hskip-2pt\ll\hskip-2pt |n(\mu)\hskip-2pt-\hskip-2pt1|$,
the electron liquid is simply that given by Gutzwiller's theory,\cite{Gutzwiller1,Gutzwiller2,Gutzwiller3} which is under SSA.
If $n(\mu)$ is kept constant, $\rho_\mu(0)$ is constant as a function of $U$ under SSA, as will be discussed later in Sect.\hskip2pt\ref{SecNatureMott}; thus, $\rho_\mu(0)$ for such a large $U$ is almost constant as a function of $U$ and as large as that for $U=0$.

Second, we consider the case of 
\begin{align}\label{EqUregion2}
t^2/ (DU^2) \lesssim |n(\mu)-1| \lesssim |t|/(DU) ,
\end{align}
where $n(\mu)\hskip-2pt\ne\hskip-2pt 1$ is also assumed; $U$ in this case is smaller than that considered in the first case, if $n(\mu)\hskip-2pt\ne\hskip-2pt 1$ are the same as each other between the two cases. 
It follows that
\begin{subequations}
\begin{align}
&1/\tilde{\phi}_1 = O\bigl(|n(\mu)-1|\bigr), 
\\ & \label{EqUcom}
O(1/U)<\rho_\mu(0)< O(1/|t|),
\\ & 
k_{\rm B}T_{\rm K} =O\bigl[t^2/(DU) \bigr] 
= O\bigl(|J|/D\bigr).
\end{align}
\end{subequations}
The RVB mechanism is crucial and the electron liquid is the RVB liquid. Since the single-site effect becomes more and more relatively crucial to the RVB mechanism as $U$ increases, $\rho_\mu(0)$ increases as $U$ increases.

Third, we consider the case of 
\begin{align}\label{EqUregion3}
|n(\mu)-1| \lesssim t^2/(DU^2) \ll |t|/(DU),
\end{align}
where $n(\mu)\hskip-2pt\ne\hskip-2pt 1$ or $n(\mu)\hskip-2pt=\hskip-2pt 1$ is assumed;
if $n(\mu)\hskip-2pt\ne\hskip-2pt 1$ are the same as each other between this case and the first or second case, $U$ in this case is smaller than that considered in the first or second case.
It follows that
\begin{subequations}\label{EqUregion31}
\begin{align} \label{EqUregion31A}
& 1/\tilde{\phi}_1 = O\left[t^2/(DU^2)\right],
\\ \label{EqUregion31B} &
\rho_\mu(0) = O(1/U), 
\\ \label{EqUregion31C} & 
k_{\rm B}T_{\rm K} =O\bigl[t^2/(DU)\bigr]=O(|J|/D).
\end{align}
\end{subequations}
The RVB mechanism is crucial and the electron liquid is a typical type of RVB liquid.
The density of states $\rho_\mu(0)$ becomes smaller as $U$ increases. Particularly if $n(\mu)\hskip-2pt =\hskip-2pt 1$, $\rho_\mu(0)\hskip-2pt\rightarrow\hskip-2pt 0$ as $U/|t|\hskip-2pt\rightarrow\hskip-2pt +\infty$.
According to Eq.\hskip2pt(\ref{EqMapDelta}), 
\begin{align}\label{EqMapCondA}
\Delta(\varepsilon) & 
= 
- \bigl| {\rm Im}\Sigma_\sigma(\varepsilon+{i}0)\bigr|
+ 1/[\pi\rho_\mu(\varepsilon)],
\end{align}
for $\varepsilon$ such that ${\rm Re}R_\sigma(\varepsilon+{\rm i}0)\hskip-2pt=\hskip-2pt 0$.
Since ${\rm Re}R_\sigma(+{\rm i}0)\hskip-2pt=\hskip-2pt 0$ for $\mu=\mu_0$ and ${\rm Im}\Sigma_\sigma(+{i}0)
\rightarrow 0$ as $T\rightarrow 0\hskip2pt$K, $\Delta(0)\hskip-2pt=\hskip-2pt O(U)$.

According to the considerations of the above three cases, the RVB mechanism is crucial, provided that
\begin{align}\label{EqCondRVB}
|n(\mu)-1| \lesssim |t|/(DU). 
\end{align}
If $n(\mu)\hskip-2pt\ne\hskip-2pt 1$ is kept constant, $\rho_\mu(0)$ is not a monotonous function of $U$; $\rho_\mu(0)$ as a function of $U$ is minimal at approximately $U$ such that $|n(\mu)\hskip-1pt -\hskip-1pt 1|\hskip-2pt =\hskip-2pt t^2/(DU^2)$. 
If $n(\mu)\hskip-2pt=\hskip-2pt 1$, $\rho_\mu(0)$ is a monotonously decreasing function of $U$: $\rho_\mu(0)\propto 1/U$.

Last, we consider a particular case of Eq.\hskip2pt(\ref{EqUregion3}): the Heisenberg limit of $U/|t|\hskip-2pt \rightarrow\hskip-2pt +\infty$ with $J\hskip-2pt =\hskip-2pt -4t^2/U$ and $n(\mu)\hskip-2pt=\hskip-2pt 1$ kept constant. 
It follows that
$1/\tilde{\phi}_1\hskip-2pt \rightarrow\hskip-2pt 0$, 
$\rho_\mu(0)\hskip-2pt \rightarrow\hskip-2pt 0$, and $k_{\rm B}T_{\rm K}\hskip-2pt =\hskip-2pt O(|J|/D)$.
Although the bandwidth of the midband is nonzero and finite, either its spectral weight or density of states is infinitesimal. 
The RVB electron liquid in the Heisenberg limit is the most typical type of RVB electron liquid; it is a quasi-spin liquid, or simply the RVB spin liquid.

%%%%%%%%%%%%%%%%%%%%%%%%%%%%%%%%%%%%%%%%%%%%%%%%%%%%%%%
\subsection{Metallic conductivity in the Heisenberg limit}
\label{SecConductivity}
We consider magnetic impurities:
\begin{align}\label{EqImpurityTerm}
\mathcal{H}^\prime &= 
- \sum_{i\sigma\sigma^\prime} J_{i}^{\prime} 
\left({\bm \sigma}^{\sigma\sigma^\prime}\hskip-5pt \cdot{\bm S}_{i}^\prime\right)
d_{i\sigma}^\dag d_{i\sigma^\prime}^{\phantom{\dag}},
\end{align}
where ${\bm S}_{i}^\prime$ is an impurity spin at the $i$th unit cell.
We consider an ensemble for $J_i^\prime$ such that $J_i^\prime$ is positive, zero, or negative and is completely random from unit cell to unit cell and from sample to sample in the ensemble:
$\bigl<\hskip-2.5pt\bigl<J_i^\prime\bigr>\hskip-2.5pt\bigr>\hskip-2pt =\hskip-2pt 0$,
$\bigl<\hskip-2.5pt\bigl<J_i^\prime J_{j}^\prime\bigr>\hskip-2.5pt\bigr>\hskip-2pt =\hskip-2pt 
\delta_{ij}\bigl<\hskip-2.5pt\bigl<\left|J_i^\prime\right|^2\bigr>\hskip-2.5pt\bigr>$, 
$\bigl<\hskip-2.5pt\bigl<J_{i_1}^\prime J_{i_2}^\prime J_{i_3}^\prime\bigr>\hskip-2.5pt\bigr> \hskip-2pt =\hskip-2pt 0$, 
and so on,
where $\left<\hskip-1.5pt\left<\cdots\right>\hskip-1.5pt\right>$
stands for the ensemble average.
The translational symmetry is restored by the ensemble average. 
In the following part, the double thermal-ensemble average is simply called an {\it average}.

Assuming that $|J_{i}^\prime|\ll k_{\rm B}T_{\rm K}$ for any $i$, we treat impurity scattering in the Born approximation. The {\it average} self-energy $\overline{\Sigma}_{\sigma}({i}\varepsilon_l)$ due to the impurity scattering is given by self-consistently solving the following two equations:
%
%\begin{subequations}
\begin{align}
&\overline{\Sigma}_{\sigma}({i}\varepsilon_l) =
\tilde{\phi}_s^2
S^\prime(S^\prime+1)
\delta_{ij}\Bigl<\hskip-4pt\Bigl<\left|J_i^\prime\right|^2\Bigr>\hskip-4pt\Bigr>
\frac1{L}\sum_{\bm k}\frac1{\tilde{\phi}_1}\overline{\mathrm{g}}_{\sigma}({i}\varepsilon_l,{\bm k}),
\\
&\overline{\mathrm{g}}_{\sigma}({i}\varepsilon_l,{\bm k}\hskip-0.5pt)=\frac1{{i}\varepsilon_l \hskip-1pt + \hskip-1pt \mu^* \hskip-1pt - \hskip-1pt \xi({\bm k}\hskip-0.5pt) \hskip-1pt - \hskip-1pt \tilde{\gamma}_{\rm K}({i}\varepsilon_l \hskip-1pt) \hskip-1pt - \hskip-1pt (1/\tilde{\phi}_1)\overline{\Sigma}_\sigma({i}\varepsilon_l \hskip-0.5pt) },
\end{align}
%\end{subequations}
%
where $S^\prime$ is the magnitude of impurity spins,
$\overline{\mathrm{g}}_{\sigma}({i}\varepsilon_l,{\bm k})$ is the {\it average} Green function multiplied by $\tilde{\phi}_1$, which corresponds to $\mathrm{g}_{\sigma}({i}\varepsilon_l,{\bm k})$ given by Eq.\hskip2pt(\ref{EqGreenKL2}) or (\ref{EqGr-KL}), and $\tilde{\gamma}_{\rm K}({i}\varepsilon_l)$ is given by Eq.\hskip2pt(\ref{EqGreenKL4}).
The {\it average} conductivity is given by\cite{kubo}
\begin{align}\label{EqKubo}
\sigma_{xx}(\omega) = 
\frac{\hbar}{{i}\omega}\Bigl[
K_{xx}(\omega+{i}0)-K_{xx}(0)\Bigr],
\end{align}
where
\begin{align}\label{EqKxx}
K_{xx}({i}\omega_l) &\hskip-1.5pt = \hskip-1.5pt
\frac1{La^D}\hskip-2pt \int_{0}^{\beta} \hskip-5pt d\tau
e^{{i}\omega_l\tau} \Bigl< \hskip-1.5pt e^{\tau(\mathcal{H}-\mu\mathcal{N})}
\hat{j}_x \hskip1pt e^{-\tau(\mathcal{H}-\mu\mathcal{N})} \hat{j}_x \hskip-1pt \Bigr>
\nonumber \\ 
&= 
\frac{e^2}{\hbar^2}\frac{(2t)^2}{D a^{D-2}}
 \Pi_{xx}({i}\omega_l),
\end{align}
where $\beta\hskip-2pt =\hskip-2pt1/(k_{\rm B}T)$, $\hat{j}_x\hskip-2pt =\hskip-2pt -(e/\hbar)\sum_{{\bm k}\sigma}
[(\partial/\partial k_1)E({\bm k})]\hat{n}_{\bm k}$, with
%
%\begin{align}
$\hat{n}_{{\bm k}\sigma} \hskip-2pt=\hskip-2pt (1/L)\hskip-1pt
\sum_{ii^\prime} e^{{i}{\bm k}\cdot\left({\bm R}_i-{\bm R}_{i^\prime}\right)} d_{i\sigma}^\dag d_{i^\prime\sigma}^{\phantom{\dag}}$,
%\end{align}
and
\begin{align}\label{EqPixx}
&\Pi_{xx}({i}\omega_l) \hskip-1pt =\hskip-1pt
\frac1{L}\hskip-1pt
\sum_{{\bm k}{\bm p}}\sum_{\sigma\sigma^\prime}
\sin(k_1 a)\sin(p_1 a) \hskip-3pt
\int_{0}^{\beta} \hskip-5pt d\tau
e^{{i}\omega_l\tau} 
\nonumber \\ & \hskip50pt \times \hskip-2pt
\Bigl<\hskip-1pt 
e^{\tau(\mathcal{H}-\mu\mathcal{N})}\hskip1pt \hat{n}_{{\bm k}\sigma}\hskip1pt e^{-\tau(\mathcal{H}-\mu\mathcal{N})}
\hskip1pt \hat{n}_{{\bm p}\sigma^\prime}\hskip-1pt
\Bigr>.
\end{align}
Here, $\left<\cdots\right>$ stands for the {\it average}.
In order to satisfy the Ward relation,\cite{ward} 
the ladder type of vertex correction due to the superexchange interaction has to be considered; the vertex correction due to the impurity scattering to be considered is also of the ladder type, but it vanishes. Then,
%
%\begin{subequations}
\begin{align}\label{EqUpperPiJ}
\Pi_{xx}({i}\omega_l) =
\frac{1}{\tilde{\phi}_1^2}
\frac{2\pi_{xx}({i}\omega_l)}
{\displaystyle 1 + 3\tilde{W}_s^2J\pi_{xx}({i}\omega_l)/(4D)}, 
\end{align}
where $\tilde{W}_s$ is the Wilson ratio and
\begin{align}\label{EqLowerPiJD} 
\pi_{xx}({i}\omega_l)\hskip-1pt = \hskip-1pt
- \frac{k_{\rm B}T}{L}\hskip-1pt\sum_{n{\bm k}}
\sin^2(k_x a)\overline{\mathrm{g}}_\sigma({i}\varepsilon_l,{\bm k})\overline{\mathrm{g}}_\sigma({i}\varepsilon_l \hskip-1pt + \hskip-1pt {i}\omega_l,{\bm k}).
\end{align}

The $\omega$-linear term of $K_{xx}(\omega+{i}0)$ or $\Pi_{xx}(\omega+{i}0)$ contributes to the static conductivity $\sigma_{xx}(0)$. It follows that
\begin{align}\label{EqTwoVerJ1}
\frac{d \Pi_{xx}({i}\omega_l\hskip-1pt )}
{d({i}\omega_l)} &=
%\frac{2}{\tilde{\phi}_1^2} 
\frac{(2/\tilde{\phi}_1^2)
\bigl[d\pi_{xx}({i}\omega_l)/d({i}\omega_l )\bigr]}
{\displaystyle \bigl[
1 + 3 \tilde{W}_s^2J\pi_{xx}({i}\omega_l ) /(4D) \bigr]^2}.
\end{align}
%\end{subequations}
%
If Eq.~(\ref{EqXiPi}) is used, it is easy to show that
\begin{align}\label{EqWardtdtd*}
\bigl[1 + 3\tilde{W}_s^2J\pi_{xx}(0)/(4D) \bigr]\tilde{\phi}_1 t^*=t,
\end{align}
with $t^*$ given by Eq.\hskip2pt(\ref{EqGr-KL3}).
Then, it follows that
\begin{align}\label{EqTwoVer2}
& \hskip-70pt
\sigma_{xx}(0) =
\frac{e^2}{\hbar^2}\frac{(2t^*)^2}{Da^{D-2}}
S_{xx}(0),
\end{align}
where $S_{xx}(\hskip-0.5pt 0\hskip-0.5pt ) \hskip-2pt = \hskip-2pt 
(2\hbar/i)\bigl[(d /d\omega) \pi_{xx}(\omega+{i}0)\bigr]_{\omega=0}$, or
\begin{align}\label{EqPJ1} 
S_{xx}(\hskip-0.5pt 0\hskip-0.5pt ) & \hskip-2pt =\hskip-2pt
\frac{2\hbar}{\pi L}\hskip-2pt \sum_{\bm k}
\sin^2(\hskip-0.5pt k_1 \hskip-0.5pt a\hskip-0.5pt) \hskip-4pt
\int_{-\infty}^{+\infty} \hskip-14pt d\varepsilon 
\hskip-1pt \left[\hskip-1pt -\hskip-1pt
\frac{f_{\hskip-1pt+\hskip-1pt}(\hskip-0.5pt \varepsilon\hskip-0.5pt)}{d \varepsilon }\hskip-1pt\right]\hskip-4.5pt
%\nonumber \\ & \qquad \times 
\Bigl[{\rm Im} \hskip1pt 
\overline{\mathrm{g}}_\sigma\hskip-0.5pt(\hskip-0.5pt\varepsilon\hskip-1.5pt + \hskip-1.5pt
{i}0,\hskip-0.5pt {\bm k})\hskip-1.5pt\Bigr]^2 \hskip-1pt .
\end{align}
If $2\sin^2(k_1a) = 1-\cos(2 k_1a)$ is used and the term that includes $\cos(2k_1a)$ is ignored, it follows that
\begin{align}\label{EqPJ2}
S_{xx}(0) & =
\frac{\hbar}{\pi} \hskip-2pt \int_{-\infty}^{+\infty} \hskip-12pt dE \hskip-2pt \left[\tilde{\phi}_1\rho_\mu(E)\right]\hskip-2pt
\int_{-\infty}^{+\infty} \hskip-10pt d\varepsilon \hskip-2pt
\left[-\frac{f_+(\varepsilon)}{d \varepsilon }\right]
\nonumber \\ & \hskip-10pt\times
\hskip-3pt \left[\hskip-1pt {\rm Im} \hskip1pt 
\frac1{\varepsilon \hskip-2pt - \hskip-2pt E
\hskip-2pt - \hskip-2pt \tilde{\gamma}_{\rm K}
(\hskip-0.5pt\varepsilon \hskip-2pt + \hskip-2pt {i}0 \hskip-0.5pt) 
\hskip-2pt - \hskip-2pt
\bigl(\hskip-1pt 1/\tilde{\phi}_1 \hskip-1pt\bigr)
\overline{\Sigma}_\sigma(\hskip-0.5pt \varepsilon \hskip-2pt + \hskip-2pt {i}0 \hskip-0.5pt) }\hskip-1pt \right]^2 \hskip-3pt .
\end{align}
We define two relaxation times, $\tau_{\rm K}$ and $\tau_s$, by
\begin{align}
\frac{\hbar}{2\tau_{\rm K}} &=
\int_{-\infty}^{+\infty}\hskip-12pt d\varepsilon
\left[-\frac{df_+(\varepsilon)}{d\varepsilon} \right]
{\rm Im} \bigl[-\tilde{\gamma}_{\rm K}(\varepsilon+{i}0)\bigr]
\nonumber \\ &=
\left(\frac{\pi^2}{3}\tilde{W}_{21}+\tilde{W}_{22}\right)
\frac{\tilde{\phi}_1}{\pi\Delta(0)}\left(k_{\rm B}T\right)^2,
\\ 
\frac{\hbar}{2\tau_s} &=
\int_{-\infty}^{+\infty}\hskip-12pt d\varepsilon
\left[-\frac{df_+(\varepsilon)}{d\varepsilon} \right]
\frac1{\tilde{\phi}_1}{\rm Im}\left[-\overline{\Sigma}_\sigma(\varepsilon+{i}0)\right],
\end{align}
respectively.
If $-{i}\hbar/(2\tau_{\rm K})$ and $-{i}\hbar/(2\tau_s)$ are used for $\tilde{\gamma}_{\rm K}(\varepsilon +{i}0)$ and $(1/\tilde{\phi}_1)\overline{\Sigma}_\sigma(\varepsilon + {i}0)$, respectively, and the energy dependence of $\rho_\mu(E)$ is ignored in Eq.\hskip2pt(\ref{EqPJ2}), it follows that
\begin{align}\label{EqSigmaXX}
\sigma_{xx}(0) &=
\frac{e^2}{\hbar^2}\frac{8|t^*|^2}{Da^{D-2}}
\frac{\tilde{\phi}_1\rho_\mu(0)}{(1/\tau_{\rm K})+ (1/\tau_s)}.
\end{align}

Since $|t^*|\hskip-2pt=\hskip-2pt O(|J|/D)$ and $\tilde{\phi}_1\rho_\mu(0)\hskip-2pt =\hskip-2pt O(1/|t^*|)$, $\sigma_{xx}(0)$ is nonzero even in the limit $1/\tilde{\phi}_1\rightarrow 0$; $\rho_\mu(0)\hskip-2pt \rightarrow\hskip-2pt 0$ in the limit $1/\tilde{\phi}_1\rightarrow 0$. 
In an {\it absolutely} clean system, $\hbar/\tau_s\hskip-2pt=\hskip-2pt 0$. 
If no symmetry is broken or no complete gap opens even at $T\hskip-2pt=\hskip-2pt +0\hskip2pt$K in the {\it absolutely} clean system, $\sigma_{xx}(0)$ diverges as $T\hskip-2pt\rightarrow\hskip-2pt 0\hskip2pt$K because $1/\tau_{\rm K}\hskip-2pt \rightarrow\hskip-2pt 0$ as $T\hskip-2pt\rightarrow\hskip-2pt 0\hskip2pt$K.
If the impurity scattering is sufficiently weak,
the RVB electron liquid shows metallic conductivity at a sufficiently low $T$ such that $k_{\rm B}T_c\hskip-2pt <\hskip-2pt k_{\rm B}T\hskip-2pt \ll\hskip-2pt |J|/D$ in not only three dimensions or higher but also, if the Anderson localization\cite{abrahams} can be ignored, one and two dimensions, and even in the Heisenberg limit, in which $\rho_\mu(0)\rightarrow 0$.

%%%%%%%%%%%%%%%%%%%%%%%%%%%%%%%%%%%%%%%%%%%%%%%%%%%%%%%%%%%%%%
\subsection{Adiabatic continuation}
\label{SecAdiabatic}
In the Heisenberg limit, the Hubbard model is reduced to
\begin{align}\label{EqHeisenberg}
\mathcal{H}_S \hskip-1pt = \hskip-1pt
L\epsilon_d 
\hskip-1pt- \hskip-1pt\frac1{2}\frac{J}{D} 
\hskip-1pt\sum_{\left<ij\right>} \hskip-1pt
\left({\bm S}_i \hskip-1pt\cdot\hskip-1pt{\bm S}_{j}\right)
\hskip-1pt-\hskip-1pt 2 \hskip-1pt \sum_{i} 
J_{i}^{\prime}\left({\bm S}_{i} \hskip-1pt\cdot\hskip-1pt
{\bm S}_i^\prime\right),
\end{align}
where the Hilbert space is constrained within the subspace where no empty or double occupancy is allowed, and
${\bm S}_i = 
(1/2)\sum_{\sigma\sigma^\prime}{\bm \sigma}^{\sigma\sigma^\prime}
d_{i\sigma}^\dag d_{i\sigma^\prime}^{\phantom{\dag}}$.
The last term in Eq.\hskip2pt(\ref{EqHeisenberg}) is the impurity term given by Eq.\hskip2pt(\ref{EqImpurityTerm}).
Since ${\bm S}_i$'s satisfy the commutation relation for spin within the constrained Hilbert subspace, if the impurity term is excluded, $\mathcal{H}_S$ is the Heisenberg model.
The local gauge symmetry does not exist in the Hubbard model but exists in the Heisenberg model:
$\bigl[\mathcal{H}, n_{i\uparrow}+n_{i\downarrow} \bigr]\ne 0$ and 
$\bigl[\mathcal{H}_S, n_{i\uparrow}+n_{i\downarrow} \bigr]= 0$
for any $i$.

The role of the superexchange interaction is dual: the cause and suppression of magnetic instability.
Since no suppression occurs in infinite dimensions, 
the mean-field approximation is rigorous for the Heisenberg model in infinite dimensions;
the N\'{e}el temperature is as high as $T_{\rm N}\hskip-2pt=\hskip-2pt|J|/(2k_{\rm B})$ in not only the Heisenberg model but also the Heisenberg limit of the Hubbard model, as will be shown later in Eq.\hskip2pt(\ref{EqCW-Law2}). 
The N\'{e}el temperature $T_{\rm N}$ is suppressed by two mechanisms:
critical spin fluctuations and the RVB mechanism.\cite{fazekas} 
The RVB mechanism stabilizes or prefers an unordered electron liquid in the Hubbard model and an unordered spin liquid in the Heisenberg model, rather than the N\'{e}el state.
The stabilization energy due to the RVB mechanism is $O(|J|/D)$ per pair of nearest neighbors or per unit cell; the RVB mechanism is $O(1/D)$.
If $D$ is sufficiently small such that no $T_{\rm N}$ exists or, if it exists, $T_{\rm N}\hskip-2pt \ll\hskip-2pt |J|/(Dk_{\rm B})$, electron and spin liquids in the Hubbard and Heisenberg models at $T$ such that $T_{\rm N}\hskip-2pt<\hskip-2pt T\hskip-2pt\ll\hskip-2pt |J|/(Dk_{\rm B})$ are the RVB electron and spin liquids, respectively.

The local gauge symmetry is a peculiar symmetry such that
it cannot be spontaneously broken nor restored.\cite{elitzur} 
In the reduction of the Hubbard model into the Heisenberg model, the local gauge symmetry is {\it not spontaneously} restored but is {\it forced} to be restored by constraining the Hilbert space within the subspace.
The difference in the local gauge symmetry between two phases can never deny the possibility of the adiabatic continuation between them, as discussed below.

The relationship between the Hubbard and Heisenberg models is similar to that between the Anderson and $s$-$d$ models.
In the $s$-$d$ limit, the Anderson model is reduced to the $s$-$d$ model.
The local gauge symmetry does not exist in the Anderson model but exists in the $s$-$d$ model.
The density of states for $d$ electrons can be defined in the Anderson model.
If it is denoted by $\tilde{\rho}(\varepsilon)$, $\tilde{\rho}(\varepsilon)>0$ at least for a sufficiently small $\varepsilon$, which means that low-energy single-particle excitations are possible, as in the Hubbard model. 
On the other hand, $\tilde{\rho}(\varepsilon)$ cannot be defined in the $s$-$d$ model, which means that no single-particle excitation is possible, as in the Heisenberg model. 
On the other hand, according to Nozi\`{e}res' theory,\cite{nozieres} it is possible to describe the spin liquid in the $s$-$d$ model as the normal Fermi liquid. 
The Fermi-liquid theory for the spin liquid in the $s$-$d$ model is exactly equivalent to that for the Fermi liquid in the Anderson model in the $s$-$d$ limit.\cite{yamada1,yamada2}
Thus, the spin liquid in the $s$-$d$ model adiabatically continues to the Fermi liquid in the Anderson model.

The conductivity of the RVB spin liquid is zero, but that of the RVB electron liquid can be divergent at $T=+0\hskip2pt$K even in the Heisenberg limit if no impurity exists and no symmetry is broken even at $T=+0\hskip2pt$K, as studied in Sect.\hskip2pt\ref{SecConductivity}.
This extreme difference in the conductivity cannot exclude the possibility of the adiabatic continuation between the RVB spin and electron liquids either, as discussed below.

The strength of magnetic impurities can be used as an adiabatic parameter. 
Here, it is assumed that $-\infty< J_i^{\prime} <+\infty$ and 
$0< \bigl<\hskip-2.5pt\bigl<\left|J_i^\prime\right|^2\bigr>\hskip-2.5pt\bigr><+\infty$.
Clean and dirty limits are defined by the limit 
$\bigl<\hskip-2.5pt\bigl<\left|J_i^\prime\right|^2\bigr>\hskip-2.5pt\bigr>\rightarrow 0$ and the limit
$\bigl<\hskip-2.5pt\bigl<\left|J_i^\prime\right|^2\bigr>\hskip-2.5pt\bigr>\rightarrow +\infty$,
respectively.
In the dirty-limit Hubbard model, an electron is localized almost within a unit cell, so that the local gauge symmetry is almost {\it restored} and the conductivity is almost zero.
It is certain that every physical property of the dirty-limit Hubbard model in the Heisenberg limit is the same as that of the dirty-limit Heisenberg model. Thus, the electron state in the dirty-limit Hubbard model in the Heisenberg limit and the spin state in the dirty-limit Heisenberg model adiabatically continue to each other.

According to the scaling theory for the Anderson localization,\cite{abrahams}
there is no critical point between the metallic and insulating phases, or between the itinerant and localized states, or between the clean and dirty limits; there is also no lower limit of the metallic conductivity or minimum metallic conductivity.
Therefore, the RVB electron liquid in the clean-limit Hubbard model adiabatically continues to the electron state in the dirty-limit Hubbard model.
It is obvious that the RVB spin liquid in the clean-limit Heisenberg model adiabatically continues to the spin state in the dirty-limit Heisenberg model. Thus, the RVB electron liquid in the clean-limit Hubbard model and the RVB spin liquid in the clean-limit Heisenberg model adiabatically continue to each other.

%
%%%%%%%%%%%%%%%%%%%%%%%%%%%%%%%%%%%%%%%%%%%%%%%%%%
\section{Discussion}
\label{SecDiscussion}
\subsection{Nature of the Mott insulator}
\label{SecNatureMott}
Either the Mott insulator or Lieb and Wu's insulator competes with the RVB liquid. It is desirable to examine its nature in order to confirm that the RVB liquid is stable against it.
Here, we examine the nature of the Mott insulator. 
We assume that $U/|t|\hskip-2pt\gg\hskip-2pt 1$ and $\mu\hskip-2pt=\hskip-2pt\mu_0$, unless otherwise noted.

If $T\hskip-2pt>\hskip-2pt T_c$, no symmetry is broken. 
The Kondo energy $k_{\rm B}T_{\rm K}$ is the energy scale of quantum spin fluctuations in the RVB liquid as well as a measure of its stabilization energy: $k_{\rm B}T_{\rm K}\hskip-2pt=\hskip-2pt O\bigl[t^2/(DU)\bigr]$.\cite{comT-depTK}
We assume that $D$ is so small that $T_c\hskip-2pt \ll\hskip-2pt T_{\rm K}$. 
If $T$ is so low that $T_c\hskip-2pt<\hskip-2pt T\hskip-2pt\ll\hskip-2pt T_{\rm K}$, quantum spin fluctuations are more dominant than thermal spin fluctuations, the entropy is small, and electrons are itinerant. The RVB liquid is stabilized in the low-$T$ phase of $T_c\hskip-2pt<\hskip-2pt T\hskip-2pt\ll\hskip-2pt T_{\rm K}$ in sufficiently low dimensions. 
If $T$ is so high that $T\hskip-2pt\gg\hskip-2pt T_{\rm K}$, thermal spin fluctuations are more dominant than quantum spin fluctuations, the entropy is as large as $k_{\rm B}\ln2$ per unit cell, and electrons behave as local moments.
Since the high-$T$ phase where $T\hskip-2pt\gg\hskip-2pt T_{\rm K}$ behaves as an insulator, it is simply the Mott insulator.
An MI transition or crossover occurs at $T\hskip-2pt\simeq\hskip-2pt T_{\rm K}$ as a function of $T$ between the RVB liquid and the Mott insulator.
Then, a high-$T$ phase where $T\hskip-2pt\gtrsim\hskip-2pt T_{\rm K}$, rather than $T\hskip-2pt\gg\hskip-2pt T_{\rm K}$, is also the Mott insulator.
If $D$ is so large that $T_c\hskip-2pt \gtrsim\hskip-2pt T_{\rm K}$, a high-$T$ phase where $T\hskip-2pt>\hskip-2pt T_c$ is the Mott insulator because $T\hskip-2pt \gtrsim\hskip-2pt T_{\rm K}$.
The Mott insulator is stabilized as a high-$T$ phase with a large entropy in low or high dimensions, while the RVB liquid is stabilized as a low-$T$ phase with a small entropy only in sufficiently low dimensions.
The Mott insulator and the RVB liquid never contradict each other.

If $T\hskip-2pt\le\hskip-2pt T_c$, the symmetry is broken, except for one dimension.
If $T_c\gtrsim T_{\rm K}$, not only the N\'{e}el state but also the type of insulator proposed by Slater\cite{slater} is possible below $T_c$ in a multiband model, as will be discussed later in Sect.\hskip2pt\ref{SecMIinActual}.

Since the RVB mechanism is a multisite effect, it never appears under S$^3$A; it is expected that $k_{\rm B}T_{\rm K}\hskip-2pt=\hskip-2pt 0$ will be possible for $U/|t|\gg 1$. Since any type of the conventional Weiss mean field is also a multisite effect, it never appears under S$^3$A; no symmetry can be broken. It is expected that the Mott insulator, which is a high-$T$ phase with a large entropy, will be possible even at $T\hskip-2pt=\hskip-2pt 0\hskip2pt$K under S$^3$A.

According to a numerical study\cite{RevModDMFT} based on DMFT or S$^3$A, the Mott insulator is possible even at $T\hskip-2pt=\hskip-2pt0\hskip2pt$K. 
An MI transition with hysteresis occurs below the critical temperature $T_0$.
The hysteresis is characterized by two transition lines of $U_{c1}(T)$ and $U_{c2}(T)$ in the $T$-$U$ phase diagram. The model used in the numerical study is different from that used in the present study. If the absolute bandwidth is denoted by $W$, $U_{c1}(T)\hskip-2pt\simeq\hskip-2pt W$, $U_{c2}(T)\hskip-2pt\simeq\hskip-2pt W$, and $U_{c1}(T)\hskip-2pt<\hskip-2pt U_{c2}(T)$ for $T\hskip-2pt<\hskip-2pt T_0$; $U_{c1}(T_0)\hskip-2pt=\hskip-2pt U_{c2}(T_0)$.
When $U$ increases, an MI transition occurs at $U\hskip-2pt=\hskip-2pt U_{c2}(T)$; when $U$ decreases, it occurs at $U\hskip-2pt=\hskip-2pt U_{c1}(T)$.
The insulating phase at $T\hskip-2pt=\hskip-2pt 0\hskip2pt$K for $U\hskip-2pt \ge\hskip-2pt U_{c1}(0\hskip2pt{\rm K})$ or $U\hskip-2pt \ge\hskip-2pt U_{c2}(0\hskip2pt{\rm K})$, depending on the decreasing or increasing process of $U$, is a typical type of the Mott insulator, i.e., {\it the Mott insulator in which a complete gap opens}.

According to our previous paper,\cite{FJO-MottIns} either under or beyond S$^3$A, a complete gap opens if and only if 
$\tilde{\Sigma}_\sigma(\varepsilon+{i}0)$ and/or $\tilde{\Gamma}(\varepsilon+{i}0)$ of the Anderson model has a pole at $\varepsilon\hskip-2pt =\hskip-2pt 0$; when $\mu\hskip-2pt=\hskip-2pt\mu_0$, e.g., a gap as large as $\epsilon_0$ opens if and only if
\begin{subequations}\label{EqGapScnario}
\begin{align}\label{EqGapScnario1}
&\tilde{\Sigma}_\sigma(\hskip-0.5pt 
\varepsilon \hskip-1.5pt + \hskip-1.5pt {i}0\hskip-0.5pt)
\hskip-2pt =\hskip-2pt
\frac{U}{2} \hskip-1.75pt +\hskip-1.75pt 
\frac{|\lambda_\Sigma|}{\varepsilon \hskip-1.5pt + \hskip-1.5pt {i}0} 
%\nonumber \\ & \quad
\hskip-1.5pt -\hskip-2pt
\frac1{\pi}\hskip-2.5pt \left[\hskip-1pt \int_{-\infty}^{-\epsilon_{0}\hskip-0.5pt/\hskip-0.5pt2}
\hskip-23pt d\epsilon \hskip7pt + \hskip-4pt
\int_{\hskip-0.5pt +\epsilon_{0}\hskip-0.5pt/\hskip-0.5pt 2}^{+\infty}
\hskip-18pt d\epsilon \hskip6pt
\right] \hskip-3.5pt \frac{{\rm Im} 
\tilde{\Sigma}_\sigma(\hskip-0.75pt \epsilon \hskip-1.5pt + \hskip-1.5pt{i}0 \hskip-0.5pt)}
{\varepsilon \hskip-1pt - \hskip-1pt \epsilon \hskip-1pt +\hskip-1pt {i}0},
%%%%%
\\ \label{EqGapScnario2} &
\tilde{\Gamma}(\hskip-0.5pt 
\varepsilon \hskip-1.5pt + \hskip-1.5pt {i}0\hskip-0.5pt)
\hskip-2pt = 
\frac{|\lambda_\Gamma|}{\varepsilon \hskip-1.5pt + \hskip-1.5pt {i}0} 
%\nonumber \\ & \quad
\hskip-1.5pt -\hskip-2pt
\frac1{\pi}\hskip-2.5pt \left[\hskip-1pt \int_{-\infty}^{-\epsilon_{0}\hskip-0.5pt/\hskip-0.5pt2}
\hskip-23pt d\epsilon \hskip6pt + \hskip-4pt
\int_{\hskip-0.5pt +\epsilon_{0}\hskip-0.5pt/\hskip-0.5pt 2}^{+\infty}
\hskip-18pt d\epsilon \hskip6pt
\right] \hskip-3.5pt \frac{{\rm Im} 
\tilde{\Gamma}(\hskip-0.75pt \epsilon \hskip-1.5pt + \hskip-1.5pt{i}0 \hskip-0.5pt)}
{\varepsilon \hskip-1pt - \hskip-1pt \epsilon \hskip-1pt +\hskip-1pt {i}0},
\end{align}
\end{subequations}
with $|\lambda_\Sigma|\hskip-2pt > \hskip-2pt0$ or $|\lambda_\Gamma|\hskip-2pt >\hskip-2pt 0$, are satisfied.
If $|\lambda_\Sigma|> 0$, or if $\tilde{\Sigma}_\sigma(\varepsilon\hskip-1pt +\hskip-1pt {i}0)$ has a pole at $\varepsilon\hskip-2pt =\hskip-2pt 0$, $k_{\rm B}T_{\rm K}\hskip-2pt =\hskip-2pt 0$ and the residual entropy of the Hubbard model is $k_{\rm B}\hskip-1pt \ln 2$ per unit cell.
If $\lambda_\Sigma\hskip-1pt =\hskip-1pt 0$ and $|\lambda_\Gamma|\hskip-2pt >\hskip-2pt 0$, or if only $\tilde{\Gamma}(\varepsilon\hskip-1pt +\hskip-1pt {i}0)$ has a pole at $\varepsilon\hskip-2pt =\hskip-2pt 0$, the residual entropy is zero;
the argument in our previous paper\cite{FJO-MottIns} that if $|\lambda_\Gamma|\hskip-2pt >\hskip-2pt 0$ then $|\lambda_\Sigma|$ has to be nonzero is irrelevant.

Since $\tilde{\Gamma}(\varepsilon+{i}0)$ can have no pole at $\varepsilon\hskip-2pt=\hskip-2pt 0$ under S$^3$A,\cite{FJO-MottIns}
$\tilde{\Sigma}_\sigma(\varepsilon+{i}0)$ has to have a pole at $\varepsilon\hskip-2pt=\hskip-2pt 0$ for the Mott insulator at $T=0\hskip2pt$K. Then, $k_{\rm B}T_{\rm K}\hskip-2pt=\hskip-2pt 0$ and the residual entropy is $k_{\rm B}\ln 2$ per unit cell, i.e., the third law of thermodynamics is broken. 
The Mott insulator at $T=0\hskip2pt$K has to be regarded as a high-$T$ phase rather than a low-$T$ phase. 
Either under or beyond S$^3$A, the Mott insulator can be stabilized only as a high-$T$ phase with a large entropy but can never be stabilized as a low-$T$ phase with a small entropy, even if it is stabilized at $T\hskip-1pt=\hskip-1pt0\hskip2pt$K.

If symmetry breaking is ignored, it is easy to extend the analysis based on KLT to $T\hskip-2pt \le\hskip-2pt T_c$, except for $T\hskip-2pt=\hskip-2pt0\hskip2pt$K.
If $T\hskip-2pt>\hskip-2pt0\hskip2pt$K, $\rho_\mu(\varepsilon)\hskip-2pt>\hskip-2pt0$ even if the ground state is an insulator. Then, $k_{\rm B}T_{\rm K}\hskip-2pt=\hskip-2ptO\bigl[t^2/(DU)\bigr]$.
If $T\hskip-2pt\gtrsim\hskip-2pt T_{\rm K}$, the Mott insulator is stabilized; if $0\hskip2pt{\rm K}\hskip-2pt<\hskip-2pt T\hskip-2pt \ll\hskip-2pt T_{\rm K}$, the RVB liquid is stabilized.
An MI transition or crossover occurs at $T\hskip-2pt\simeq\hskip-2pt T_{\rm K}$ as a function of $T$ between them.
The analysis for $T\hskip-2pt>\hskip-2pt0\hskip2pt$K never denies the possibility that $\rho_\mu(0)\hskip-2pt=\hskip-2pt0$ for $T\hskip-2pt=\hskip-2pt0\hskip2pt$K, i.e., the ground state is an insulator.\cite{com3rdLaw}

Since no higher-order term in $1/D$ is included in S$^3$A, strictly speaking, it is not a theory for $1/D\hskip-2pt \rightarrow\hskip-2pt 0$ but a theory for exactly $1/D\hskip-2pt=\hskip-2pt0$.
It should be determined whether, if symmetry breaking is ignored, S$^3$A is equivalent to KLT in the limit $1/D\hskip-1pt \rightarrow\hskip-1pt 0$ and rigorous in the limit $1/D\hskip-1pt \rightarrow\hskip-1pt 0$.

% or for infinite dimensions.

Beyond S$^3$A, or in KLT, $k_{\rm B}T_{\rm K}\hskip-2pt=\hskip-2ptO\bigl[t^2/(DU)\bigr]$ is more or less nonzero.
In the limit of $1/D\hskip-2pt\rightarrow\hskip-2pt 0$ followed by $T\hskip-2pt\rightarrow\hskip-2pt 0\hskip2pt$K, $T\hskip-2pt\gg\hskip-2pt T_{\rm K}$ and
the Mott insulator is stabilized; the numerical study\cite{RevModDMFT} based on DMFT or S$^3$A is consistent with KLT. 
In the limit of $T\hskip-2pt\rightarrow\hskip-2pt 0\hskip2pt$K followed by $1/D\hskip-2pt\rightarrow\hskip-2pt 0$, $T\hskip-2pt\ll\hskip-2pt T_{\rm K}$ and the RVB liquid is stabilized; the numerical study is inconsistent with KLT.
The inconsistency means or implies that S$^3$A is not rigorous in the limit $1/D\rightarrow 0$.

In the limit $U/|t|\hskip-2pt\rightarrow\hskip-2pt 0$, if $T$ is so low that $k_{\rm B}T\hskip-2pt \ll\hskip-2pt |t|$, a metal is stabilized, because $k_{\rm B}T_{\rm K}\hskip-2pt=\hskip-2pt O(|t|)$ and $T\hskip-2pt\ll\hskip-2pt T_{\rm K}$. 
In the limit $U/|t|\hskip-2pt \rightarrow\hskip-2pt +\infty$, if $T$ is nonzero, the Mott insulator is stabilized, because $k_{\rm B}T_{\rm K}\hskip-2pt=\hskip-2pt O[t^2/(DU)]\hskip-2pt \rightarrow \hskip-2pt0$ as $U/|t|\hskip-2pt \rightarrow\hskip-2pt +\infty$ and $T\hskip-2pt\gg\hskip-2pt T_{\rm K}$.
If $T$ is nonzero and sufficiently low such that $0<k_{\rm B}T\hskip-2pt \ll\hskip-2pt |t|$, an MI crossover or transition occurs as a function of $U$.
Since $k_{\rm B}T_{\rm K}\hskip-2pt=\hskip-2pt O[t^2/(DU)]$ is nonzero for any $|t|/U\hskip-2pt>\hskip-2pt 0$ unless $1/D\hskip-2pt=\hskip-2pt 0$, it is expected that no transition but only a crossover will occur beyond S$^3$A.
If this expectation is true, the transition in the numerical study\cite{RevModDMFT} is inconsistent with the crossover beyond S$^3$A.
Against the expectation,
we assume that a transition is possible at a sufficiently low $T$ beyond S$^3$A.
We consider the model used in the numerical study;
its absolute bandwidth, which is finite, is also denoted by $W$, as in the discussion above. The Kondo energy is given by $k_{\rm B}T_{\rm K}\hskip-2pt=\hskip-2pt O[W^2/(DU)]$ in the limit $T\hskip-2pt\rightarrow \hskip-2pt0\hskip2pt$K.
Since no transition is possible as a function of $T$ at least in the limit $U/W\hskip-2pt\rightarrow\hskip-2pt 0$, a critical point has to exist in the $T$-$U$ phase diagram. 
If hysteresis exists, $U_{c1}(+0\hskip2pt{\rm K})\hskip-2pt<\hskip-2pt U_{c2}(+0\hskip2pt{\rm K})$; if not, $U_{c1}(+0\hskip2pt{\rm K})\hskip-2pt=\hskip-2pt U_{c2}(+0\hskip2pt{\rm K})$.
Since $k_{\rm B}T_{\rm K}\hskip-2pt=\hskip-2pt O[W^2/(DU)]$ is more or less nonzero for any $W/U\hskip-2pt >\hskip-2pt 0$ unless $1/D\hskip-2pt=\hskip-2pt 0$, $U_{c2}(T)\hskip-2pt\rightarrow\hskip-2pt +\infty$ as $T\hskip-2pt\rightarrow \hskip-2pt 0\hskip2pt$K for any finite $D$; i.e., at least $U_{c2}(+0\hskip2pt{\rm K})$ is infinite beyond S$^3$A, regardless of whether hysteresis exists or not.
In the numerical study, on the other hand, hysteresis appears and either $U_{c1}(0\hskip2pt{\rm K})$ or $U_{c2}(0\hskip2pt{\rm K})$ is finite and $O(W)$,\cite{comPhaseDiagram} as discussed above.
The finite $U_{c2}(0\hskip2pt{\rm K})$ or $U_{c2}(+0\hskip2pt{\rm K})$ in the numerical study is inconsistent with the infinite $U_{c2}(+0\hskip2pt{\rm K})$ beyond S$^3$A.
The $T$-$U$ phase diagram of the numerical study is inconsistent with that beyond S$^3$A, regardless of whether an MI crossover or transition occurs beyond S$^3$A.
This finding also means or implies that S$^3$A is not rigorous even in the limit $1/D\hskip-2pt\rightarrow\hskip-2pt 0$.

We consider an electron state in the limit $T\rightarrow 0\hskip2pt$K, never at $T=0\hskip2pt$K. 
If $k_{\rm B}T_{\rm K}>0$, 
the electron state is a metal.
Thus, ${\rm Im}\Sigma_\sigma(+{i}0,{\bm k})\rightarrow 0$ as $T\rightarrow 0\hskip2pt{\rm K}$. Then, 
\begin{subequations}\label{EqConstRho}
\begin{align}
\rho_\mu(0) =\frac1{L}\sum_{\bm k}
\delta\bigl[\mu-E({\bm k})-{\rm Re}\Sigma_\sigma(+{i}0,{\bm k})\bigr],
\end{align}
and, according to the Fermi-surface sum rule,\cite{Luttinger1,Luttinger2} 
\begin{align}
n(\mu) = \frac1{L}\sum_{{\bm k}\sigma}
\theta\bigl[\mu-E({\bm k})-{\rm Re}\Sigma_\sigma(+{i}0,{\bm k})\bigr],
\end{align}
\end{subequations}
% \cite{Gutzwiller1,Gutzwiller2,Gutzwiller3} 
%
where $\theta(x)\hskip-2pt=\hskip-2pt(1\hskip-2pt+\hskip-2pt x/|x|)/2$.
Under S$^3$A or SSA, the self-energy $\Sigma_\sigma(+{i}0,{\bm k})$ does not depend on ${\bm k}$.
It is easy to show that $\rho_\mu(0)$ is constant as a function of $U$ if $n(\mu)$ is kept constant; $\rho_\mu(0)$ is simply given by that for $U\hskip-2pt =\hskip-2pt 0$.
The constant $\rho_\mu(0)$ as a function of $U$ is a property peculiar to the metallic phase under SSA.
If the Mott transition occurs, the decrease in $\rho_\mu(0)$ with increasing $U$ is necessarily discontinuous from the constant $\rho_\mu(0)$ of the metal to the zero $\rho_\mu(0)$ of the Mott insulator. 
On the other hand, if the RVB mechanism is considered and if Eq.\hskip2pt(\ref{EqAsympFinal}) or $1/\tilde{\phi}_1\hskip-2pt=\hskip-2pt O[t^2/(DU)]$ is used, 
$\Delta\Sigma_\sigma^{\rm (RVB)}(+{i}0,{\bm k})\hskip-2pt \propto\hskip-2pt (1/D)^0 U \varphi_D({\bm k})$.\cite{comKLT1/D}
If $U/|t|\hskip-2pt\gg\hskip-2pt 1$, the dispersion of $E({\bm k})$ can be ignored in Eq.\hskip2pt(\ref{EqConstRho}). Then, $\rho_\mu(0) \hskip-2pt=\hskip-2pt O(1/U)$ for $\mu\hskip-2pt=\hskip-2pt \mu_0$ or $n(\mu)\hskip-2pt=\hskip-2pt 1$.
The continuous decrease in $\rho_\mu(0)$ with increasing $U$ is a property peculiar to the RVB liquid with the half filling. 
The inconsistency in the $U$ dependence of $\rho_\mu(0)$ between under and beyond S$^3$A also means or implies that S$^3$A is not rigorous even in the limit $1/D\hskip-2pt\rightarrow\hskip-2pt 0$.

Any of the three inconsistencies discussed above between the numerical study\cite{RevModDMFT} and the present paper is simply 
because the RVB mechanism cannot be considered under S$^3$A but can be considered beyond S$^3$A. 
Beyond S$^3$A, $k_{\rm B}T_{\rm K}\hskip-2pt=\hskip-2pt O\bigl[t^2/(DU)\bigr]$ is nonzero.
If {\it the Mott insulator in which a completely opens}, which is characterized by $k_{\rm B}T_{\rm K}\hskip-2pt=\hskip-2pt0$, is possible at $T\hskip-2pt=\hskip-2pt 0\hskip2pt$K under S$^3$A, 
S$^3$A is not necessarily equivalent to KLT in the limit $1/D\hskip-2pt\rightarrow\hskip-2pt 0$ nor necessarily rigorous in the limit $1/D\hskip-2pt\rightarrow\hskip-2pt 0$.
The rigorousness of S$^3$A in the limit $1/D\hskip-2pt\rightarrow\hskip-2pt 0$ is examined in Appendix\ref{AppDMFT}.

%%%%%%%%%%%%%%%%%%%%%%%%%% 
\subsection{Nature of Lieb and Wu's insulator}
\label{SecNatureLW}
If the fact that $\rho_\mu(0)\hskip-2pt>\hskip-2pt0$ for $T\hskip-2pt>\hskip-2pt0\hskip2pt$K is seriously considered, the RVB-TL liquid is stabilized in the low-$T$ phase of $0\hskip2pt{\rm K}\hskip-2pt<\hskip-2pt T\hskip-2pt\ll\hskip-2pt |J|/k_{\rm B}$ in one dimension; the possibility is not denied that $\rho_\mu(0)\hskip-2pt=\hskip-2pt0$ for $T\hskip-2pt=\hskip-2pt0\hskip2pt$K, i.e., the ground state is an insulator.\cite{com3rdLaw}
According to the treatment in Sect.\hskip2pt\ref{SecRigidity}, the ground state for $|\mu\hskip-1pt-\hskip-1pt\mu_0|\hskip-2pt<\hskip-2pt(1/2)\epsilon_{\rm G}(U)$ in the grand canonical ensemble is simply Lieb and Wu's insulator itself, which is given by the Bethe-ansatz solution for the canonical ensemble.
Here, we examine the nature of this insulator.
We assume that $D\hskip-2pt=\hskip-2pt 1$, $T\hskip-2pt=\hskip-2pt 0\hskip2pt{\rm K}$, and $U/|t|\hskip-2pt\gg\hskip-2pt 1$, unless otherwise noted.

If the Bethe-ansatz solution is used, it is possible to determine all the physical properties in the grand canonical ensemble with essentially the same treatment as that in Sect.\hskip2pt\ref{SecRigidity}, in principle; then, 
it is also possible to determine the mapped Anderson model, in principle.
We assume that the Anderson model is determined and solved.
Since a complete gap opens in Lieb and Wu's insulator and its residual entropy per unit cell is zero or infinitesimal in the thermodynamic limit, the scenario that 
$\lambda_\Sigma\hskip-2.5pt=\hskip-2.5pt0$ and $|\lambda_\Gamma|\hskip-2.5pt>\hskip-2.5pt 0$ in Eq.\hskip1.5pt(\ref{EqGapScnario}) is only possible for the insulator: $\tilde{\Sigma}_\sigma(\varepsilon \hskip-2pt+\hskip-2pt {i}0)$ is analytic at $\varepsilon\hskip-2pt=\hskip-2pt 0$, but $\tilde{\Gamma}(\varepsilon\hskip-2pt+\hskip-2pt{i}0)$ has a pole at $\varepsilon\hskip-2pt=\hskip-2pt 0$, or just on $\mu$.
If the insulator is {\it rigid} against the movement of $\mu$, as discussed in Sect.\hskip2pt\ref{SecRigidity}, the pole of $\tilde{\Gamma}(\varepsilon\hskip-2pt+\hskip-2pt{i}0)$ moves as $\mu$ moves. If $\mu\hskip-2pt\ne\hskip-2pt \mu_0$, $\tilde{\Gamma}(\varepsilon\hskip-2pt+\hskip-2pt{i}0)$ has a pole at $\varepsilon\hskip-2pt\ne\hskip-2pt 0$ on the real axis. Since $\Delta(\varepsilon)\hskip-2pt=\hskip-2pt-{\rm Im}\tilde{\Gamma}(\varepsilon\hskip-2pt+\hskip-2pt{i}0)$, as shown in Eq.\hskip2pt(\ref{EqAndersonD}), $\Delta(\varepsilon)\hskip-2pt=\hskip-2pt0$ for $\varepsilon\hskip-2pt\simeq\hskip-2pt 0$.
If so, there is no Fermi surface in the Anderson model, so that the ground state of the Anderson model is not the normal Fermi liquid and $\tilde{\Sigma}_\sigma(\varepsilon\hskip-2pt+\hskip-2pt{i}0)$ is not analytic at $\varepsilon\hskip-2pt=\hskip-2pt0$.
There is inconsistency between the possibility and {\it rigidity} of Lieb and Wu's insulator.

Three explanations are possible for this inconsistency:
One is the pinning of the chemical potential, as in Wilson's insulator.
If not only the long-range Coulomb interaction but also the formation of an electric double layer between Wilson's insulator and its reservoir is considered, the gap center is pinned to the chemical potential of the reservoir.
If a similar or different type of pinning is possible between Lieb and Wu's insulator and its reservoir, it is possible that the band center $\epsilon_d$ is adjusted in such a way that the gap center $\mu_0\hskip-2pt=\hskip-2pt\epsilon_d\hskip-2pt+\hskip-2pt(1/2)U$ is pinned to the chemical potential of the reservoir.
However, it seems plausible that there is no appropriate pinning mechanism in one dimension. 
Another is that Lieb and Wu's insulator is so singular that it cannot be treated by KLT. Either the gap function or the ground-state energy is singular at $U\hskip-2pt=\hskip-2pt 0$ as a function of $U$.\cite{lieb-wu,takahashi}
This means that Lieb and Wu's insulator cannot be treated by a simple perturbation in terms of $U$.
If the gap-opening or Eq.\hskip2pt(\ref{EqGapScnario}) is assumed from the beginning,\cite{ComFeynmanMethod} KLT may treat Lieb and Wu's insulator for $\mu\hskip-2pt=\hskip-2pt \mu_0$; however, 
KLT can never treat Lieb and Wu's insulator for $\mu\hskip-2pt\ne\hskip-2pt \mu_0$. In other words, the Bethe-ansatz solution for Lieb and Wu's insulator may be a self-consistent solution of KLT for $\mu\hskip-2pt=\hskip-2pt \mu_0$ but can never for $\mu\hskip-2pt\ne\hskip-2pt \mu_0$.
This finding means or implies that Lieb and Wu's insulator is unstable or impossible in the grand canonical ensemble, at least for $\mu\hskip-2pt\ne\hskip-2pt \mu_0$.
Then, the most probable explanation is that the ground state for $|\mu\hskip-2pt-\hskip-2pt\mu_0|\hskip-2pt<\hskip-2pt(1/2)\epsilon_{\rm G}(U)$ is not Lieb and Wu's insulator itself, which is an eigenstate of $\mathcal{N}$, but an electron state that is no eigenstate of $\mathcal{N}$,
a type of insulator different from Lieb and Wu's insulator 
or simply the RVB-TL electron liquid in the limit $T\hskip-2pt\rightarrow\hskip-2pt 0\hskip2pt$K.
Not only an excited state but also the ground state has to be {\it nonrigid} against the movement of $\mu$ in the grand canonical ensemble, except for Wilson's band insulator; the ground state can be {\it nonrigid} at least if it is more or less a linear combination or mixture of different $N$ states, e.g., because of an electron reservoir.

We assume the {\it explicit} presence of an electron reservoir in $D\hskip-2pt\ge\hskip-2pt 1$ dimensions.
A many-body eigenstate is no eigenstate of $\mathcal{N}$. 
We define the {\it efficiency} of the reservoir by\cite{comDeltaN} 
\begin{align}\label{EqDeltaN}
\delta N = \sqrt{\left<(\mathcal{N}-\left<\mathcal{N}\hskip1pt\right>)^2\right>}.
\end{align}
It is plausible that $\delta N\hskip-2pt\gg \hskip-2pt 1$ and $\delta N/L \hskip-2pt\rightarrow\hskip-2pt 0$ as $L\hskip-2pt\rightarrow\hskip-2pt +\infty$, if the reservoir is appropriate.
If more or less $\delta N\hskip-2pt>\hskip-2pt 0$, $\left<\mathcal{N}\right>$ can be not only an integer but also an irrational number; $\left<\mathcal{N}\right>$ is continuous as a function of $\mu$.
If no symmetry breaking occurs,
it is likely that the nature of electron correlation is continuous as a function of the continuous variable $\left<\mathcal{N}\right>$; then,
it is unlikely that there exists a critical deviation $\delta N_{c}$ from the half filling such that the ground state is an insulator for $|\hskip-2pt\left<\mathcal{N}\right> \hskip-1pt-\hskip-1pt L|\hskip-2pt < \hskip-2pt \delta N_{c}$ but is a metal for $|\hskip-2pt\left<\mathcal{N}\right> \hskip-1pt-\hskip-1pt L|\hskip-2pt > \hskip-2pt \delta N_{c}$.
This argument implies that neither Lieb and Wu's insulator nor the Mott insulator in which a complete gap opens is possible.

In our previous paper,\cite{toyama} the reservoir effect is considered with a simple model in $D\hskip-2pt\ge\hskip-2pt 1$ dimensions in which the translational symmetry is restored by the ensemble average. If symmetry breaking is ignored, the Green function is given by
\begin{align}\label{EqGreenGammma}
G_\sigma({i}\varepsilon_l, {\bm k})\hskip-2pt =\hskip-2pt
\frac1{{i}\varepsilon_l \hskip-2pt +\hskip-2pt \mu 
\hskip-2pt - \hskip-2pt E({\bm k}) \hskip-2pt - \hskip-2pt \Sigma_\sigma({i}\varepsilon_l,{\bm k})
\hskip-2pt - \hskip-2pt \Gamma_{\rm R}({i}\varepsilon_l)},
\end{align}
where $\Gamma_{\rm R}({i}\varepsilon_l)$ is due to hybridization with the reservoir. From the mapping condition, it follows that\cite{toyama}
\begin{align}
\Delta (\varepsilon) \ge -{\rm Im}\Gamma_{\rm R}(\varepsilon \hskip-1pt + \hskip-1pt{i}0).
\end{align}
If the reservoir is appropriate, $-{\rm Im}\Gamma_{\rm R}(\varepsilon \hskip-1pt + \hskip-1pt{i}0)\hskip-2pt =\hskip-2pt +0^+$.
Since $\Delta (\varepsilon)$ is nonzero, $k_{\rm B}T_{\rm K}$ has to be more or less nonzero and the single-site $\tilde{\Sigma}_\sigma(\varepsilon+{i}0)$ has to be more or less normal.
If $\Delta(0)\hskip-2pt<\hskip-2pt+\infty$ is assumed, it is easy to extend the analysis in Sect.\hskip2pt\ref{SecRVB} to this model, regardless of $T$. 
If $|n(\mu)\hskip-1pt -\hskip-1pt 1|\hskip-2pt \lesssim\hskip-2pt |t|/(DU)$, e.g., the ground state is the RVB liquid with $k_{\rm B}T_{\rm K}\hskip-2pt=\hskip-2pt O(|J|/D)$, $\rho_\mu(0)\hskip-2pt=\hskip-2pt O(1/U)$, and $\Delta(0)\hskip-2pt=\hskip-2pt O(U)$; the eventual $\Delta(0)$ is consistent with the assumption of $\Delta(0)\hskip-2pt<\hskip-2pt+\infty$.
The liquid is not {\it rigid} against the movement of $\mu$; e.g., the three-peak structure with the midband between the Hubbard bands varies with $\mu$.
Since $\Delta (\varepsilon)\hskip-2pt >\hskip-2pt 0$ for $T\hskip-2pt =\hskip-2pt 0\hskip2pt$K is no assumption, it is definite that
the Mott insulator at $T\hskip-2pt =\hskip-2pt 0\hskip2pt$K in which a complete gap opens is impossible. 
It is reasonable that the Mott insulator at $T\hskip-2pt =\hskip-2pt 0\hskip2pt$K is unstable even against the infinitesimal $\Gamma_{\rm R}(\varepsilon \hskip-1pt + \hskip-1pt{i}0)$, because it is infinitely degenerate. 
Since $\Delta(0)\hskip-2pt<\hskip-2pt+\infty$ or $\tilde{\Gamma}(\varepsilon\hskip-1pt + \hskip-1pt{i}0)$ with no pole at $\varepsilon\hskip-2pt =\hskip-2pt 0$ for $T\hskip-2pt =\hskip-2pt 0\hskip2pt$K is assumed, the possibility cannot be denied that an MI transition occurs at $T\hskip-2pt =\hskip-2pt 0\hskip2pt$K and the ground state is an insulator in which the third law of thermodynamics holds. The insulator, if possible, has to be a mixture of different $N$ states and cannot be Lieb and Wu's insulator itself, which is an eigenstate of $\mathcal{N}$; it is desirable to determine the critical $\delta N_c$ defined above and the critical $\mu$ corresponding to $N_c/L$ in order to confirm that either the transition or insulator at $T\hskip-2pt =\hskip-2pt 0\hskip2pt$K is actually possible, although either is of no physical significance.\cite{com3rdLaw}

If the reservoir is appropriate, the modification of many-body eigenstates of the Bethe-ansatz solution by the reservoir has to be small. 
It is, therefore, expected that 
\begin{subequations}\label{EqRigid}
\begin{align}
& \hskip10pt
n\left[\mu_0 \pm (1/2)\epsilon_{\rm G}(U)\right]=
1 \pm O(\delta N/L), 
\\ & 
\bigl[\chi_c(0,0)\bigr]_{|\mu-\mu_0|<(1/2)\epsilon_{\rm G}(U)}
\hskip-2pt =\hskip-2pt O\bigl[(\delta N/L)/\epsilon_{\rm G}(U)\bigr],
\end{align}
\end{subequations}
regardless of the ground state.
If $\delta N\hskip-2pt> \hskip-2pt 0$ once, metallic configurations with $N\hskip-2pt\ne\hskip-2pt L$ contribute to any statistical average in the grand canonical ensemble.
It is expected that more or less $\rho_\mu(0)\hskip-2pt>\hskip-2pt0$ regardless of $T$. 
If so, it is easy to extend the analysis for $T\hskip-2pt >\hskip-2pt 0\hskip2pt$K in Sect.\hskip2pt\ref{SecRVB} to $T\hskip-2pt =\hskip-2pt 0\hskip2pt$K.
The ground state is the RVB-TL liquid even for $|\mu-\mu_0|\hskip-2pt < \hskip-2pt (1/2)\epsilon_{\rm G}(U)$; it has to be simply that in the limit $T\hskip-2pt\rightarrow\hskip-2pt 0\hskip2pt$K.
If Eq.\hskip2pt(\ref{EqRigid}) is satisfied, many physical properties of the liquid cannot depend on $\mu$: e.g.,
$n(\mu)=1$, $\chi_c(0,0)\hskip-2pt = \hskip-2pt 0$,\cite{comInsComp} 
$k_{\rm B}T_{\rm K}\hskip-2pt =\hskip-2pt O(|J|/D)$, $\rho_\mu(0)\hskip-2pt =\hskip-2pt O(1/U)$, and so on. 
Few properties can depend on $\mu$: e.g., the three peak structure with the midband between the Hubbard bands. 
The liquid is not {\it rigid} against the movement of $\mu$.
It is expected that the physical properties of the liquid will be the same as those of Lieb and Wu's insulator, expect for those closely related to the itineracy of electrons.
On the other hand, the possibility cannot be denied that $\rho_\mu(0)\hskip-2pt=\hskip-2pt0$ even for $\delta N\hskip-2pt> \hskip-2pt 0$. 
In this case, the ground state is a type of insulator different from Lieb and Wu's insulator, because it cannot be {\it rigid} against the movement of $\mu$.
It is desirable to determine which is the ground state for $|\mu\hskip-1pt -\hskip-1pt \mu_0|\hskip-2pt<\hskip-2pt(1/2)\epsilon_{\rm G}(U)$ in the grand canonical ensemble, Lieb and Wu's insulator, a type of insulator different from it, or the RVB-TL liquid,\cite{com3rdLaw} particularly in the {\it explicit} presence of a realistic and appropriate electron reservoir, in which there is no translational symmetry, $\delta N\hskip-2pt\gg \hskip-2pt 1$, and $\delta N/L \hskip-2pt\rightarrow\hskip-2pt 0$ as $L\hskip-2pt\rightarrow\hskip-2pt +\infty$.

%the insulator is so peculiar that it is only possible for $N\hskip-2pt=\hskip-2pt L$ but never for $N\hskip-2pt\ne\hskip-2pt L$ and a complete gap opens in it even if no symmetry is broken in it.
%Lieb and Wu's insulator is originally the half-filled ground state in the canonical ensemble. If Lieb and Wu's insulator itself is also the ground state in the grand canonical ensemble, it is doubtful whether the conventional perturbative theory based on the Feynman-diagram method can treat the insulating ground state.
%
%It is doubtful whether the conventional perturbative theory based on the Feynman-diagram method can treat Lieb and Wu's insulator.
 
%If $T\hskip-2pt>\hskip-2pt0\hskip2pt$K once, a complete gap can never open and more or less $\rho_\mu(0)\hskip-2pt >\hskip-2pt 0$, even if a complete gap opens in the ground state.
%
%At least if more or less $\rho_\mu(0)\hskip-2pt >\hskip-2pt 0$, regardless of $T\hskip-2pt=\hskip-2pt0\hskip2pt$K or $T\hskip-2pt>\hskip-2pt0\hskip2pt$K, there is no doubt on the validity of the conventional perturbative theory, on which KLT is based.
%regardless of $T\hskip-2pt=\hskip-2pt0\hskip2pt$K or $T\hskip-2pt>\hskip-2pt0\hskip2pt$K, 

%
\subsection{RVB liquid in low dimensions}
\label{SecDiscussionRVB}
If $U/|t|\hskip-2pt\gg\hskip-2pt 1$ in sufficiently low-$D$ dimensions, $T_c\hskip-2pt\ll\hskip-2pt |J|/(k_{\rm B}D)$ can be satisfied.
The low-$T$ phase of $T_c\hskip-2pt<\hskip-2pt T\hskip-2pt\ll\hskip-2pt |J|/(k_{\rm B}D)$ is mainly studied in the present paper.
Since $\rho_\mu(0)$ of this phase is necessarily more or less nonzero,
there is no doubt that KLT can treat the phase, even if KLT cannot treat Lieb and Wu's insulator itself.
If $|n(\mu)\hskip-1pt -\hskip-1pt 1|\hskip-2pt \lesssim\hskip-2pt |t|/(DU)$,
the RVB liquid is stabilized in the low-$T$ phase.

% at $T\hskip-2pt=\hskip-2pt 0\hskip2pt$K in one dimension

First, we consider one dimension, for which $T_c\hskip-2pt=\hskip-2pt 0\hskip2pt$K.
The RVB liquid in one dimension is also the RVB-TL liquid.
The {\it intermediate} phase with $|\mu\hskip-1pt-\hskip-1pt\mu_0|\hskip-2pt\le\hskip-2pt(1/2)\epsilon_{\rm G}(U)$ is also simply the RVB-TL liquid.
Since Eq.\hskip2pt(\ref{EqRigid}) has to be satisfied in the limit $T\hskip-2pt\rightarrow\hskip-2pt 0\hskip2pt$K,
$n(\mu)\hskip-2pt\rightarrow\hskip-2pt 1$ and
$\chi_c(0,0)\hskip-2pt\rightarrow\hskip-2pt 0$ as $T\hskip-2pt\rightarrow\hskip-2pt 0\hskip2pt$K for the {\it intermediate} phase; $n(\mu)\hskip-2pt\simeq\hskip-2pt 1$ and $\chi_c(0,0)\hskip-2pt\simeq\hskip-2pt 0$ for the phase at $T\hskip-2pt>\hskip-2pt 0\hskip2pt$K.
On the other hand, if $|\mu\hskip-1pt-\hskip-1pt\mu_0|\hskip-2pt >\hskip-2pt (1/2)\epsilon_{\rm G}(U)$, $\chi_c(0,0)$ is nonzero regardless of $T$. 
It is expected that a metal-metal (MM) transition or crossover will occur at $\mu\hskip-2pt\simeq \hskip-2pt\mu_0\hskip-1pt\pm(1/2)\hskip-1pt\epsilon_{\rm G}(U)$ as a function of $\mu$ between the RVB-TL liquid with nonzero $\chi_c(0,0)$ and the {\it intermediate} phase with zero or small $\chi_c(0,0)$; if an MM crossover occurs, the crossover has to be very sharp at a sufficiently low $T$ and almost a transition in the limit $T\hskip-2pt\rightarrow\hskip-2pt 0\hskip2pt$K.
It is also expected that the conductivity of the {\it intermediate} phase will be metallic or of a {\it bad} metal due to $2k_{\rm F}$ and $4k_{\rm F}$ fluctuations peculiar to one dimension but never of an activation type whose activation energy is $O\left[\epsilon_{\rm G}(U)\right]$.
% even if the ground state is an insulator.
%
It is expected that the physical properties of the {\it intermediate} phase will be almost the same as those of Lieb and Wu's insulator in the canonical ensemble, except for those closely related to the itineracy of electrons.

We consider the Heisenberg limit at $T\hskip-2pt =\hskip-2pt +0\hskip2pt$K.
If only the RVB self-energy is considered beyond S$^3$A, the spectrum of a single-particle excitation of the RVB-TL liquid is given by
\begin{align}
\xi(k) &=
c_J J \cos(ka)-\mu^*,
\end{align}
where $c_J\simeq 1$ and $\mu^*=0$.
The spectrum of an electron-hole pair excitation is given by
\begin{align}
\omega(q) &= \xi(k+q)- \xi(k) 
\nonumber \\ &
= c_J |J|\bigl\{ \cos[(k+q)a] - \cos(ka)\bigr\},
\end{align}
where $\cos[(k+q)a] \hskip-2pt > \hskip-2pt 0$ and $\cos(ka)<0$.
Then, 
\begin{align}
c_J |J\sin(qa)| \le \omega(q) \le 2c_J |J\sin(qa/2)|.
\end{align}
In the Heisenberg limit, low-energy charge fluctuations are almost completely depressed; thus, the pair excitation $\omega(q)$ is almost a spin excitation.
The {\it spin-excitation} spectrum $\omega(q)$ in the Hubbard model is similar to the spin-excitation spectrum in the Heisenberg model.\cite{cloiseaux} 
The similarity of the {\it spin-excitation} spectrum is evidence of the adiabatic continuation between the RVB-TL electron and spin liquids.

According to previous papers,\cite{TL1,TL2,TL3} the spin liquid in the Heisenberg model is the TL spin liquid; i.e., it is the RVB-TL spin liquid.
This is also evidence of the adiabatic continuation between the RVB-TL electron and spin liquids.

The ground-state energy of Lieb and Wu's insulator as a function of a complex $z\hskip-2pt=\hskip-2pt t/U$ has no singularity at $z\hskip-2pt=\hskip-2pt 0$.\cite{takahashi} This fact means or implies that Lieb and Wu's insulator adiabatically continues to the RVB-TL spin liquid; if so, the insulator also adiabatically continues to the RVB-TL electron liquid. 
It is certain that the RVB mechanism is also crucial for the stabilization of Lieb and Wu's insulator.
The RVB-TL electron liquid in the {\it intermediate} phase in the grand canonical ensemble and Lieb and Wu's insulator in the canonical ensemble are never contradictory to each other.
The similarity and difference between them have to be similar to those between the RVB-TL electron and spin liquids.

Next, we consider two dimensions. 
The Hubbard model on the square lattice is also of particular interest; no symmetry can be broken at $T\hskip-2pt>\hskip-2pt 0\hskip2pt$K,\cite{mermin} $\rho_\mu(\varepsilon)$ for $U\hskip-2pt =\hskip-2pt 0$ diverges logarithmically as $\varepsilon\hskip-2pt\rightarrow\hskip-2pt 0$ because of the saddle-point van Hove singularity peculiar to two dimensions, and the Fermi surface for $U\hskip-2pt =\hskip-2pt 0$ shows a perfect nesting for ${\bm q}\hskip-2pt=\hskip-2pt{\bm Q}$, where ${\bm Q}\hskip-2pt=\hskip-2pt(\pm 1, \pm 1)(\pi/a)$, in the half-filled case.
We assume that $U/|t|\hskip-2pt\gg\hskip-2pt 1 $ and $\mu \hskip-2pt=\hskip-2pt\mu_0$.

The static susceptibility of the Anderson model is approximately given by Eq.\hskip2pt(\ref{EqSusAM}): $\tilde{\chi}_s(0;T)\hskip-2pt\simeq\hskip-2pt 4\tilde{\phi}_1\rho_\mu(0)$.
If $T$ is nonzero, the logarithmic divergence of $\tilde{\phi}_1\rho_\mu(\varepsilon)$ as $\varepsilon\hskip-2pt\rightarrow\hskip-2pt 0$ is suppressed by the imaginary part of the self-energy.
%If $T$ is sufficiently high such that $k_{\rm B}T\simeq |J|/2$, e.g., $\tilde{\chi}_s(0;T) \hskip-2pt =\hskip-2ptO[1/(2|J|)]$.
%
Since the suppression disappears in the limit $T\hskip-2pt\rightarrow\hskip-2pt 0\hskip2pt$K, it is expected that $\tilde{\chi}_s(0;T)\hskip-2pt\rightarrow\hskip-2pt +\infty$ as $T\hskip-2pt\rightarrow \hskip-2pt0\hskip1pt$K.\cite{Com2D}
The Fermi surface shows a sharp nesting for ${\bm Q}$, at least for $0\hskip2pt{\rm K}\hskip-2pt<T\hskip-2pt\ll\hskip-2pt |J|/(2k_{\rm B})$.
The superexchange interaction $J_s(0,{\bm q})$ is maximum at ${\bm q}\hskip-2pt=\hskip-2pt{\bm Q}$. Thus, the half-filled ground state is presumably the N\'{e}el state with the ordering vector ${\bm Q}$.
The low-$T$ phase of $0\hskip2pt{\rm K}\hskip-2pt<T\hskip-2pt\ll\hskip-2pt |J|/(2k_{\rm B})$ is the RVB electron liquid in the critical region.

We consider the $T$ dependence of the static homogeneous susceptibility; it can be described as 
%
%\begin{subequations}
\begin{align}\label{Eq1/Chi-2}
& 1/\chi_s(0,0; T) =
1/\chi_s(0,{\bm Q};T) 
\nonumber \\ & \hskip45pt + 
(1/4)\bigl[\Delta_s + \Delta_Q(T) - \Delta_\Gamma(T)\bigr],
\\ & \hskip15pt
\Delta_s = J_s(0,{\bm Q})- J_s(0,0)=4|J|, 
\\ & \hskip15pt
\Delta_Q(T) = J_Q(0,{\bm Q};T)- J_Q(0,0;T), 
\\ & \hskip15pt
\Delta_\Gamma(T) = \Lambda_\Gamma(0,{\bm Q};T)- \Lambda_\Gamma(0,0;T),
\end{align}
%\end{subequations}
%
from Eq.\hskip2pt(\ref{EqChiKondo}).
The $T$ dependence of $\Delta_s$ can be ignored.
It is expected that an anomaly will appear in $\chi_s(0,0; T)$ of the RVB liquid in the critical region, as discussed bolow.

%%It follows that $\Delta_s(T) \hskip-2pt= \hskip-2pt4|J|$.
%$J_s(0,{\bm Q};T)\hskip-2pt =\hskip-2pt 2|J|$, $J_s(0,0;T)\hskip-2pt=\hskip-2pt-2|J|$, and

As preliminary, we consider $D$ dimensions.
According to previous papers,\cite{FJO-supJ,miyai}
if $\mu$ lies in the vicinity of one of the band edges and $\rho_\mu(\varepsilon)$ has a sharp peak in the vicinity of $\mu$, $J_Q(0,{\bm q};T)$ is ferromagnetic; i.e., $J_Q(0,0;T)$ is positively large and $J_Q(0,0;T)$ increases almost linearly against $T$ as $T\hskip-2pt\rightarrow\hskip-2pt 0\hskip2pt$K. 
On the other hand, if $\mu$ lies around the band center and the Fermi surface shows a sharp nesting, it is antiferromagnetic; i.e., $J_Q(0,{\bm Q}_{\rm N};T)$ is positively large and $J_Q(0,{\bm Q}_{\rm N};T)$ increases almost linearly against $T$ as $T\hskip-2pt\rightarrow\hskip-2pt 0\hskip2pt$K, where ${\bm Q}_{\rm N}$ is the nesting wave number. 
The $T$ dependence of $J_Q(0,{\bm q};T)$ is a mechanism of the Curie-Weiss (CW) law, as will be discussed later in Sect.\hskip2pt\ref{SecItinerantLocal}.

We consider two dimeansions.
The Fermi surface shows a sharp nesting for ${\bm Q}\hskip-2pt=\hskip-2pt(\pm 1, \pm 1)(\pi/a)$, so that $J_Q(0,{\bm Q};T)$ is positively large at $T\ll |J|/(2k_{\rm B})$ and $J_Q(0,{\bm Q};T)$ increases as $T\hskip-2pt\rightarrow\hskip-2pt 0\hskip2pt$K.
The density of states $\rho_\mu(\varepsilon)$ has a logarithmic peak at the band center, so that $J_Q(0,0;T)$ is also positive at $T\ll |J|/(2k_{\rm B})$ and $J_Q(0,0;T)$ increases as $T\hskip-2pt\rightarrow\hskip-2pt 0\hskip2pt$K.
Since both $\mu$ and the peak of $\rho_\mu(\varepsilon)$ are at the band center, the nesting effect is larger than the logarithmic-peak effect. Then, $\Delta_Q(T)\hskip-2pt >\hskip-2pt0$
and the $T$ dependence of $J_Q(0,{\bm Q};T)$ is much stronger than that of $J_Q(0,0;T)$. 
The $T$ dependence of $\Delta_Q(T)$ is large; $\Delta_Q(T)$ increases as $T\hskip-2pt\rightarrow\hskip-2pt 0\hskip2pt$K.
The N\'{e}el temperature $T_{\rm N}$ cannot be nonzero because of critical fluctuations, or $\Lambda(0,{\bm q};T)$; this means that the $T$ dependence of $\Lambda(0,{\bm q};T)$ is large.
In general, the ${\bm q}$ dependence of the mode-mode coupling term $\Lambda(0,{\bm q};T)$ is small. Then, $\Delta_\Gamma(T)$ has to be small and the $T$ dependence of the small $\Delta_\Gamma(T)$ is also small.
In the critical region, the CW law is suppressed by $\Lambda(0,{\bm q};T)$, and $\chi_s(0,{\bm Q};T)$ is almost constant as a function of $T$.
Thus, according to Eq.\hskip2pt(\ref{Eq1/Chi-2}), the $T$ dependence of $1/\chi_s(0,0; T)$ resembles that of $\Delta_Q(T)$.
Since $\Delta_Q(T)$ increases as $T\hskip-2pt\rightarrow\hskip-2pt 0\hskip2pt$K, $1/\chi_s(0,0; T)$ increases as $T\hskip-2pt\rightarrow\hskip-2pt 0\hskip2pt$K, i.e., $\chi_s(0,0; T)$ decreases as $T\hskip-2pt\rightarrow\hskip-2pt 0\hskip2pt$K.
The decrease in or suppression of $\chi_s(0,0; T)$ as $T\hskip-2pt\rightarrow \hskip-2pt0\hskip2pt$K also occurs in the Heisenberg model on the square lattice. \cite{HZ2-1,HZ2-2}
This similarity of the suppression of $\chi_s(0,0; T)$ is evidence of the adiabatic continuation between the RVB electron and spin liquids.

It is easy to extend the study in the present paper to a different type of Hubbard model, i.e., one on a different symmetry or type of lattice and/or with not only $t$ between nearest neighbors but also $t^\prime$, $t^{\prime\prime}$, and so on between other neighbors.
The superexchange interaction appears between not only nearest neighbors but also other neighbors: $J\hskip-2pt\propto\hskip-2pt |t|^2/U$, $J^\prime\hskip-2pt\propto\hskip-2pt |t^\prime|^2/U$, $J^{\prime\prime}\hskip-2pt\propto\hskip-2pt |t^{\prime\prime}|^2/U$, and so on. Since the RVB mechanism is of the first order in the superexchange interaction, the eventual stabilization energy or $k_{\rm B}T_{\rm K}$ is the sum of the contributions of $J$, $J^\prime$, $J^{\prime\prime}$, and so on.
Since $k_{\rm B}T_{\rm K}$ is nonzero for a finite $D$,
the same qualitative conclusion as that for the Hubbard model of Eq.\hskip2pt(\ref{EqHubbardModel}) can be drawn for a different type of Hubbard model.

The most interesting extension is that to the triangular lattice.
If $T\hskip-2pt>\hskip-2pt 0\hskip2pt$K, no symmetry is broken \cite{mermin}.
The electron state at $0\hskip2pt{\rm K}\hskip-2pt<\hskip-2pt T \hskip-2pt\ll\hskip-2pt |J|/(Dk_{\rm B})$ in the Hubbard model is a frustrated electron liquid, and the spin state at $0\hskip2pt{\rm K}< T \ll |J|/(Dk_{\rm B})$ in the Heisenberg model is the RVB spin liquid proposed by Fazekas and Anderson.\cite{fazekas}
We propose that the frustrated electron liquid in the Hubbard model is simply the RVB electron liquid, and that the RVB electron and spin liquids in the Hubbard and Heisenberg models on the triangular lattice adiabatically continue to each other.

In three dimensions and higher, it is possible that $T_{\rm N}\hskip-2pt \ll\hskip-2pt |J|/(Dk_{\rm B})$, at least, if frustration or quasi-low dimensionality is sufficient in the Hubbard and Heisenberg models.
If $T_{\rm N}\hskip-2pt\ll\hskip-2pt |J|/(Dk_{\rm B})$, it is interesting to study how magnetic properties at $T_{\rm N}\hskip-2pt<\hskip-2pt T\hskip-2pt \ll\hskip-2pt |J|/(Dk_{\rm B})$ resemble each other between an electron liquid in the Hubbard model in the strong-coupling region, which is the RVB electron liquid, and a spin liquid in the Heisenberg model, which is the RVB spin liquid.

\subsection{Itinerant electrons versus local moments}
\label{SecItinerantLocal}
Electrons behave as local moments at $T\hskip-2pt\gg\hskip-2pt T_{\rm K}$, or local moments form at $T\hskip-2pt\gg\hskip-2pt T_{\rm K}$.
We consider the half-filled case in infinite dimensions or in the limit $1/D\hskip-2pt\rightarrow\hskip-2pt 0$ as the most typical case.
Assuming that $T\hskip-2pt>\hskip-2pt T_{\rm N}$, where $T_{\rm N}$ is the N\'{e}el temperature to be determined, 
we consider $\chi_s(0,{\bm q})$ given by Eq.\hskip2pt(\ref{EqChiKondo}) with $I_s(0,{\bm q})$ given by Eq.\hskip2pt(\ref{EqThreeJ}). 
As $1/D\hskip-2pt\rightarrow\hskip-2pt 0$, $k_{\rm B}T_{\rm K}\hskip-2pt=\hskip-2pt O(|J|/D)\rightarrow 0$ and $\Lambda(0,{\bm q})\hskip-2pt\rightarrow\hskip-2pt 0$, because they are of higher order in $1/D$. Since $T/T_{\rm K}\hskip-2pt\rightarrow\hskip-2pt +\infty$, $J_Q(0,{\bm q})\hskip-2pt\rightarrow\hskip-2pt 0$ and $\tilde{\chi}_s(0)\hskip-2pt =\hskip-2pt 1/(k_{\rm B}T)$. Then,
\begin{align} \label{EqCW-Law1}
&\chi_s(0,{\bm q}) = 
1/\bigl[k_{\rm B}T - (1/4)J_s(0,{\bm q}) \bigr],
\\ \label{EqCW-Law2} & \hskip3pt
T_{\rm N} = J_s(0,{\bm Q})/(4k_{\rm B})
=|J|/(2k_{\rm B}),
\end{align}
where ${\bm Q}\hskip-2pt=\hskip-2pt(\pm 1, \pm 1, \cdots, \pm 1)(\pi/a)$. 
The static susceptibility $\chi_s(0,{\bm q})$ obeys the CW law of local-moment magnetism due to the $T$ dependence of the single-site $\tilde{\chi}_s(0)$. 
Both $\chi_s(0,{\bm q})$ and $T_{\rm N}$ are in agreement with those in the mean-field approximation for the Heisenberg model. 
These agreements are reasonable because 
either the $1/D$ expansion theory or the mean-field approximation is rigorous in the limit $1/D\hskip-2pt\rightarrow\hskip-2pt0$.

If $T>T_{\rm N}$ and $T\gg T_{\rm K}$, $\chi_s(0,{\bm q})$ is approximately given by Eq.\hskip2pt(\ref{EqCW-Law1}) even for a finite $D$.
If $T\gg T_{\rm K}$, local moments form in either low or high dimensions.

The Kondo energy $k_{\rm B}T_{\rm K}$ is also the energy scale of quantum spin fluctuations. 
If $T\hskip-2pt\ll\hskip-2pt T_{\rm K}$, the ensemble of electrons or spins behaves as a liquid, i.e., an electron or spin liquid. If $T\hskip-2pt\gg\hskip-2pt T_{\rm K}$, electrons or spins behave as local moments.
Itinerant-electron magnetism and local-moment magnetism are characterized by $T_{\rm N}\hskip-2pt\ll\hskip-2pt T_{\rm K}$ and $T_{\rm N}\hskip-2pt\gg\hskip-2pt T_{\rm K}$, respectively.
In sufficiently low dimensions, $T_{\rm N}\hskip-2pt\ll\hskip-2pt T_{\rm K}$ and itinerant-electron magnetism appears in either an electron model such as the Hubbard model or a spin model such as the Heisenberg model. Magnetism in the RVB electron or spin liquid is a typical type of itinerant-electron magnetism.
In sufficiently high dimensions, $T_{\rm N}\hskip-2pt\gg\hskip-2pt T_{\rm K}$ and local-moment magnetism appears in either an electron or spin model.
Magnetism in infinite dimensions is a typical type of local-moment magnetism.

According to Eqs.\hskip2pt(\ref{EqChiKondo}) and (\ref{EqThreeJ}), the possible mechanisms of the CW law are only the $T$ dependences of $\tilde{\chi}_s(0)$, $J_Q(0,{\bm q})$, and $\Lambda(0,{\bm q})$.
According to the self-consistent renormalization theory (SCR) of spin fluctuations,\cite{SCR1,SCR2,moriya} the mode-mode coupling term becomes smaller as $T$ decreases in certain cases. 
If $\Lambda(0,{\bm q})$ deceases linearly against $T$ as $T$ decreases, the $T$ dependence of $\Lambda(0,{\bm q})$ gives the CW law; the ${\bm q}$ dependence of $\Lambda(0,{\bm q})$ has to be small. 
If critical fluctuations develop as $T$ decreases, $\Lambda(0,{\bm q})$ increases as $T$ decreases and the $T$ dependence of $\Lambda(0,{\bm q})$ suppresses the CW law, as in the Hubbard model on the square lattice. 
It is interesting to examine which occurs because of $\Lambda(0,{\bm q})$ in each actual system, the CW law or the suppression of the CW law.

If $T_N\hskip-2pt \ll\hskip-2pt T_{\rm K}$ and $T_N\hskip-2pt<T\hskip-2pt\ll\hskip-2pt T_{\rm K}$, $J_Q(0,{\bm q})$ increases almost linearly as $T$ deceases in the two cases of ${\bm q}\hskip-2pt=\hskip-2pt0$ and ${\bm q}\hskip-2pt=\hskip-2pt{\bm Q}_N$ discussed above. 
The $T$ dependence of $J_Q(0,{\bm q})$ gives the CW law of itinerant-electron magnetism,\cite{FJO-supJ,miyai} which holds only for particular ${\bm q}$'s in the vicinity of the ordering wave number.
If $T_N\hskip-2pt \gg\hskip-2pt T_{\rm K}$ and $T\hskip-2pt>\hskip-2pt T_N$, the $T$ dependence of $\tilde{\chi}_s(0)$ gives the CW law of local-moment magnetism, in which the Curie constant does not depend on ${\bm q}$ but the Weiss temperature depends on ${\bm q}$, as shown in Eq.\hskip2pt(\ref{EqCW-Law1}).

Either the mechanism due to the $T$ dependence of $J_Q(0,{\bm q})$ for particular ${\bm q}$'s or that of the single-site $\tilde{\chi}_s(0)$ for any ${\bm q}$ is of the zeroth order in $1/D$.
The mechanism due to the $T$ dependence of $\Lambda(0,{\bm q})$ is of higher order in $1/D$.

\subsection{Metal-insulator transitions in actual compounds}
\label{SecMIinActual}
In a multiband model, not only antiferromagnetic order but also orbital order is possible.
We assume that $T_{\rm c}\gg T_{\rm K}$, where $T_c$ is the critical temperature of antiferromagnetic or orbital order.
If $T> T_{\rm c}$, then $T\gg T_{\rm K}$, the entropy is $O(k_{\rm B}\ln 2)$ per unit cell, and the static susceptibility obeys the CW law of local-moment magnetism.
The paramagnetic phase at $T> T_{\rm c}$ is simply the Mott insulator. 
The ordered phase at $T\le T_{\rm c}$ is the N\'{e}el state of local moment magnetism or the type of insulator proposed by Slater.\cite{slater}

It is possible that $T_{\rm K}$ substantially depends on the symmetry or type of lattice and the lattice constant.
An MI transition is possible in conjunction with such a lattice effect on $T_{\rm K}$ as a function of an appropriate parameter, such as $T$, $n(\mu)$ or doping, pressure, substitution, and so on, 
between a high-$T_{\rm K}$ metallic phase with $T_{\rm K}\gg T$ on a type of lattice and a low-$T_{\rm K}$ insulating phase with $T_{\rm K}\lesssim T$ on a different type of lattice or the same type of lattice but with a different lattice constant. 
If the antiferromagnetic or orbital order appears in the low-$T_{\rm K}$ insulating phase, it is the N\'{e}el state of local moment magnetism or the type of insulator proposed by Slater;\cite{slater} if not, it is the Mott insulator.
It is expected that this type of MI transition can explain MI transitions observed in many compounds.\cite{tokura}
Similarly, an MM or insulator-insulator (I\hskip1pt I) transition is also possible as a function of an appropriate parameter due to the dependence of $T_{\rm K}$ on the parameter in conjunction with the lattice effect. 

%%\mbox{(I\hskip1pt I)}

\subsection{Normal state for studying low-$T$ ordered phases}
The RVB electron liquid studied in the present research is simply the {\it normal} state at $T>T_c$ for studying possible low-$T$ ordered phases at $T\le T_c$, such as the N\'{e}el state of itinerant-electron magnetism, which is of the zeroth order in $1/D$, and an anisotropic superconducting state, which is of higher order in $1/D$, and so on.
The {\it normal} state proposed by Anderson\cite{Anderson-SC} for high-temperature superconductivity in cuprate oxides has to be the RVB electron liquid studied in the present paper, rather than an exotic Fermi liquid.
The study in the present research confirms the relevance of the normal state assumed in a previous theory of high-temperature superconductivity in cuprate oxides\cite{FJO-SC1,FJO-SC2,FJO-SC3,FJO-SC4} 
and in previous theories of itinerant-electron ferromagnetism \cite{FJO-supJ} and antiferromagnetism;\cite{FJO-AF1,FJO-AF2}
the Kondo energy or the effective Fermi energy in the previous theories has to be understood as that enhanced by the RVB mechanism, if the mechanism is crucial.

%%%%%%%%%%%%%%%%%%%%%%%%%%%%%%%%%%%%%%%%%%%%%%%%%%%%
\section{Conclusions}
\label{SecConclusion}
The Hubbard model is studied.
Every irreducible physical property of the Hubbard model is decomposed into single-site and multisite properties.
The single-site property can be mapped to a local property of the Anderson model that is self-consistently determined to satisfy the mapping condition.
Every single-site property is equal to its corresponding property of the Anderson model.
Certain local properties that are not single-site properties are also equal to their corresponding properties of the Anderson model; e.g.,
the density of state per unit cell of the Hubbard model, which is denoted by $\rho_\mu(\varepsilon)$, is equal to that of the Anderson model.

In the field theory, the superexchange interaction arises from the virtual exchange of a pair excitation of an electron in the upper Hubbard band and a hole in the lower Hubbard band.
If the on-site repulsion is $U$ and the transfer integral between nearest neighbors is $-t/\sqrt{D}$, where $D$ is the dimensionality,
the exchange interaction constant between nearest neighbors is $J/D$, where $J=-4t^2/U$. This $J$ is in agreement with that given by the conventional derivation.
The superexchange interaction is a multisite effect and is a higher-order effect in $1/D$.

The Kondo-lattice theory (KLT) is a perturbative theory based on the mapping to the Anderson model to include multisite terms in terms of intersite mutual interactions. 
%e.g., the superexchange interaction. 
Its {\it unperturbed} state is constructed through the mapping to the Anderson model; in principle, all the single-site terms are rigorously considered in it.
The Kondo temperature or energy, $T_{\rm K}$ or $k_{\rm B}T_{\rm K}$, is defined through the Anderson model.
If $T_{\rm K}\hskip-2pt >\hskip-2pt 0\hskip2pt$K and $T$ is so low that $T\ll T_{\rm K}$, the {\it unperturbed} state is the normal Fermi liquid.
Since every single-site term is of the zeroth order in $1/D$ and multisite terms are of higher order in $1/D$ except for certain types of the conventional Weiss mean field,
KLT is also the $1/D$ expansion theory.

Since a gap never opens at a nonzero temperature $T$ such that $T\hskip-2pt>\hskip-2pt T_c$, where $T_c$ is $0\hskip2pt$K for $D\hskip-2pt =\hskip-2pt 1$ and is the highest critical temperature among possible ones for $D\hskip-2pt\ge \hskip-2pt2$, the density of states $\rho_\mu(\varepsilon)$ is more or less nonzero at $T\hskip-2pt>\hskip-2pt T_c$.
If more or less $\rho_\mu(\varepsilon)\hskip-2pt>\hskip-2pt 0$ or $\rho_\mu(0)\hskip-2pt>\hskip-2pt 0$ is seriously considered, $k_{\rm B}T_{\rm K}$ is also more or less nonzero. 
Nonzero $k_{\rm B}T_{\rm K}$ and $\rho_\mu(0)$ have to be self-consistently determined with multisite effects to satisfy the mapping condition.

Exactly and almost half-filled cases in the strong-coupling region of $U/|t| \hskip-2pt\gg\hskip-2pt 1$ in the grand canonical ensemble are studied on the basis of KLT.
The number of electrons per unit cell is denoted by $n$.
If $|n\hskip-1pt -\hskip-1pt 1| \hskip-2pt\lesssim \hskip-2pt |t|/(DU)$, the resonating-valence-bond (RVB) mechanism is crucial. 
The Fock-type self-energy due to the superexchange interaction is the RVB self-energy. The Kondo energy is substantially enhanced by the RVB self-energy, so that $k_{\rm B}T_{\rm K}\hskip-2pt =\hskip-2pt O(|J|/D)$. 
If the dimensionality $D$ is so mall that $k_{\rm B}T_c \hskip-2pt \ll\hskip-2pt |J|/D$, the low-$T$ phase of $T_c\hskip-2pt <T\hskip-2pt \ll\hskip-2pt T_{\rm K}$ is the RVB electron liquid, which is stabilized by the Kondo effect in conjunction with the RVB mechanism; the RVB electron liquid in one dimension is also the Tomonaga-Luttinger (TL) liquid or the RVB-TL liquid.
The density of states $\rho_\mu(\varepsilon)$ of the RVB electron liquid has a three-peak structure with a narrow midband between the upper and lower Hubbard bands, which corresponds to the three-peak structure with the Kondo peak between two subpeaks in the Anderson model. 
The midband is on the chemical potential within the Hubbard pseudo-gap.
The bandwidth of the midband is $O(k_{\rm B}T_{\rm K})$, $O(|J|/D)$, or $O[t^2/(DU)]$; its spectral weight is $O\bigl[t^2/(DU^2)\bigr]$ per unit cell; and $\rho_\mu(0)\hskip-2pt =\hskip-2pt O(1/U)$. 
Since the midband almost disappears in the Heisenberg limit, the RVB electron liquid in the Heisenberg limit is a quasi-spin liquid.
The quasi-spin liquid shows metallic conductivity.

According to previous studies of the Kondo effect,
the local electron liquid in the Anderson model and the local spin liquid in the $s$-$d$ model adiabatically continue to each other, although the local gauge symmetry does not exist in the Anderson model but exists in the $s$-$d$ model.
According to the scaling theory for the Anderson localization,
if no symmetry breaking occurs at a metal-insulator transition, 
the metallic and insulating phases adiabatically continue to each other. 
On the basis of these previous studies 
and the study in the present paper,
it is proposed that
the RVB electron liquid in the Hubbard model and the RVB spin liquid in the Heisenberg model adiabatically continue to each other, although the local gauge symmetry does not exist in the Hubbard model and the conductivity of the RVB electron liquid is metallic, while the local gauge symmetry exists in the Heisenberg model and the conductivity of the RVB spin liquid is zero.

If $T\hskip-1pt\gtrsim\hskip-1pt T_{\rm K}$, thermal spin fluctuations are more dominant than quantum spin fluctuations, local moments forms, and the entropy is as large as $k_{\rm B}\ln2$ per unit cell. The high-$T$ phase where $T\hskip-1pt\gtrsim\hskip-1pt T_{\rm K}$ is the Mott insulator.
The Mott insulator, which is a high-$T$ phase with a large entropy, never contradicts the RVB liquid, which is a low-$T$ phase with a small entropy.

According to the Bethe-ansatz solution, the half-filled ground state in the canonical ensemble in one dimension is Lieb and Wu's insulator.
The insulating ground state never contradicts 
the RVB-TL liquid at $0\hskip2pt{\rm K}<T\ll |J|/k_{\rm B}$ in the grand canonical ensemble.
It is desirable to determine which is the half-filled ground state in the grand canonical ensemble in one dimension, Lieb and Wu's insulator, a type of insulator different from it, or the RVB-TL liquid.

%The N\'{e}el temperature is denoted by $T_{\rm N}$. Itinerant-electron magnetism and local-meoment magnetism are characterized by $T_{\rm K}\hskip-1pt\gg\hskip-1pt T_{\rm N}$ and $T_{\rm K}\hskip-1pt \ll\hskip-1pt T_{\rm N}$, respectively. Magnetism of the RVB electron or spin liquid, where $T_{\rm K}\hskip-1pt \gg\hskip-1pt T_{\rm N}$, is a typical type of itinerant-electron magnetism. Magnetism in infinite dimensions, where $T_{\rm K}\hskip-1pt \ll\hskip-1pt T_{\rm N}$, is a typical type of local-meoment magnetism.

%The RVB electron liquid studied in the present research is simply the {\it normal} state at $T>T_c$ for studying possible low-$T$ ordered phases at $T\le T_c$, such as the N\'{e}el state of itinerant-electron magnetism, which is of the zeroth order in $1/D$, and an anisotropic superconducting state, which is of higher order in $1/D$, and so on.

\appendix
%\section{}
%%%%%%%%%%%%%%%%%%%%%%%%%%%%%%%%%%%%%%%%%%%%%%%
\section{Sum Rule for $\Delta(\varepsilon)$ }
\label{AppSumRule}
We consider
\begin{align}\label{EqF1}
F(\varepsilon \hskip-1pt + \hskip-1pt {i}0) 
\hskip-2pt = \hskip-2pt \bigl[\varepsilon 
\hskip-1pt + \hskip-1pt \mu \hskip-1pt -\hskip-1pt \epsilon_d 
\hskip-1pt - \hskip-1pt \Sigma_\sigma(\varepsilon\hskip-1pt + \hskip-1pt{i}0) \hskip-1pt \bigr] 
\hskip-2pt - \hskip-2pt 1/R_{\sigma}(\varepsilon \hskip-1pt + \hskip-1pt {i}0).
\end{align}
According to the mapping condition of Eq.\hskip2pt(\ref{EqMapDelta}), 
\begin{align}\label{EqF2}
{\rm Im}\hskip1pt F(\varepsilon+{i}0)=
- \Delta(\varepsilon).
\end{align} 
Since $\Delta\Sigma_\sigma(\varepsilon+{i}0,{\bm k})\rightarrow 0$ as $\varepsilon\rightarrow\pm\infty$,
\begin{align}\label{EqF3}
\lim_{\varepsilon\rightarrow \pm \infty}F(\varepsilon+{i}0) &
=(2t^2/\varepsilon) + O\bigl(1/\varepsilon^2\bigr).
\end{align}
Since $F(\varepsilon+{i}0)$ is analytic in the upper-half complex plane, 
according to Eqs.\hskip2pt(\ref{EqF2}) and (\ref{EqF3}),
\begin{align}\label{EqSumRuleA}
\int_{-\infty}^{+\infty} \hskip-10pt d\varepsilon \Delta(\varepsilon)
= 2\pi t^2 .
\end{align}
%
%This is the sum rule for $\Delta(\varepsilon)$ of the mapped Anderson model.

%$\rho_\mu(\varepsilon)\hskip-2pt=\hskip-2pt\rho_\mu(-\varepsilon)$, $\Delta(\varepsilon)\hskip-2pt=\hskip-2pt\Delta(-\varepsilon)$, 
%%%%%%%%%%%%%%%%%%%%%%%%%%%%%%%%%%%%%%%%%%%%%%%%%%%%%%%%%%%%%%
\section{Theoretical Constraint for $1/\tilde{\phi}_1$ }
\label{AppAppSumRule}
\subsection{Lower limit of $1/\tilde{\phi}_1$}
\label{AppAppSumRule1}
We assume that 
$U/|t|\hskip-2pt\gg\hskip-2pt 1$, $\mu\hskip-2pt=\hskip-2pt\mu_0$, and $T_c\hskip-2pt<\hskip-2pt T\hskip-2pt\ll\hskip-2pt T_{\rm K}$. 
The density of states $\rho_\mu(\varepsilon)$ has a symmetric three-peak structure with the midband between the Hubbard bands.
% $\Delta(\varepsilon)$ also has a symmetric three-peak structure.

We consider the contribution from the midband to the integration in Eq.\hskip2pt(\ref{EqSumRuleA}).
According to Eq.\hskip2pt(\ref{EqExpansionAM}), $|{\rm Im}\Sigma_\sigma(\hskip-1pt+{i}0)|\hskip-2pt=\hskip-2pt O\bigl(\tilde{\phi}_1k_{\rm B}T^2/T_{\rm K}\hskip-1pt\bigr)$; it can be ignored for $T\hskip-2pt\ll\hskip-2pt T_{\rm K}$.
According to Eq.\hskip2pt(\ref{EqMapCondA}), since ${\rm Re\hskip1pt}R_{\sigma}(+{i}0)\hskip-2pt=\hskip-2pt 0$, $\Delta(0)\hskip-1pt=\hskip-1pt 1/[\pi\rho_\mu(0)]$.
According to Eqs.\hskip2pt(\ref{EqRhoNum1}) and (\ref{EqUregion31C}),
$1/\rho_\mu(0)\hskip-2pt=\hskip-2pt O\bigl(\tilde{\phi}_1k_{\rm B}T_{\rm K}\bigr)$ and
$k_{\rm B}T_{\rm K}\hskip-2pt =\hskip-2pt O\bigl[t^2/(DU)\bigr]$,
respectively. Then, 
\begin{align}\label{EqPeakHeighgt}
\Delta(0) &
=O\bigl[(\tilde{\phi}_1 t^2 )/(DU)\bigr].
\end{align}
Since $\tilde{\phi}_1\gg 1$, $\Delta(0)$ is large; $\Delta(\varepsilon)$ has to have a peak at $\varepsilon=0$ in order to satisfy Eq.\hskip2pt(\ref{EqSumRuleA}).
Since the peak width is $O(k_{\rm B}T_{\rm K})$ or $O[t^2/(DU)]$ and the peak height is given by (\ref{EqPeakHeighgt}), 
the contribution from the midband is as large as
\begin{align}\label{EqSumRuleMid}
\int_{-O(k_{\rm B}T_{\rm K})}^{+O(k_{\rm B}T_{\rm K})}\hskip-20pt
 d \varepsilon \Delta(\varepsilon)
&= O\bigl[\tilde{\phi}_1 t^4/(DU)^2\bigr].
\end{align}
According to the sum rule of Eq.\hskip2pt(\ref{EqSumRuleA}), Eq.\hskip2pt(\ref{EqSumRuleMid}) has to be smaller than $2\pi t^2$. 
Then, 
\begin{align}\label{EqUpperLimit}
1/\tilde{\phi}_1 \ge O[t^2/(DU)^2]. 
\end{align}

%This is a theoretical lower limit for $1/\tilde{\phi}_1$.

We consider the contribution from the Hubbard bands to the integration in Eq.\hskip2pt(\ref{EqSumRuleA}).
Since $\rho_\mu(\varepsilon)\hskip-2pt=\hskip-2pt -(1/\pi){\rm Im}R_{\sigma}(\varepsilon+{i}0)$ has peaks at $\varepsilon\hskip-2pt\simeq\hskip-2pt\pm U/2$,
${\rm Re}\hskip1ptR_{\sigma}(\varepsilon+{i}0)\hskip-2pt=\hskip-2pt0$ for $\epsilon\hskip-2pt=\hskip-2pt\epsilon_{\pm}\hskip-2pt\simeq\hskip-2pt \pm U/2$.
The peak height and bandwidth are
$\rho_\mu(\epsilon_{\pm})\hskip-2pt = \hskip-2ptO(1/|t|)$ and
$W_{\rm H}\hskip-2pt=\hskip-2pt O(|t|)$, respectively.\cite{hubbard3}
According to Eq.\hskip2pt(\ref{EqMapCondA}),
since $\bigl| {\rm Im}\Sigma_\sigma(\epsilon_{\pm}\hskip-1pt+\hskip-1pt{i}0)\bigr|\hskip-2pt=\hskip-2pt O(|t|)$,
$\Delta(\epsilon_{\pm}) \hskip-2pt =\hskip-2pt O(|t|)$.
The contribution from the Hubbard bands is as large as
$\Delta(\epsilon_{\pm}) W_{\rm H}\hskip-2pt =\hskip-2pt O\left(t^2\right)$,
which is consistent with Eq.\hskip2pt(\ref{EqSumRuleA}).

%%%%%%%%%%%%%%%%%%%%%%%%%%%%%%%%%%%%%%%%%%%%%%%%%%%%%%%%%%%%
\subsection{Asymptotic behavior of $1/\tilde{\phi}_1$}
\label{AppAppSumRule2}
We consider $\Delta(\varepsilon)$ defined by Eq.\hskip2pt(\ref{EqAndersonD}). It is given in terms of $V_{\bm k}$ and $E_{c}({\bm k})$.
Within KLT, it can be assumed without the loss of generality that $V_{\bm k}$ is constant: $V_{\bm k}=V$. Then,
\begin{align}
& \Delta(\varepsilon) = \pi |V|^2 \rho_c(\varepsilon),
\\
& \rho_c(\varepsilon)=
\frac1{\tilde{L}}\sum_{\bm k} \delta\left[\varepsilon+\tilde{\mu}-E_{c}({\bm k})\right].
\end{align}
According to the sum rule of Eq.\hskip2pt(\ref{EqSumRuleA}), 
\begin{align}
\pi |V|^2=2t^2.
\end{align}

In the $s$-$d$ or Heisenberg limit of $U/|t|\rightarrow +\infty$, with $J=-4t^2/U$ kept constant, the Anderson model can be mapped to the $s$-$d$ model with the $s$-$d$ exchange interaction constant of
\begin{align}
\tilde{J}_{s\mbox{-}d}= -4|V|^2/U 
= - (8/\pi)(t^2/U).
\end{align}
The dimensionless coupling constant is given by
\begin{align}\label{EqG-s-d}
\tilde{\mathrm{g}}(\varepsilon)= \tilde{J}_{s\mbox{-}d}\tilde{\rho}_c(\varepsilon)
=- [4/(\pi U)] \Delta(\varepsilon).
\end{align}
If $\tilde{\mathrm{g}}(\varepsilon)$ is constant as a function of $\varepsilon$, $T_{\rm K}$ is given by
\begin{align}\label{EqMostDivTK} 
k_{\rm B}T_{\rm K} = W_c e^{-1/|\tilde{\mathrm{g}}(0)|},
\end{align}
in the most-divergent approximation\cite{abrikosov},
where $W_c$ is a half of the conduction bandwidth.

The energy dependence of $\tilde{\mathrm{g}}(\varepsilon)$ has to be seriously considered in the mapped $s$-$d$ model.
According to the scaling theory for the $s$-$d$ model,\cite{poorman,wilsonKG}
high-energy processes substantially renormalize fixed-point or eventual low-energy properties, but they can cause no symmetry breaking.
Thus, whether the eventual $k_{\rm B}T_{\rm K}$ is zero or nonzero depends on whether the bare $\tilde{\mathrm{g}}(0)$ is zero or nonzero. If $\tilde{\mathrm{g}}(0)>0$, the eventual $k_{\rm B}T_{\rm K}$ is nonzero; if $\tilde{\mathrm{g}}(0)=0$, the eventual $k_{\rm B}T_{\rm K}$ is zero.

According to Eqs.\hskip2pt(\ref{EqPeakHeighgt}) and (\ref{EqG-s-d}), it follows that
\begin{align}
|\tilde{\mathrm{g}}(0)|=
O\bigl[\tilde{\phi}_1t^2/\bigl(DU^2\bigr)\bigr].
\end{align}
If $|\tilde{\mathrm{g}}(0)|\rightarrow 0$ as $U/|t|\rightarrow +\infty$, it is inconsistent with nonzero $k_{\rm B}T_{\rm K}$ in the Heisenberg limit; if $|\tilde{\mathrm{g}}(0)|\rightarrow +\infty$ as $U/|t|\rightarrow +\infty$, it is inconsistent with Eq.\hskip2pt(\ref{EqUpperLimit}).
Since $k_{\rm B}T_{\rm K}\hskip-2pt =\hskip-2pt O\bigl(|J|/D\bigr)$, according to Eq.\hskip2pt(\ref{EqMostDivTK}), $\tilde{\mathrm{g}}(0)$ has to be the zeroth order in $1/D$. Since $|\tilde{\mathrm{g}}(0)|$ has to be nonzero and finite in the limit of either $U/|t|\rightarrow+\infty$ or $1/D\hskip-2pt\rightarrow\hskip-2pt 0$, 
\begin{align}\label{EqAsympFinal}
1/\tilde{\phi}_1 
=O\bigl[t^2/(D U^2)\bigr].
\end{align}

%%%%%%%%%%%%%%%%%%%%%%%%%%%%%%%%%%%%%%%%%%%%%%%%%%%%%%%%%%%%%%
\section{Proof of an Equality}
\label{AppEq}
There is a useful relationship between $\Xi_D$ defined by Eq.\hskip2pt(\ref{EqXiD}) and $\pi_{xx}(0)$ defined by Eq.\hskip2pt(\ref{EqLowerPiJD}), as studied below. In the presence of magnetic impurities,
%$\Xi_D$ and $\pi_{xx}(0)$ are given by
%
\begin{align}\label{EqIntC1}
\Xi_D & =
-\frac1{\pi L} \sum_{{\bm k}}
\varphi_D({\bm k}) \hskip-2pt
\int_{-\infty}^{+\infty}\hskip-12pt d\epsilon 
f_+(\epsilon)
\hskip1pt{\rm Im}\hskip1pt \overline{\mathrm{g}}_\sigma(\epsilon+{i}0,{\bm k})
\nonumber \\ &=
-\frac{\sqrt{D}}{\pi L} \sum_{{\bm k}} \cos(k_1a) \hskip-2pt
\int_{-\infty}^{+\infty}\hskip-12pt d\epsilon 
f_+(\epsilon)
\hskip1pt{\rm Im}\hskip1pt \overline{\mathrm{g}}_\sigma(\epsilon+{i}0,{\bm k}),
\end{align}
\begin{align}\label{EqStaticPi}
\pi_{xx}(0) &= 
\frac{2}{\pi L} \sum_{\bm k} \sin^2(k_1 a) 
\int_{-\infty}^{+\infty} \hskip-12pt d\epsilon 
f_+(\epsilon) \bigl[{\rm Im}\hskip1pt \overline{\mathrm{g}}_\sigma(\epsilon+{i}0,{\bm k})\bigr] 
%%%%%%%%%%%%%%%%%%%%%%%%%%%%%%%%%%%%%
\nonumber \\ & \hskip10pt\times
\bigl[{\rm Re} \hskip1pt \overline{\mathrm{g}}_\sigma(\epsilon+{i}0,{\bm k})\bigr].
\end{align}
%
%where $k_x$ is denoted by $k_1$.
Equation\hskip2pt(\ref{EqIntC1}) is also given in the integration form by
\begin{align}\label{EqIntC2}
\Xi_D &=
\frac{\sqrt{D}a^D}{\pi(2\pi)^D} \hskip-1pt
\hskip-2pt\int_{-\pi/a}^{+\pi/a} \hskip-15pt dk_1 
\cdots \hskip-3pt \int_{-\pi/a}^{+\pi/a} \hskip-15pt dk_D \hskip0pt
\cos(k_1a) \hskip-2pt
%%%%%%%%%%%%%%%%%%%%%%%%%%%%%%%%%%%%%
\nonumber \\ & \hskip10pt \times
\int_{-\infty}^{+\infty} \hskip-12pt d\epsilon f_+(\epsilon)
\hskip1pt{\rm Im}\hskip1pt \overline{\mathrm{g}}_\sigma(\epsilon+{i}0,{\bm k}).
\end{align}
By the partial integration of Eq.~(\ref{EqIntC2}) with respect to $k_1$, 
\begin{align}\label{EqIntS1}
\Xi_D &= 
2t^*
\frac{a^D}{\pi (2\pi)^D} \hskip-2pt\int_{-\pi/a}^{+\pi/a} \hskip-20pt dk_1 \cdots \hskip-2pt\int_{-\pi/a}^{+\pi/a} \hskip-20pt dk_D\hskip2pt
\hskip-2pt\sin^2(k_1a)\hskip-2pt 
%%%%%%%%%%%%%%%%%%%%%%%%%%%%%%%%%%%%%
\nonumber \\ & \hskip-10pt \times \hskip-3pt
\int_{-\infty}^{+\infty} \hskip-15pt d\epsilon 
f_+(\varepsilon) 
\bigl[ {\rm Im}\hskip1pt \overline{\mathrm{g}}_\sigma(\epsilon \hskip-1pt + \hskip-1pt {i}0,{\bm k})\bigr]
\bigl[ {\rm Re} \hskip1pt \overline{\mathrm{g}}_\sigma(\epsilon \hskip-1pt + \hskip-1pt {i}0,{\bm k})\bigr].
\end{align}
This is also given in the sum form by
\begin{align}
\hskip-2pt
\Xi_D &=
2t^*\frac{2}{\pi L}\sum_{\bm k} \sin^2(k_1a) \hskip-2pt
\int_{-\infty}^{+\infty} \hskip-15pt d\epsilon 
f_+(\varepsilon) 
\bigl[ {\rm Im}\hskip1pt \overline{\mathrm{g}}_\sigma(\epsilon \hskip-1pt + \hskip-1pt {i}0,{\bm k})\bigr]
%%%%%%%%%%%%%%%%%%%%%%%%%%%%%%%%%%%
\nonumber \\ & \hskip10pt \times
\bigl[ {\rm Re} \hskip1pt \overline{\mathrm{g}}_\sigma(\epsilon \hskip-1pt + \hskip-1pt {i}0,{\bm k})\bigr].
\end{align}
It immediately follows that
\begin{align}\label{EqXiPi}
\Xi_D &
=2 t^* \pi_{xx}(0) .
\end{align}

%\noindent Check: 
%\[ \pi_{xx}(0) = \frac{(\tilde{\phi}_1/2) \Xi_D}
%{t - (3/8)\tilde{\phi}_1 \tilde{W}_s^2 \Xi_{D}(J/D)} \]

%\[ 1 + \frac{3J}{4D}\tilde{W}_s^2\pi_{xx}(0) = 1 + \frac{3J}{4D}\tilde{W}_s^2
%\frac{(\tilde{\phi}_1/2)\Xi_D}{t - \frac{3}{8}\tilde{\phi}_1 \tilde{W}_s^2 \Xi_{D}\frac{J}{D}} \]

%\[\frac{t}{t - \frac{3}{8}\tilde{\phi}_1 \tilde{W}_s^2 \Xi_{D}\frac{J}{D}}
%= \frac1{\tilde{\phi}_1} \frac{t}{t/\tilde{\phi}_1 - \frac{3}{8} \tilde{W}_s^2 \Xi_{D}\frac{J}{D}}
%= \frac{t}{\tilde{\phi}_1 t^*} \]

\section{On the Rigorousness of S$^3$A}
\label{AppDMFT}
Not only all the single-site terms but also four types of the conventional Weiss mean field, which are multisite terms, 
are of the zeroth order in $1/D$: spin density wave or magnetism and orbital order,\cite{comOrbital} which are possible for $U/|t|\hskip-2pt >\hskip-2pt 0$, and charge density wave and isotropic $s$-wave or BCS superconductivity, which are possible for $U/|t|\hskip-2pt<\hskip-2pt 0$.
All the other multisite terms are of higher order in $1/D$.
Since S$^3$A can treat no conventional Weiss mean field or no symmetry breaking, 
S$^3$A is not necessarily rigorous even in the limit $1/D\hskip-2pt\rightarrow\hskip-2pt 0$; it is expected that S$^3$A will be rigorous in the limit $1/D\hskip-2pt\rightarrow\hskip-2pt 0$, if the Hilbert space is constrained within the subspace where no symmetry is allowed to be broken.
The purpose of this Appendix is to examine whether S$^3$A is rigorous in the limit $1/D\hskip-2pt\rightarrow\hskip-2pt 0$ within the Hilbert subspace.
We consider the Hubbard model of Eq\hskip2pt(\ref{EqHubbardModel}); the absolute bandwidth of it is $4\sqrt{D}|t|$.
We denote $\tilde{\phi}_1$ under S$^3$A by $\tilde{\phi}_{{\rm S}^3{\rm A}}$, $\tilde{\phi}_{{\rm S}^3{\rm A}}(T)$, or $\tilde{\phi}_{{\rm S}^3{\rm A}}(D,T)$, depending on necessity or sufficiency; 
$\tilde{\phi}_1$ is used for that beyond S$^3$A or that of KLT.

If $n(\mu)\hskip-2pt\ne \hskip-2pt1$, the Mott insulator at $T\hskip-2pt =\hskip-2pt 0\hskip2pt$K is impossible either under or beyond S$^3$A; $1/\tilde{\phi}_{{\rm S}^3{\rm A}}\hskip-2pt >\hskip-2pt 0$ and $1/\tilde{\phi}_{1}\hskip-2pt >\hskip-2pt 0$.
Since $1/\tilde{\phi}_1\hskip-2pt \rightarrow\hskip-2pt 1/\tilde{\phi}_{{\rm S}^3{\rm A}}\hskip-2pt >\hskip-2pt 0$ as $1/D \rightarrow 0$, S$^3$A is equivalent to KLT in the limit $1/D\hskip-2pt\rightarrow\hskip-2pt 0$ and rigorous in the limit $1/D\hskip-2pt\rightarrow\hskip-2pt 0$ within the Hilbert subspace.
Then, the key issue is whether the Mott insulator is possible at $T\hskip-2pt =\hskip-2pt 0\hskip2pt$K for the half filling.
In the following part, 
we assume the half filling in the canonical or grand canonical ensemble: $N\hskip-2pt=\hskip-2pt L$ or $n(\mu)\hskip-2pt=\hskip-2pt 1$.

%for $N\hskip-2pt=\hskip-2pt L$ or $n(\mu)\hskip-2pt=\hskip-2pt1$.
% the study cannot deny the possibility that the ground state is an insulator.\cite{com3rdLaw}

There is a well-known scenario for the Mott insulator and transition, which is based on previous studies.\cite{mott,hubbard1,hubbard3,brinkman}
If the absolute bandwidth is $4\sqrt{D}|t|$,
the ground state is a metal for $U\hskip-2pt\lesssim\hskip-2pt 4\sqrt{D}|t|$ and is the Mott insulator at least for $U\hskip-2pt>\hskip-2pt 4\sqrt{D}|t|$.
If $T\hskip-2pt=\hskip-2pt 0\hskip2pt$K, e.g.,
the Mott transition occurs at $U\hskip-2pt\simeq\hskip-2pt 4\sqrt{D}|t|$ as a function of $U$.
A complete gap opens in the Mott insulator at $T\hskip-2pt=\hskip-2pt 0\hskip2pt$K; the gap is as large as $U\hskip-2pt-\hskip-2pt 4\sqrt{D}|t|$.
The numerical study,\cite{RevModDMFT} which is under S$^3$A, confirms this scenario; in addition, it shows that hysteresis appears in the Mott transition.
On the other hand, the study in Sect.\hskip2pt\ref{SecNatureMott}, which is beyond S$^3$A, shows that what is stabilized in the nonzero and low-$T$ phase of $0\hskip2pt{\rm K} \hskip-2pt<\hskip-2pt T\hskip-2pt\ll |J|/(k_{\rm B}D)$ for $U/|t|\hskip-2pt\gg\hskip-2pt 1$ is not the Mott insulator but the RVB liquid. 
The scenario is not relevant beyond S$^3$A, at least for $0\hskip2pt{\rm K} \hskip-2pt<\hskip-2pt T\hskip-2pt\ll |J|/(k_{\rm B}D)$. 
Since the RVB mechanism, which appears only beyond S$^3$A, is never considered in any stage or aspect of the scenario,
it is expected that the scenario will be relevant only under S$^3$A, even if it is relevant.
In the following part, we assume that $U\hskip-2pt>\hskip-2pt 4\sqrt{D}|t|$.

If the Mott insulator is possible at $T\hskip-2pt=\hskip-2pt 0\hskip2pt$K in the canonical ensemble, it is possible only for $N\hskip-2pt = \hskip-2pt L$ but never for $N\hskip-2pt \ne \hskip-2pt L$.
If so, and if the reservoir effect is only {\it implicitly} treated through the statistical average, the Mott insulator is also possible at $T\hskip-2pt=\hskip-2pt 0\hskip2pt$K in the grand canonical ensemble; the insulator is {\it rigid} against the movement of $\mu$.
However, since the single-site $\Sigma_\sigma(\varepsilon\hskip-0pt+\hskip-0pt{\rm i0})$ has to have a pole at $\varepsilon\hskip-2pt=\hskip-2pt 0$ or just on $\mu$, as discussed in Sect.\hskip2pt\ref{SecNatureMott}, 
the Mott insulator cannot be {\it rigid} against the movement of $\mu$.
If the Mott insulator at $T\hskip-2pt=\hskip-2pt0\hskip2pt$K is possible, 
there is inconsistency between its possibility and {\it rigidity}. 
The inconsistency means or implies that the Mott insulator at $T\hskip-2pt=\hskip-2pt0\hskip2pt$K is impossible at least in the grand canonical ensemble, even if it is possible in the canonical ensemble. 
The critical argument given above is almost in parallel with that given in Sect.\hskip2pt\ref{SecNatureLW} against the possibility of Lieb and Wu's insulator.
In the following part, we assume the grand canonical ensemble.

%against the possibility of the Mott insulator at $T\hskip-2pt=\hskip-2pt0\hskip2pt$K

The Mott insulator at $T\hskip-2pt=\hskip-2pt 0\hskip2pt$K is infinitely degenerate.\cite{FJO-MottIns}
In general, the ground state that is infinitely degenerate is unstable even against an infinitesimal perturbation;
the third law of thermodynamics is mainly based on this fact.
It is doubtful whether the Mott insulator at $T\hskip-2pt=\hskip-2pt 0\hskip2pt$K is stable, particularly in the {\it explicit} presence of an electron reservoir.
This critical argument can be confirmed at least in the simple reservoir model, as studied in Sect.\hskip2pt\ref{SecNatureLW}.

% in Sect.\hskip2pt\ref{SecNatureLW} of the present paper, e.g., it is shown that the Mott insulator at $T\hskip-2pt=\hskip-2pt 0\hskip2pt$K is not stable in the simple reservoir model, in which the translational symmetry is restored.

Either of the two critical arguments above casts doubt that the well-known scenario is not relevant for $T\hskip-2pt=\hskip-2pt 0\hskip2pt$K, even under S$^3$A.
We do not exclude another scenario that the Mott insulator is impossible at $T\hskip-2pt=\hskip-2pt 0\hskip2pt$K, even under S$^3$A.

If $T\hskip-2pt >\hskip-2pt 0\hskip2pt$K, more or less $\rho_\mu(0)\hskip-2pt >\hskip-2pt 0$; the single-site $\Sigma_\sigma(\varepsilon\hskip-2pt+\hskip-2pt{i}0)$ can have no pole on the real axis and can be expanded in the form of Eq.\hskip2pt(\ref{EqExpansionAM}).
Thus, more or less $1/\tilde{\phi}_{{\rm S}^3{\rm A}}(T)\hskip-2pt>\hskip-2pt0$ for $T\hskip-2pt >\hskip-2pt 0\hskip1pt$.
If the ground state is the Mott insulator under S$^3$A, $1/\tilde{\phi}_{{\rm S}^3{\rm A}}(T)\hskip-2pt\rightarrow \hskip-2pt 0$ as $T\hskip-2pt\rightarrow\hskip-2pt 0\hskip2pt{\rm K}$ and $1/\tilde{\phi}_{{\rm S}^3{\rm A}}(0\hskip2pt{\rm K})\hskip-2pt=\hskip-2pt0$.
If the ground state is a metal under S$^3$A, $1/\tilde{\phi}_{{\rm S}^3{\rm A}}(T)\hskip-2pt>\hskip-2pt 0$ for $T\ge 0\hskip2pt{\rm K}$. 
On the other hand, if the RVB mechanism is considered, $1/\tilde{\phi}_1\hskip-2pt=\hskip-2pt O[t^2/(D U^2)]$, 
as shown in Eq.\hskip2pt(\ref{EqAsympFinal}).
As the asymptotic behavior of $1/\tilde{\phi}_1$ as $U/|t|\hskip-2pt \rightarrow \hskip-2pt+\infty$ and $1/D\hskip-2pt\rightarrow\hskip-2pt 0$ beyond S$^3$A,
it is reasonable to assume that 
\begin{align}\label{EqPhiKLT}
& 1/\tilde{\phi}_1 = \max\bigl\{ O\bigl[t^2/(D U^2)\bigr], 1/\tilde{\phi}_{{\rm S}^3{\rm A}}(T)\bigr\}.
\end{align}
%
%at least if $T\hskip-2pt > \hskip-2pt 0\hskip2pt$K or $1/\tilde{\phi}_{{\rm S}^3{\rm A}}(T)\hskip-2pt > \hskip-2pt 0$.
We call the limit of $1/D\hskip-2pt \rightarrow\hskip-2pt 0$ and $T\hskip-2pt\rightarrow\hskip-2pt 0\hskip2pt{\rm K}$ a double limit.
The issue to be studied is simply whether the double limit is unique; S$^3$A for $T\hskip-2pt =\hskip-2pt 0\hskip2pt$K corresponds to the limit of $1/D\hskip-2pt\rightarrow\hskip-2pt 0$ followed by $T\hskip-2pt\rightarrow\hskip-2pt 0\hskip2pt$K beyond S$^3$A.

First, we assume that the Mott insulator is possible at $T=0\hskip2pt$K under S$^3$A, following the well-known scenario: $1/\tilde{\phi}_{{\rm S}^3{\rm A}}(T)\hskip-2pt \rightarrow \hskip-2pt0$ as $T\hskip-2pt\rightarrow\hskip-2pt0\hskip2pt$K.
If the double limit is taken in such a way that $t^2/(D U^2)\hskip-2pt \ll \hskip-2pt1/\tilde{\phi}_{{\rm S}^3{\rm A}}(T)$ is kept satisfied, KLT is reduced to S$^3$A; the Mott insulator is stabilized.
If the double limit is taken in such a way that $t^2/(D U^2)\hskip-2pt \gg \hskip-2pt 1/\tilde{\phi}_{{\rm S}^3{\rm A}}(T)$ and $t^2/(D U^2)\hskip-2pt \gg \hskip-2pt k_{\rm B}T$ are kept satisfied, KLT is not reduced to S$^3$A; the RVB liquid is stabilized.
The liquid is an extremely bad metal with $k_{\rm B}T_{\rm K}(T)\hskip-2pt=\hskip-2pt O[t^2/(DU)]\hskip-2pt \rightarrow \hskip-2pt 0$, $1/\tilde{\phi}_1\hskip-2pt=\hskip-2pt O[t^2/(D U^2)]\hskip-2pt \rightarrow \hskip-2pt 0$, and $\rho_{\mu}(0)\hskip-2pt=\hskip-2pt O(1/U)\hskip-2pt > \hskip-2pt 0$.
% it is almost an insulator. an abnormal metal
Since there is a slight difference between the two types of limit, the double limit is not unique. 
Thus, S$^3$A is not necessarily equivalent to KLT in the limit $1/D\hskip-2pt\rightarrow\hskip-2pt 0$ nor necessarily rigorous even in the limit $1/D\hskip-2pt\rightarrow\hskip-2pt 0$ within the Hilbert subspace.

Next, we assume that the Mott insulator is impossible at $T\hskip-2pt=\hskip-2pt 0\hskip2pt$K under S$^3$A, against the well-known scenario: $1/\tilde{\phi}_{{\rm S}^3{\rm A}}(D,T)\hskip-2pt > \hskip-2pt 0$ for $T\hskip-2pt \ge \hskip-2pt 0\hskip2pt$K. 
We consider, e.g., the case of $T\hskip-2pt = \hskip-2pt 0\hskip2pt$K.
According to Eq.\hskip2pt(\ref{EqPhiKLT}),
a crossover $D_c$ is defined by 
$t^2/(D_c U^2)\hskip-2pt =\hskip-2pt 1/\tilde{\phi}_{{\rm S}^3{\rm A}}(D_c,0\hskip2pt{\rm K})$; if $\tilde{\phi}_{{\rm S}^3{\rm A}}(D,0\hskip2pt{\rm K})$ depends on the reservoir efficiency $\delta N$ defined by Eq.\hskip2pt(\ref{EqDeltaN}), $D_c$ also depends on it, as will be discussed later.
If $D\hskip-2pt\ll\hskip-2pt D_c$, the ground state is the RVB liquid; $k_{\rm B}T_{\rm K}(0\hskip2pt{\rm K})\hskip-2pt=\hskip-2pt O[t^2/(DU)]$ and $\rho_{\mu}(0)\hskip-2pt=\hskip-2pt O(1/U)$.
If $D\hskip-2pt\gg\hskip-2pt D_c$, the ground state is the normal Fermi liquid in which the RVB mechanism is not crucial; $k_{\rm B}T_{\rm K}(0\hskip2pt{\rm K})\hskip-2pt =\hskip-2pt O\bigl[|t|/\tilde{\phi}_{{\rm S}^3{\rm A}}(D,0\hskip2pt{\rm K})\bigr]$ and $\rho_\mu(0)$ is simply given by that for $U\hskip-2pt=\hskip-2pt 0$.
As a function of $U$, $\rho_\mu(0)$ is not monotonous and is minimal at approximately $U$ such that $t^2/(D U^2)\hskip-2pt =\hskip-2pt 1/\tilde{\phi}_{{\rm S}^3{\rm A}}(D,0\hskip2pt{\rm K})$;
as a function of $D$, $\rho_\mu(0)$ is almost constant for $D\hskip-2pt\lesssim\hskip-2pt D_c$, an increasing function for $D$ such that $D\hskip-2pt\gtrsim\hskip-2pt D_c$ but not $D\hskip-2pt\gg\hskip-2pt D_c$, and almost constant for $D\hskip-2pt\gg\hskip-2pt D_c$. 
It is easy to show that the double limit is unique.
Thus, S$^3$A is equivalent to KLT in the limit $1/D\hskip-2pt\rightarrow\hskip-2pt 0$ and rigorous in the limit $1/D\hskip-2pt\rightarrow\hskip-2pt 0$ within the Hilbert subspace.

If the reservoir effect is only {\it implicitly} treated through the statistical average, $\delta N\hskip-2pt=\hskip-2pt 0$ for $T\hskip-2pt=\hskip-2pt 0\hskip2pt$K, unless the ground state is degenerate between different $N$ states.
If the Mott insulator is possible at $T\hskip-2pt=\hskip-2pt 0\hskip2pt$K for $\delta N\hskip-2pt=\hskip-2pt 0$ under S$^3$A, $1/\tilde{\phi}_{{\rm S}^3{\rm A}}(0\hskip2pt{\rm K})\hskip-2pt >\hskip-2pt 0$ for $\delta N \hskip-2pt>\hskip-2pt 0\hskip2pt$ and $1/\tilde{\phi}_{{\rm S}^3{\rm A}}(0\hskip2pt{\rm K})\hskip-2pt \rightarrow\hskip-2pt 0$ as $\delta N \hskip-2pt\rightarrow\hskip-2pt 0\hskip2pt$ in the presence of an electron reservoir.
If so, the crossover $D_c$ depends on $\delta N$ in such a way that $D_c\hskip-2pt\rightarrow\hskip-2pt +\infty$ as $\delta N\hskip-2pt\rightarrow\hskip-2pt 0$;
S$^3$A is rigorous in the limit $1/D\hskip-2pt\rightarrow\hskip-2pt 0$ within the Hilbert subspace for $\delta N\hskip-2pt>\hskip-2pt 0$ but not for $\delta N\hskip-2pt=\hskip-2pt 0$.

% as demonstrated in the numerical study,\cite{RevModDMFT}

%According to the main body of the present paper, what is stabilized in the half-filled case at $0\hskip2pt{\rm K} \hskip-2pt<\hskip-2pt T\hskip-2pt\ll |J|/(k_{\rm B}D)$, even in the limit $T\hskip-2pt\rightarrow\hskip-2pt 0\hskip2pt$K, beyond S$^3$A is not the Mott insulator but the RVB liquid.

We conclude this Appendix. 
Two critical arguments are given against the possibility of the Mott insulator at $T\hskip-2pt=\hskip-2pt 0\hskip2pt$K, in addition to those given in Sect.\hskip2pt\ref{SecDiscussion}.
Whether S$^3$A is rigorous in the limit $1/D\hskip-2pt\rightarrow\hskip-2pt 0$ within the constrained Hilbert subspace where no symmetry is allowed to be broken depends on whether the Mott insulator is possible at $T\hskip-2pt=\hskip-2pt0\hskip2pt$K for the half filling under S$^3$A.
If possible, S$^3$A is not necessarily rigorous; if not, S$^3$A is rigorous.
It is desirable to determine whether the Mott insulator is possible at $T\hskip-2pt=\hskip-2pt0\hskip2pt$K under S$^3$A, or the third law of thermodynamics can be broken under S$^3$A, particularly in the {\it explicit} presence of an electron reservoir.

%%%%%%%%%%%%%%%%%%%%%%%%%%%%%%%%%%%%%%%%%
%\newpage


\begin{thebibliography}{}
\bibitem{ComKramers}
If $N$ is odd, e.g., the ground state can never be a singlet and the residual entropy can never be zero.
\bibitem{comWidth}
If both $U$ and $t$ are nonzero, it is expected that 
the tails of $D_{N\alpha}(\varepsilon)$, $\rho_{N}(\varepsilon)$, and $\rho_{\mu}(\varepsilon)$ will be infinitely long, because there is no upper limit in many-body eigenenergy $E_{N\alpha}$, as shown in Eq.\hskip2pt(\ref{EqEa-Region}).
\bibitem{comFermiSF}
In Wilson's band insulator, the volume of the Fermi surface is zero in one of the particle and hole pictures but is as large as the volume of the Brillouin zone in the other.
\bibitem{ComFeynmanMethod}
The gap function of a symmetry-broken state below $T_c$ is singular at $U=0$ or $J=0$, depending on the perturbative scheme in terms of $U$ or $J=-4t^2/U$, as $\epsilon_{\rm G}(U)$ of Lieb and Wu's insulator as a function of $U$. This singularity or the symmetry-broken state can be treated by the Feynman-diagram method, if an infinitesimal symmetry-breaking field is introduced, or if the opening of a gap is assumed from the beginning.
\bibitem{comInequlaities}
It is probable that neither $S_1(\varepsilon,{\bm k})$ nor $S_2(\varepsilon,{\bm k})$ is constant as a function of ${\bm k}$.
Thus, it is probable that either Eq.\hskip2pt(\ref{EqInequalityY}) or (\ref{EqInequalityZ}) is positive and no equality holds in either Eq.\hskip2pt(\ref{EqY1P}) or (\ref{EqZ1P}).
\bibitem{FJO-supJ}
F. J. Ohkawa\mbox{,}
Phys. Rev. B {\bf 65}, 174424 (2002).
%Phys. Rev. B {\bf 65} (2002), 174424.
According to this paper, the superexchange interaction can be ferromagnetic if the Hund coupling is large in a multiband model. It is likely that itinerant-electron ferromagnetic instability is possible only in a multiband model.
\bibitem{comT-depTK}
Since $\Xi$ given by Eq.\hskip2pt(\ref{EqXiD}) is a decreasing function of $T$, $k_{\rm B}T_{\rm K}$ is a decreasing function of $T$. The $T$ dependence of $k_{\rm B}T_{\rm K}$ can be ignored in the quantitative discussion in Sect.\hskip2pt\ref{SecDiscussion}.
\bibitem{com3rdLaw}
According to another expression of the third law of thermodynamics, the absolute zero Kelvin $T=0\hskip2pt$K can never be reached. It is plausible that $T>0\hskip2pt$K is implicitly assumed in the statistical average; e.g., the factor of $1/(k_{\rm B}T)$, which appears in many places, is implicitly assumed to be one such that $0\hskip-1pt <\hskip-1pt 1/(k_{\rm B}T)\hskip-1pt <\hskip-1pt +\infty$. If the ground state, which is the state at $T=0\hskip2pt$K, is different from the state in the limit $T\rightarrow 0\hskip2pt$K, the ground state is of no physical significance.
\bibitem{comPhaseDiagram}
If the absolute bandwidth becomes infinite as $1/D\hskip-2pt\rightarrow\hskip-2pt 0$, as in the model of the present paper,
$U_{c2}(0\hskip2pt{\rm K})\hskip-2pt\rightarrow\hskip-2pt +\infty$ as $1/D\hskip-2pt\rightarrow\hskip-2pt 0$ even under S$^3$A.
\bibitem{comKLT1/D} 
The RVB self-energy $\Delta\Sigma_\sigma^{\rm (RVB)}(+{i}0,{\bm k})$ is formally $O(1/D)$ for almost all ${\bm k}$'s. Because of the $D$ dependence of $\tilde{\phi}_1\hskip-2pt =\hskip-1pt O(D)$, it is practically $O\bigl[(1/D)^0\bigr]$ for almost all ${\bm k}$'s. Since self-consistent $\tilde{\phi}_1$ depends on $D$, KLT is not a simple expansion scheme in terms of $1/D$.
\bibitem{comDeltaN}
If $T\hskip-2pt >\hskip-2pt 0\hskip2pt$K, necessarily $\Delta N\hskip-2pt >\hskip-2pt 0$, even if the reservoir effect is implicitly treated through the statistical average.
\bibitem{comInsComp}
An infinitesimally small compressibility is never proof that the ground state is Lieb and Wu's insulator or the Mott insulator.
\bibitem{Com2D}
Since $\tilde{\chi}_s(0)\hskip-2pt \rightarrow\hskip-2pt +\infty$ as $T\rightarrow 0\hskip2pt$K or $k_{\rm B}T_{\rm K}\rightarrow 0$ as $T\rightarrow 0\hskip2pt$K is simply due to the divergence of $\tilde{\phi}_1\rho_\mu(0)$ as $T\hskip-2pt\rightarrow\hskip-2pt 0\hskip2pt$K, the infinitesimal $k_{\rm B}T_{\rm K}$ never means that the condensation energy of the RVB liquid becomes zero as $T\hskip-2pt\rightarrow \hskip-2pt0\hskip2pt$K. The condensation energy is $O(|J|/2)$ per unit cell.
\bibitem{comOrbital}
The orbital order is possible only in a multiband model.
\end{thebibliography}
\end{document}